\documentclass[journal]{IEEEtran}
\usepackage{subfigure}
\usepackage{amsmath}	
\usepackage{graphicx}	
\usepackage{amssymb}
\usepackage{amsfonts}
\usepackage{makecell}
\usepackage{cite}
\usepackage{algpseudocode}  
\usepackage[]{hyperref}
\usepackage{authblk}
\usepackage{amsmath}
\usepackage{amsthm}
\usepackage{setspace}
\usepackage{color, soul}
\usepackage{xcolor}
\usepackage{tcolorbox}
\usepackage{bm}
\usepackage{multirow}

\newtheorem{remark}{Remark}

\newcommand{\RNum}[1]{\uppercase\expandafter{\romannumeral #1\relax}}

\usepackage{stfloats}

\makeatletter  
\newif\if@restonecol  
\makeatother

\usepackage[linesnumbered,ruled,vlined]{algorithm2e}

\newcommand \footnoteONLYtext[1]
{
	\let \mybackup \thefootnote
	\let \thefootnote \relax
	\footnotetext{#1}
	\let \thefootnote \mybackup
	\let \mybackup \imareallyundefinedcommand
}

\begin{document}
	%
	
	\title{Semantics-Empowered Space-Air-Ground-Sea Integrated Network: New Paradigm, Frameworks, and Challenges}
	
	\author{
		Siqi~Meng,~
		Shaohua~Wu,~\textit{Member},~\textit{IEEE},~
		Jiaming~Zhang,~
		Junlan~Cheng,~
		Haibo~Zhou,~\textit{Senior Member},~\textit{IEEE},~
		and~Qinyu~Zhang,~\textit{Senior Member},~\textit{IEEE}
		\thanks{
			This work has been supported in part by the National Key Research and Development Program of China under Grant no. 2020YFB1806403, and in part by the Guangdong Basic and Applied Basic Research Foundation under Grant no. 2022B1515120002, and in part by the National Natural Science Foundation of China under Grant no. 62027802, and in part by the Major Key Project of PCL under Grant no. PCL2024A01, and in part by the Shenzhen Science and Technology Program under Grant no. ZDSYS20210623091808025. \textit{(Corresponding authors: Shaohua Wu; Qinyu Zhang.)}
			
			Siqi Meng, Jiaming Zhang, and Junlan Cheng are with the School of Electronics Engineering, Harbin Institute of Technology (Shenzhen), Shenzhen 518055, China (e-mail: mengsiqi@stu.hit.edu.cn; hitzhangjiaming@163.com; hitsz.chengjunlan@foxmail.com).
			
			Shaohua Wu and Qinyu Zhang are with the Guangdong Provincial Key Laboratory of Aerospace Communication and Networking Technology, Harbin Institute of Technology (Shenzhen), Shenzhen 518055, China, and also with the Peng Cheng Laboratory, Shenzhen 518055, China (e-mail: hitwush@hit.edu.cn; zqy@hit.edu.cn).
			
			Haibo Zhou is with the School of Electronic Science and Engineering, Nanjing University, Nanjing 210023, China (e-mail: haibozhou@nju.edu.cn).

		}
	}
	\maketitle
	\begin{abstract}
		In the coming sixth generation (6G) communication era, to provide seamless and ubiquitous connections, the space-air-ground-sea integrated network (SAGSIN) is envisioned to address the challenges of communication coverage in areas with difficult conditions, such as the forest, desert, and sea. Considering the fundamental limitations of the SAGSIN including large-scale scenarios, highly dynamic channels, and limited device capabilities, traditional communications based on Shannon information theory cannot satisfy the communication demands. Moreover, bit-level reconstruction  is usually redundant for many human-to-machine or machine-to-machine applications in the SAGSIN.  
		Therefore, it is imperative to consider high-level communications  towards semantics exchange,  called semantic communications. 
		In this survey, according to the interpretations of the term ``semantics", including ``significance", ``meaning", and ``effectiveness-related information", we review state-of-the-art works on semantic communications  from three perspectives, which are 1) significance representation and protection, 2) meaning similarity measurement and meaning enhancement, and 3) ultimate effectiveness and effectiveness yielding. Sequentially, three types of semantic communication systems can be correspondingly introduced, namely the significance-oriented, meaning-oriented, and effectiveness/task-oriented  semantic communication systems. Implementation of the above three types of systems in the SAGSIN necessitates a new  perception-communication-computing-actuation-integrated paradigm (PCCAIP), where all the available perception, computing, and actuation techniques jointly facilitate significance-oriented sampling \& transmission, semantic extraction \& reconstruction, and task decision. Finally, we point out some future challenges on semantic communications in the SAGSIN. This survey provides a comprehensive review on the future semantic communications in the SAGSIN, and elaborates on the performance metrics and techniques  related to semantic communications for references and further in-depth investigations.
	\end{abstract}
	\IEEEpeerreviewmaketitle
	\section{Introduction}\label{sec1}
	\subsection{Motivation}
	
	The sixth generation (6G) networks have been envisioned to provide seamless and ubiquitous services in an extremely reliable and energy-efficient manner with ultra low latency\cite{6Gwhite,56Gsurvey,IoTsurvey}. Especially, heterogeneous communication nodes distributed in city, rural area, no man's land, as well as sea, air, and space, are envisioned to be seamlessly and continuously connected via 6G networks. However, the widely used terrestrial networks in fifth generation (5G) communications are facing severe challenges in facilitating such seamless and ubiquitous connections due to the following reasons. Firstly, with the proliferating number of connected devices, the surge of data traffic, and the rapidly growing demand of communication coverage, the terrestrial network alone cannot satisfy such massive and broad connections due to limited communication distance of terrestrial 5G networks. Secondly, almost 90 percent of the Earth area is covered by the sea, forest, and desert where no people live, and the 5G terrestrial networks are not accessible in these areas  for the harsh environment and expensive infrastructure costs. To tackle the above inherent challenges in the current 5G network, terrestrial networks should be extended by integrating the nodes deployed in the space, air, and (or) sea to form a space-air-ground(-sea) integrated network (SAGIN or SAGSIN), which has garnered a surge of interest by researchers\cite{SAGsurvey,sagsin_railway,blockchain,surveySAGSIN,NTN}. 
	
\begin{figure}[t]
		\centering
		\includegraphics[width = 0.48\textwidth]{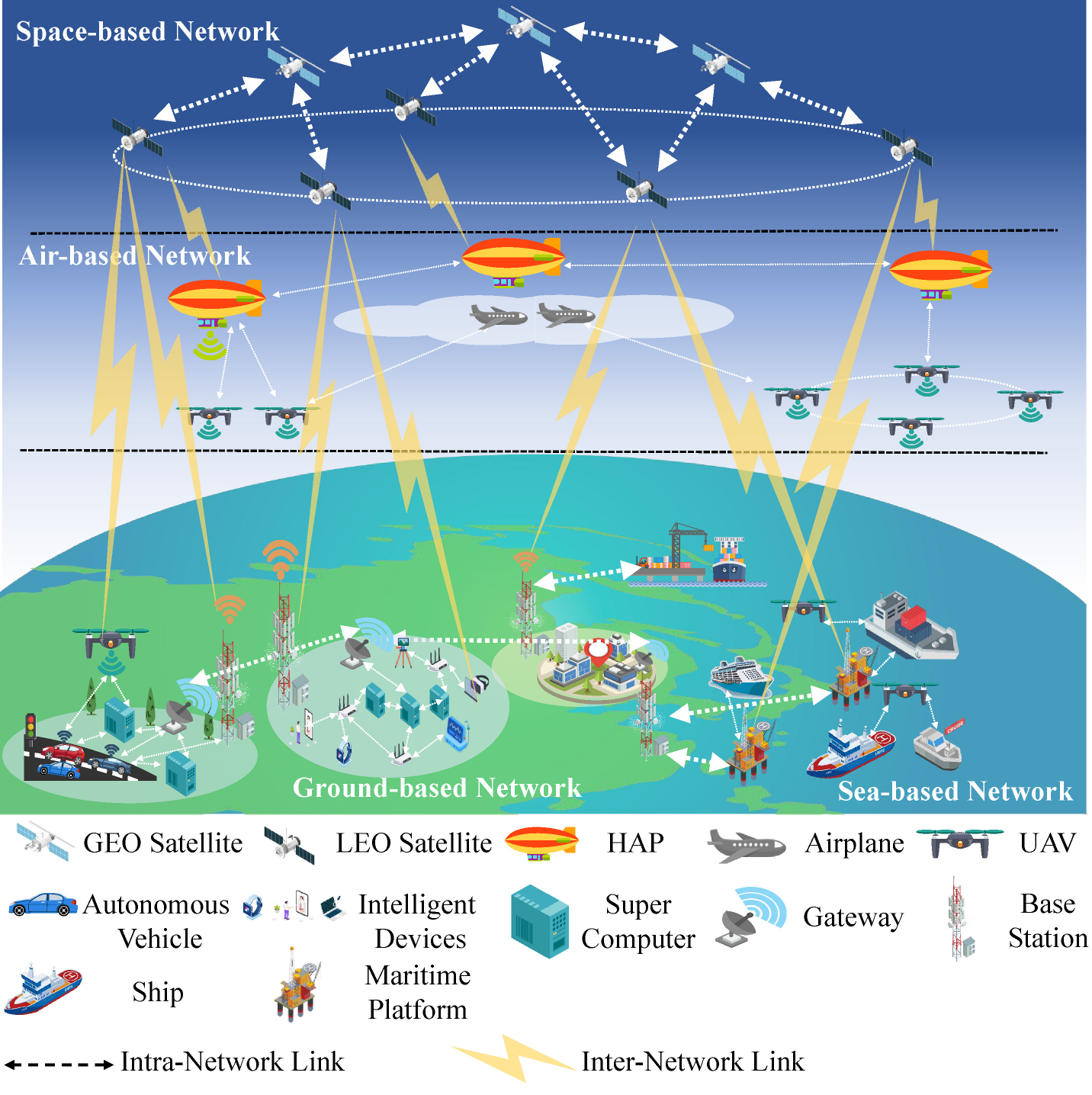}
		\caption{An Illustration of the SAGSIN.}
		\label{fig1}
	\end{figure}

	The SAGSIN is comprised of space-based network, air-based network, and sea-based network along with ground-based (terrestrial) network, as shown in Fig. \ref{fig1}. Specifically, space-based network consists of geostationary Earth orbit (GEO) satellites and low Earth orbit (LEO) satellites, which usually form constellations respectively; air-based network is composed of airplanes, unmanned aerial vehicles (UAVs), and high-altitude platforms (HAPs), which usually appear as platoons; sea-based network includes nodes such as maritime base stations, ships, and unmanned ships; ground-based network has been extensively developed with the advent of cell mobile communications, autonomous driving, and Internet of Things (IoT). By connecting the nodes within a certain part via short or long wireless links to construct the abovementioned four types of networks, and integrating the four parts via inter-network links, the SAGSIN can perfectly reap the promised benefits of 6G communications owing to its massive connections among long-distance nodes.

	However, the realization of 6G communications with extremely high reliability, very low latency, and high energy efficiency based on the SAGSIN will be affected by several fundamental limitations, which are \textit{large-scale scenarios, highly dynamic channels, and limited device capabilities.} Firstly, the communication distance between the transceiver nodes in the SAGSIN scenarios is considerably large in stark comparison with terrestrial communications. For instance, the distance from LEO satellites to ground nodes (e.g. base stations or user equipment) is 500-2,000 kilometers, which causes inherent and non-negligible propagation latency. Secondly, the communication channels over transceivers in the SAGSIN vary rapidly compared with terrestrial channels. Taking UAV-LEO-ground multi-hop communications  as an example, the LEO orbits around the Earth with significantly high velocity, leading to inherent Doppler shift in UAV-LEO and LEO-ground channels; besides, cosmic environment is so harsh that deep fading will appear due to cosmic ray and atmospheric attenuation. Thirdly, the available communication and computing resources are strictly limited for node devices in the SAGSIN. Specifically, due to hardware and battery limitation on the satellites, UAVs, and so on, transmission rate and computing speed of space/air-ground communications are remarkably lower in comparison to terrestrial communications. 
	
	The abovementioned inherent and fundamental characteristics of the SAGSIN impose severe challenges on current traditional communication technologies in facilitating the envisioned  6G communications.
	On the other hand, traditional communications based on classic Shannon information theory are reaching their limits, which may not support communications in the SAGSIN with such severe limitations. Concretely speaking, since Shannon completed his masterpiece which lays the foundation of information theory in 1948\cite{Shannon}, researchers have contributed much effort to find state-of-the-art coding and modulation schemes that can reach the channel capacity. From near Shannon limit codes such as Turbo codes\cite{turbo} and low density parity check (LDPC) codes\cite{LDPC}, to capacity-achievable channel codes such as Polar codes\cite{polar} and Spinal codes\cite{spinal}, the Shannon limit has been reached approximately by these advanced channel coding techniques. From frequency division multiple access applied in the first generation communications, to non-orthogonal multiple access (NOMA)  emerging in 5G communications, the revolution of modulation methods has also enhanced the channel capacity and been reaching the Shannon limits. Therefore, a new revolution on current communication technologies is imperative to satisfy the proliferating need on data exchange in the SAGSIN.
	
	Moreover, in many human-to-machine and machine-to-machine application scenarios in the SAGSIN, the ultimate goal of communications is to complete specific tasks at the machine terminal. Thus, perfect or approximate reconstruction of information on bit level is usually  unnecessary. For instance, in the ship detection tasks, the observer can detect the ships from an image based on only the pixels related to ships, where reconstruction of the whole image is evidently redundant. Another example is emergency monitoring and rescue tasks, where monitors are expected to only perceive and transmit urgent status information, while missed detection or error transmission of other statuses that are not very urgent are tolerable. Under  harsh conditions of the SAGSIN, pursuing perfect or approximate restoration of raw data not only will consume a huge amount of resources  but is also not feasible due to limited device capabilities. 
	
	Therefore, the above reasons necessitate an idea change from traditional communication technologies to high-level communications towards semantics exchange, named semantic communications\cite{weaver}. The reason is that semantic communications aim at  recovering only semantics at the receiver instead of reconstructing every code symbol accurately. In \cite{weaver}, Weaver identifies the three levels of communications, which are technical communications, semantic communications, and effectiveness communications. The abovementioned state-of-the-art coding and modulation schemes all make remarkable contributions to technical communications, while deliberately neglecting the semantic and effectiveness levels of communications. This is because Shannon believes that the technology of symbol transmission can be independent from the semantics that the symbols hold. However, the idea change from technical communications to semantic communications inspires us to reconsider the assumption of semantic independence and take semantic aspects of communications into consideration. This serves as the principal line of this survey. That is, by means of semantic and effectiveness communications, the semantics of the source data can be extracted at the transmitter side and utilized for task actuation at the receiver side. \footnote{It is worth noting that in this survey, we consider both semantic and effectiveness  communications as semantic communications, since they both conduct data pre-processing (i.e., extraction of semantics) before data transmission. According to \cite{weaver}, the principal difference of semantic and effectiveness communication is whether the effectiveness of terminal task actuation is considered in system design, and this is why we review task-oriented communications (a kind of effectiveness communications) in Section \ref{sec5}.}
	
	This survey is dedicated to answering the following three questions regarding to semantic communication systems deployed in the SAGSIN.
	
	\textit{Question 1. How should we interpret ``semantics" for semantic communications?}
	
	\textit{Question 2. How should we design semantic communication systems based on the interpretations of ``semantics"?}
	
	\textit{Question 3. How should we utilize the available while limited resources in the SAGSIN to implement semantic communications?}
	
	Overall, the term ``semantics" has three interpretations in state-of-the-art research on semantic communications. Firstly, from the perspective of etymology, the word ``semantics" originates from the Greek ``sēmanticós", which means ``significance". Secondly, ``semantics" has its common explanation, which is ``meaning". Thirdly, since ``semantic communications" includes both semantic and effectiveness level, this term can also be interpreted as ``effectiveness-related information".

	Based on above, we introduce three types of semantic communication systems, 
	which are respectively significance-oriented, meaning-oriented, and effectiveness-oriented semantic communication systems.
	
	(1) \textit{Significance-oriented semantic communication system.} In fact, source data are not equally significant, and thus sampling and transmitting unimportant data will cause waste in resource consumption. Taking the emergency monitoring task as an example, the status message generated later is more significant than the one generated earlier, since the former contains fresher information for the receiver; moreover, the status message which is different from the previous one is more significant, since the status variation usually implies policy or decision change for the task actuator; besides, an abnormal status message is more significant than a normal one, since urgent abnormal statuses may cause potential severe consequences. 
	Therefore, from the significance perspective, source data should be sampled, coded, and modulated in a significance-oriented manner, such that only data with more semantics (i.e., more significant data) are reserved and transmitted while those with less semantics (i.e., less significant data) get discarded. How we can measure the ``significance" will be elaborated in Section \ref{sec3}.
	
	(2) \textit{Meaning-oriented semantic communication system.} After significance-oriented sampling, although the amount of data has been preliminarily reduced, the remaining data still include considerably large amount of redundancy due to meaning-unrelated information. Therefore, semantic extraction from the data should be further adopted at the transmitter in order to eliminate the meaning-unrelated redundancy, such that the receiver can reconstruct data with similar meaning to original data. In the era of Shannon, semantic extraction and reconstruction were  tough problems, and researchers had to temporarily neglect the semantic aspects of source data and dedicate themselves to technical communication revolution. The principal reason is that the semantics of different modals of source data are distinct and complex, which cannot be mathematically expressed in a unified form. Nowadays, the thriving of artificial intelligence (AI), especially the unprecedented proliferation of deep learning (DL)  applications, will enable deep feature (semantics) extraction from various data such as texts, images, audios, and videos by massive computing resources, and utilizing DL to realize semantic communication systems is becoming a major approach. How DL-based semantic communication systems are designed for various modals of data will be discussed in Section \ref{sec4}.
	
	(3) \textit{Effectiveness-oriented semantic communication system.} After extraction of meaning-related information from the data, most of the redundancy is eliminated with only the meaning of data to be transmitted. However, not all of the extracted meaning is useful for the receiver, since there is still  redundant information for the ultimate actuation. 
	Moreover, symbol- or semantics-level reconstruction is also unnecessary from effectiveness-oriented perspective, since task actuation process may be intelligently conducted based directly on received messages. 
	Therefore, a task-oriented design for semantic communications can be adopted in order that the source data yield ultimate effectiveness more efficiently at the terminal. Specifically, a more intelligent effectiveness-related semantic extractor is introduced to replace common semantic extractor for capturing effectiveness-related information. Moreover, taking full advantage of DL-based techniques, intelligent actuator  is adopted which receives the effectiveness-related information and directly outputs the actuation results of the task without reconstruction of original data. By such task-oriented design, the reconstruction process at the receiver gets  omitted and thus computing resource consumption will decline. How such task-oriented communication system is constructed can be found in Section \ref{sec5}.

	Furthermore, in order to implement semantic communication systems in the SAGSIN from the above three perspectives, all the available perception, computing, and actuation techniques should be jointly adopted. Specifically,  significance-oriented semantic communication system adopts a perception-communication-integrated design by filtering insignificant information and protecting significant one; meaning-oriented semantic communication system utilizes perception, communication, and computing techniques through DL-based meaning extraction and reconstruction; effectiveness-oriented semantic communication system integrates all the perception, communication, computing, and actuation techniques via effectiveness-related semantic extractor and intelligent actuator without reconstruction.  Therefore, a new paradigm called \textit{Perception-Communication-Computing-Actuation Integrated Paradigm (PCCAIP)} can be proposed that guides the design of the above three types of semantic communication systems in the SAGSIN. 
	
	\subsection{Related Surveys \& Tutorials on SAGSIN}
	Since implementing SAG(S)IN in future 6G networks has become as a consensus from both industry and academia, there have been a plethora of survey papers on the topic of SAG(S)IN. Specifically, many magazine articles have contributed to the tutorial works on various applications facilitated by SAG(S)IN.  \cite{sagsin_magazine_siot,RASAG} focuses on random access technologies  in space communications as well as SAGIN-based communications. In \cite{sagsin_magazine_ntn}, challenges and opportunities of non-terrestrial network (including space, air, and sea networks in SAGSIN) in 6G era are elaborated through a case study on the millimeter wave communication among air/space nodes and ground nodes. The authors in \cite{sagsin_magazine_3D} also give a brief tutorial on new three-dimension communication architecture through SAGSIN. In terms of newly emerging technologies such as software-defined radio\cite{sagsin_magazine_software,sagsin_magazine_software2}, optical communications\cite{sagsin_magazine_fso}, and radar communications\cite{6GSAGS}, SAG(S)IN is also recommended as promising application scenarios.  Recently, with the surge of AI applications, optimizing the performance of next generation communications facilitated by SAGIN has raised much attention\cite{sagsin_magazine_ai,sagsin_magazine_ai2,sagsin_magazine_vehicle,DLSAGSIN,gushushi}. Moreover, some tutorial works on simulation platforms of SAGIN scenario can guide the communication system design of future 6G network\cite{sagsin_magazine_simu}.
	
	There are also some comprehensive survey papers that elaborate on the characteristics, design methods, and challenges of implementing SAG(S)IN. The first of such survey papers is \cite{SAGsurvey}, where the issues of system integration, protocol design, and performance analysis \& optimization are discussed in the integrated networks in order to overcome the heterogeneity (i.e., limited device capabilities), self-organization, and time-variability (i.e., highly dynamic channels) characteristics of SAGIN. Inspired by this pioneering work, \cite{sagsin_railway} gives a tutorial on how SAGIN facilitates next-generation communications in high-speed railway scenarios. Moreover, in order to address the security issues of heterogeneous SAGIN, blockchain-empowered space-air-ground IoT is proposed and the state-of-the art solutions for implementing blockchain-empowered SAGIN are reviewed in \cite{blockchain}. Security issues in SAGSIN are also the focus of \cite{surveySAGSIN}, where a comprehensive tutorial is presented on the topic of secure threats as well as attack technologies and corresponding defense methods. Also, the recent work \cite{NTN} envisions the integration of terrestrial network and non-terrestrial network in the future 6G era.
	
	However, the above tutorial works all pay close attention to only traditional communication technologies. As has been emphasized before, the inherent characteristics of SAGSIN may hinder the implementation of bit-level communications because of Shannon limits. Moreover, perfect bit reconstruction will be not that necessary in human-to-machine (or machine-to-machine) communications in future 6G networks. Hence, an idea change from traditional bit-level communication to semantic-level communication is imperative, especially in the future SAGSIN. 
	\subsection{Related Surveys \& Tutorials on Semantic Communications}
	Given that there is a recent surge of research interest on semantic communications, quite a few review and tutorial works on this topic have also emerged in the literature. In  \cite{magazinesignificance}, the authors envision a significance-oriented semantic communication system with sparse and effectiveness-aware sampling, where semantics of data is measured by timeliness metrics and end-to-end mean square error (MSE). Another magazine article \cite{magazineniukai} gives a tutorial on the framework of semantic communications by reconsidering semantic information theory. From the perspective of semantic network,  \cite{magazineshiguangming} proposes a semantic-aware network architecture by utilizing federated edge learning. Moreover,  \cite{magazinesemantic} presents a comprehensive review on DL-based semantic communications. Nevertheless, all these articles on semantic communications focus on only a certain interpretation of ``semantics" listed in the previous subsection.
	\begin{table*}[!t]\normalsize 
		\renewcommand{\arraystretch}{1.4} 
		\caption{COMPARISON OF CONTRIBUTIONS BETWEEN RELATED SURVEYS AND OUR SURVEY}\label{tablesurvey}
		\begin{tabular}[b]{|c|cc|c|} 
			\hline 
			&  \multicolumn{2}{c|}{$\textbf{Survey Papers}$} & \renewcommand{\arraystretch}{1.2}  \textbf{Contributions} \\ 
			\hline
			
			\multirow{14}{*}{\begin{tabular}[c]{@{}c@{}} \textbf{Related to}\\\textbf{SAGSIN} \end{tabular}} 
			& \multicolumn{1}{c|}{\multirow{8.2}{*}{\begin{tabular}[c]{@{}c@{}} Magazine \\ Articles \end{tabular}}} & \cite{sagsin_magazine_siot,RASAG}  & Random access technologies in SAGSIN \\ \cline{3-4}								
			& \multicolumn{1}{c|}{}  & \cite{sagsin_magazine_ntn} & Non-terrestrial network for 6G \\ \cline{3-4}
			& \multicolumn{1}{c|}{}  & \cite{sagsin_magazine_3D} &  Three-dimension communication architecture of SAGSIN\\  \cline{3-4}	
			& \multicolumn{1}{c|}{}  & \cite{sagsin_magazine_software,sagsin_magazine_software2} &  Software-defined radio facilitated by SAGSIN\\  \cline{3-4}		
			& \multicolumn{1}{c|}{}  & \cite{sagsin_magazine_fso,6GSAGS} &  Optical communications and radar communications in SAGSIN\\  \cline{3-4}	
			& \multicolumn{1}{c|}{}  & \cite{sagsin_magazine_ai,sagsin_magazine_ai2,sagsin_magazine_vehicle,DLSAGSIN,gushushi} &  AI-enabled SAGSIN\\  \cline{3-4}	
			& \multicolumn{1}{c|}{}  & \cite{sagsin_magazine_simu} &  Simulation platform establishment for SAGSIN\\  \cline{2-4}	
			&  \multicolumn{1}{c|}{\multirow{6}{*}{\begin{tabular}[c]{@{}c@{}} Comprehensive \\ Surveys  \end{tabular}}} & \cite{SAGsurvey} & {\begin{tabular}[c]{@{}c@{}}First survey on SAGSIN, discussing system integration,\\ protocol design, and performance analysis \& optimization  \end{tabular}} \\ \cline{3-4}	
			&  \multicolumn{1}{c|}{} & \cite{sagsin_railway} & Tutorial on high-speed railway facilitated by SAGSIN  \\ \cline{3-4}	
			&  \multicolumn{1}{c|}{} & \cite{blockchain}  & Blockchain-empowered SAGSIN aiming at security issues \\ \cline{3-4}	
			& \multicolumn{1}{c|}{} & \cite{surveySAGSIN} & SAGSIN security issues including attack \& defense methods \\ \cline{3-4}	
			& \multicolumn{1}{c|}{} & \cite{NTN} & Integrating terrestrial network and non-terrestrial network in 6G era \\ \cline{3-4}
			\hline
			
			\multirow{14}{*}{\begin{tabular}[c]{@{}c@{}} \textbf{Related to}\\\textbf{Semantic}\\\textbf{Communications} \end{tabular}} 
			& \multicolumn{1}{c|}{\multirow{4}{*}{\begin{tabular}[c]{@{}c@{}} Magazine \\ Articles \end{tabular}}} & \multicolumn{1}{c|}{\cite{magazinesignificance}} & Significance-oriented communications via effectiveness-aware sampling  \\ \cline{3-4}	
			& \multicolumn{1}{c|}{} & \multicolumn{1}{c|}{\cite{magazineniukai}} & Reconsidering semantic information theory \\ \cline{3-4}
			& \multicolumn{1}{c|}{} & \multicolumn{1}{c|}{\cite{magazineshiguangming}} & Federated edge learning-based semantic-aware network \\ \cline{3-4}
			& \multicolumn{1}{c|}{} & \multicolumn{1}{c|}{\cite{magazinesemantic}} & Comprehensive review on DL-based semantic communications \\ \cline{2-4}
			& \multicolumn{1}{c|}{\multirow{10}{*}{\begin{tabular}[c]{@{}c@{}} Comprehensive \\ Surveys  \end{tabular}}} & \multicolumn{1}{c|}{\cite{what2021qiao}} & \begin{tabular}[c]{@{}c@{}}First tutorial on semantic communications, reviewing human-human,\\ human-machine, and machine-machine communications\end{tabular} \\ \cline{3-4}
			& \multicolumn{1}{c|}{} & \multicolumn{1}{c|}{\cite{arxivprinciple} } & \begin{tabular}[c]{@{}c@{}}Reviewing semantic communication theory and DL-based\\ semantic communications for different sources\end{tabular} \\ 	\cline{3-4}
			& \multicolumn{1}{c|}{} & \multicolumn{1}{c|}{\cite{arxivless} } &\begin{tabular}[c]{@{}c@{}} Reasoning-driven semantic communication system and\\ semantic language for semantic representation\end{tabular} \\ 	\cline{3-4}
			& \multicolumn{1}{c|}{} & \multicolumn{1}{c|}{\cite{beyond} } & \begin{tabular}[c]{@{}c@{}}Focusing on semantic measures, semantic compression,\\ semantic transmission via JSCC, and timeliness issues\end{tabular} \\ 	\cline{3-4}
			& \multicolumn{1}{c|}{} & \multicolumn{1}{c|}{\cite{2023survey} } & Semantic-oriented, goal-oriented, and semantic-aware communications \\ 	\cline{3-4}
			& \multicolumn{1}{c|}{} & \multicolumn{1}{c|}{\cite{edgesemantic,huawei} } & Semantic communications as solutions in future networks
			\\ 	
			
			\hline
			\hline
			
			\multirow{4}{*}{\begin{tabular}[c]{@{}c@{}} \textbf{Our Survey} \end{tabular}} 
			& \multicolumn{2}{c|}{\textbf{Contributions for SAGSIN}} &  \textbf{\textbf{Contributions for Semantic communications}} \\ \cline{2-4}	
			& \multicolumn{2}{c|}{Proposes PCCAIP as a}  & $\bullet$ Reviewing SPTs for significance-oriented communications\\ 
			& \multicolumn{2}{c|}{unified framework facilitating}  &  $\bullet$ Reviewing METs for meaning-oriented communications \\ 	 
			& \multicolumn{2}{c|}{task-critical applications}  &  $\bullet$ Reviewing EYTs for effectiveness-oriented communications \\ 									
			\hline
			
		\end{tabular}\centering
	\end{table*}
	\begin{figure*}[!t]
		\centering
		\includegraphics[width = 0.96\textwidth]{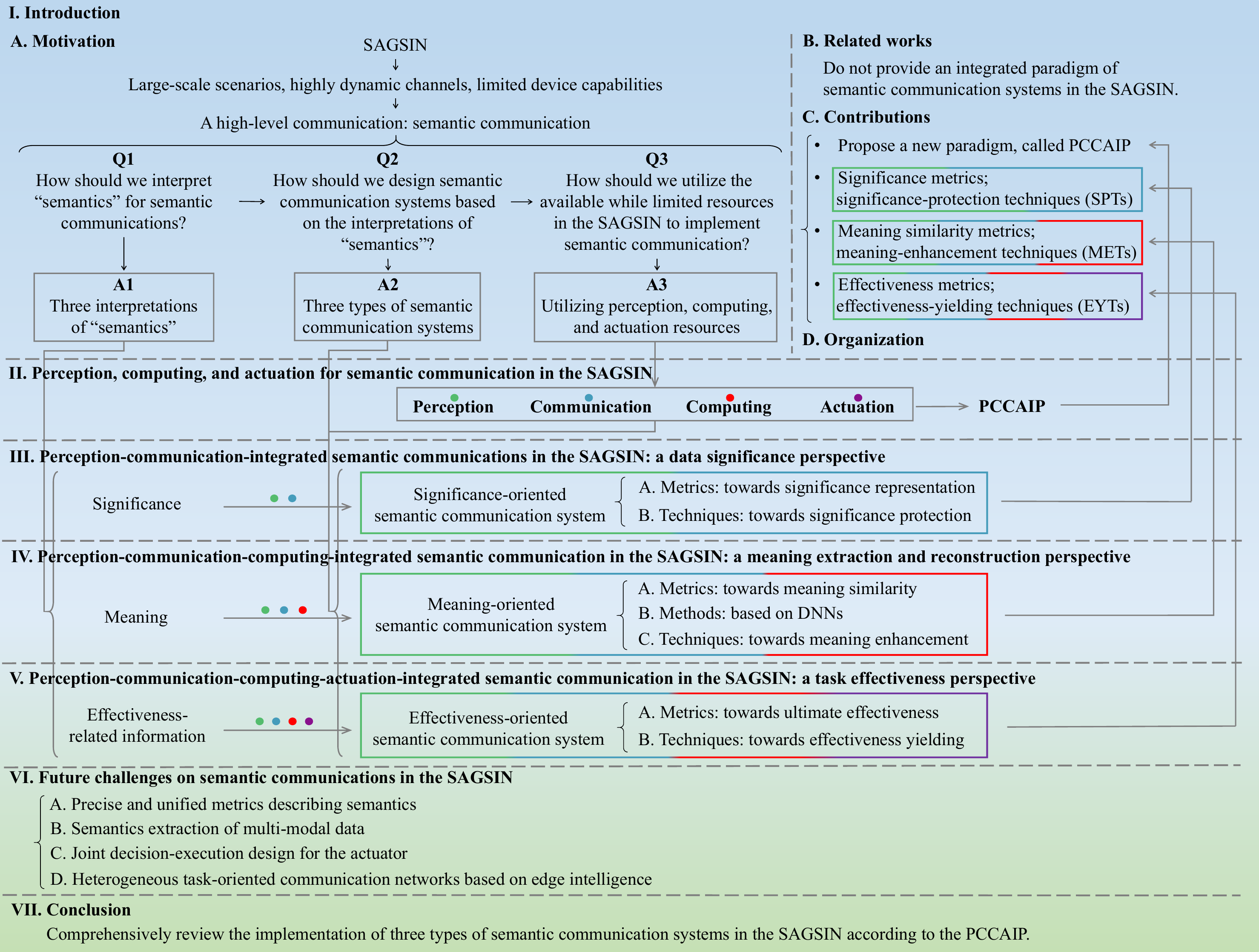}
		\caption{The architecture of this survey.}
		\label{figtotal}
	\end{figure*}
	
	Some comprehensive tutorial works on semantic communication topics can also be found in the literature. The first of such tutorial works is  \cite{what2021qiao}, where three communication modalities including human-human, human-machine, and machine-machine are addressed by semantic communications using DL-based techniques. In \cite{arxivprinciple}, semantic communication theory is reviewed, and DL-based semantic communications for different types of sources are respectively discussed. In  \cite{arxivless}, a new perspective called reasoning-driven semantic communication system is proposed, and a new semantic representation method called semantic language is studied for next generation communications.  Moreover, in the recent literature of \cite{beyond}, the authors review semantic communications from perspectives of semantic measures, semantic compression based on semantic rate-distortion theory, semantic transmission via joint source-channel coding (JSCC), and timeliness of semantic communications, respectively. In the recent tutorial work \cite{2023survey}, research on semantic communications is reviewed from three perspectives, which are semantic-oriented communications, goal-oriented communications, and semantic-awareness communications. Correspondingly, the three directions on semantic communications are pointed out, which are semantic extraction, semantic transmission, and semantic metrics. In addition to the above tutorial works, some other tutorials on edge learning\cite{edgesemantic} and new challenges of 6G\cite{huawei} also consider semantic communications as competitive solutions.
	
	However, a principal implementation of future 6G network is SAGSIN, while these studies do not provide an integrated paradigm of semantic communication system design that satisfies the needs in the SAGSIN. For instance, communication over highly dynamic channels is a core characteristic of SAGSIN, and thus robust semantic communications that resist channel fading as well as semantic noise must be taken into consideration when implementing semantic communications in SAGSIN. Also, given the limited capabilities of nodes in SAGSIN, it is crucial to prioritize semantic rate allocation and reduce computing complexity. Moreover, as has been pointed out in \cite{blockchain,surveySAGSIN}, security issues should be addressed in SAGSIN-based semantic communications. Goal-oriented communications for specific task-critical applications in SAGSIN are also in its nascent age. As a result, a comprehensive review of semantic communications in the future SAGSIN is of vital importance.
	
	\subsection{Contributions}
	
	To fill the void in literature of semantic communications for the SAGSIN, in this survey, we propose a comprehensive paradigm called PCCAIP which fully addresses fundamental limitations  in the SAGSIN. Sequentially, we  review recent literature on semantic communications from three perspectives, which are 1) significance representation and protection, 2) meaning similarity measurement and meaning enhancement, and 3) ultimate effectiveness and effectiveness yielding. The contributions of the related survey papers and ours are compared in Table \ref{tablesurvey}. Specifically, our contributions are listed as follows.
	
	\begin{itemize}
		\item We propose a new paradigm for the implementation of semantic communications in the SAGSIN called PCCAIP, in which perception, computing, and actuation techniques are jointly utilized for communications, and thus the design for the three types of state-of-the-art semantic communication systems can be guided by PCCAIP.
		\item Aiming at designing significance-oriented semantic communication systems, we introduce  the significance metrics, including content-agnostic, content-aware, and unified metrics. We then review recent works on \textit{Significance Protection Techniques (SPTs)}, including sampling, coding, and modulation.  
		\item We elaborate on a major approach of semantic communications, namely meaning-oriented semantic communication system, focusing on implementation of DL-based meaning extraction and reconstruction. We provide an idea of system design, which starts from maximizing meaning similarity metrics, and next trains DL models, and finally utilizes \textit{Meaning-Enhancement Techniques (METs)}.
		\item We discuss a newly emerging approach of semantic communications called effectiveness/task-oriented semantic communication system, which is guided by PCCAIP. We review metrics that measures the ultimate effectiveness of tasks, and discuss \textit{Effectiveness-Yielding Techniques (EYTs)} facilitating typical services in the SAGSIN.
		
	\end{itemize}

	\subsection{Organization}
	The rest of this survey is organized as follows. Section \ref{sec2} proposes the PCCAIP, where utilization of perception, computing, and actuation modules in the SAGSIN is elaborated respectively. In Section \ref{sec3}, we introduce significance metrics as semantic measures, and review SPTs including significance-oriented sampling, coding, and modulation. Section \ref{sec4} introduces meaning similarity metrics, DL models, and METs based on meaning extraction and reconstruction. The focus of Section \ref{sec5} is task-oriented communications guided by PCCAIP, where effectiveness metrics are reviewed and EYTs are discussed for specific tasks. We then envision some future challenges on semantic communication implementation in the SAGSIN in Section \ref{sec6}, and \ref{sec7} concludes the survey. The architecture of this survey is illustrated in Fig. \ref{figtotal}, and the majority of the abbreviations/acronyms used in this survey are listed in Table \ref{abbr}.
	
	\begin{table*}[!t]\normalsize
		\renewcommand{\arraystretch}{1.08} 
		\caption{SUMMARY OF THE ABBREVIATIONS/ACRONYMS USED IN THIS SURVEY}\label{abbr}
		\begin{tabular}[!t]{|c|c||c|c|} 
			\hline 
			\textbf{Abbreviation} & \textbf{Full Name} &  	\textbf{Abbreviation}&\textbf{Full Name}      \\ 
			\hline

			5G/6G& fifth/sixth generation &  SAG(S)IN& \begin{tabular}{c}space-air-ground(-sea)\\ integrated network\end{tabular}\\\hline
			LEO/GEO&low/geostationary Earth orbit  & UAV&unmanned aerial vehicle\\ \hline
			HAP&high-altitude platform&IoT/IoV&Internet of things/vehicles  \\ \hline
			LDPC&low density parity check&NOMA&non-orthogonal multiple access \\ \hline
			AI&artificial intelligence&DL&deep learning \\ \hline
			MSE/MAE&mean square/absolute error&JSCC&joint source-channel coding \\ \hline
			SPT&significance protection technique&MET&meaning-enhancement technique \\ \hline
			EYT&effectiveness-yielding technique&AR/VR/MR&augmented/virtual/mixed reality \\ \hline
			AoI&age of information &AoS&age of synchronization\\ \hline
			AoII&age of incorrect information &UoI&urgency of information\\ \hline
			GoT&goal-oriented tensor &QAoI&query age of information\\ \hline
			VoI&value of information &ACK&acknowledgment \\ \hline
			MDP&Markov decision process &(H)ARQ&(hybrid) automatic repeat request\\ \hline
			HARQ-CC&HARQ with chase combining &(HARQ-)IR&(HARQ with) incremental redundancy \\ \hline
			BSC/BEC&binary symmetric/erasure channel&SNR&signal-to-noise ratio \\ \hline
			AWGN&additive white Gaussian noise&FBL&finite block-length\\ \hline
			NACK&negative acknowledgment& OMA&orthogonal multiple access\\ \hline
			MDP&Markov decision process &PSO&particle swarm optimization \\ \hline
			ANN&artificial neural network&DNN&deep neural network\\ \hline
			NN&neural network&BER&bit error rate\\ \hline
			WER&word error rate&BLEU&bilingual evaluation understudy\\ \hline
			CIDEr&\begin{tabular}{c}consensus-based image\\ description evaluation\end{tabular}&MSS&metric of semantic similarity\\ \hline
			BERT&\begin{tabular}{c}bidirectional encoder representations\\ from Transformers\end{tabular}&PSNR&peak signal-to-noise ratio\\ \hline
			SSIM&structural similarity index&MS-SSIM&multi-scale SSIM\\ \hline
			LPIPS&\begin{tabular}{c}learned perceptual image\\patch similarity\end{tabular}&GAN&generative adversarial network\\ \hline
			FID/FDSD&Fréchet Inception/DeepSpeech distance&KID/KDSD&kernel Inception/DeepSpeech distance\\ \hline
			AKD&average keypoint distance&SDR&signal-to-distortion ratio\\ \hline
			PESQ&perceptual evaluation of speech quality&IRS&intelligent reflecting surface\\ \hline
			FFT&fast Fourier transform&MCD&Mel cepstral distortion\\ \hline
			RL&reinforcement learning&SSCC&separate source-channel coding\\ \hline
			DeepJSCC&deep-learning based JSCC&CNN&convolutional neural network\\ \hline
			RNN&recurrent neural network&ReLU&rectified linear unit\\ \hline
			FCN&fully convolutional network&LSTM&long short-term memory\\ \hline
			GRU&gated recurrent unit&NLP&natural language processing\\ \hline
			GPU&graphics processing unit&GCN&graph convolutional network\\ \hline
			RLN&reinforcement learning network&DeepSSCC&DL-based SSCC\\ \hline
			CSI&channel state information&BPSK&binary phase shift keying\\\hline
			NTSCC&nonlinear transform source-channel coding&RS (codes)&Reed-Solomon (codes)\\ \hline
			KG&knowledge graph&KB&knowledge base\\ \hline
			ProbLog&probabilistic logic programming language&ML&machine learning\\ \hline
			TP/TN&true positive/negative&FP/FN&false positive/negative\\ \hline
			IoU&intersection over union&WNCS&wireless network control system\\ \hline
			AoTI&age of task information&SNN&spiking neural network\\ \hline
			(Dec-)POMDP&(decentralized) partially observable MDP&VQA&visual question answering\\ \hline
			
		\end{tabular}\centering
	\end{table*}
	
	\section{Perception, Computing, and Actuation for Semantic communications in the SAGSIN}\label{sec2}
\begin{figure*}[ht]
		\centering
		\includegraphics[width = 0.96\textwidth]{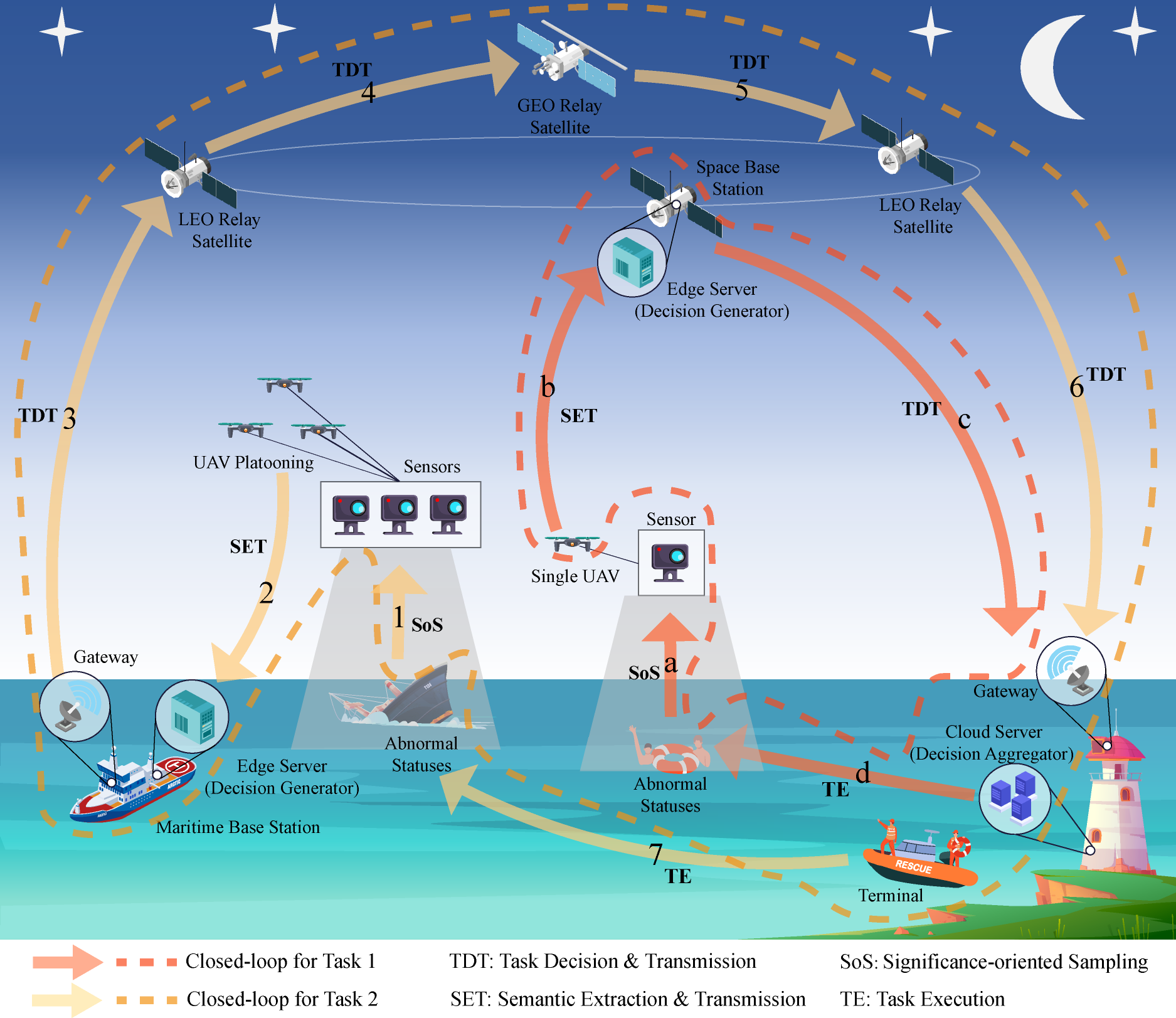}
		\caption{A scenario of remote emergency rescue task in SAGSIN which is facilitated by PCCAIP. The numbers and letters represent the closed-loop information flows. For task 1, the letters represent the following processes: a. significance-oriented sampling of abnormal statuses from a drowning person; b. semantic extraction and transmission from a  single UAV to the space base station; c. task decision transmission from the base station to the cloud server; d. decision aggregation (if necessary) and task execution (the terminal will send rescue to the drowning person). For task 2, the numbers represent the following processes: 1. significance-oriented sampling of abnormal statuses from a sinking ship; 2. semantic extraction and transmission from UAV platooning to the maritime base station; 3. task decision transmission from the base station to an LEO satellite; 4. relay from an LEO satellite to the GEO satellite; 5. relay from the GEO satellite to another LEO satellite; 6. task decision transmission from the LEO satellite to the cloud server; 7. decision aggregation (if necessary) and task execution (the terminal will send rescue to the sinking ship).}
		\label{fig2}
	\end{figure*}

	As mentioned in Section \ref{sec1}, semantic communication systems play a crucial role in the SAGSIN by enabling efficient information exchange and intelligent actuation. However, relying solely on traditional communication techniques may not meet the requirements of complex tasks due to the limitations in network resources.
		Therefore, perception, computing and actuation techniques facilitate semantic communications  to form the PCCAIP, by which a "target-perception-communication-actuation-target" closed-loop framework facilitated by intelligent computing is constructed. Specifically, perception techniques are utilized to observe the  targets (e.g., abnormal statuses shown in Fig. \ref{fig2}) in the physical world. Next, communication techniques help transmit the observed target information to remote nodes (e.g., edge servers in Fig. \ref{fig2}) for further processing. Sequentially, actuation techniques will guide the actuator (e.g., both edge servers and cloud servers in Fig. \ref{fig2}) to generate the best task decision commands. Finally, the commands will be executed by the terminal, and the status of the target will be changed due to task execution. In the following discussion, we will firstly address these techniques, and then provide a detailed introduction to the PCCAIP. 

	\subsection{Perception}
	The integration of space-based, air-based, ground-based, and sea-based networks results in a significant increase in business volume. Multiple types of sensors deployed in these networks enable real-time perception of various targets. Meanwhile, since the targets may belong to different physical processes in the real world, a substantial amount of multi-modal data are collected, which poses a great challenge in supporting services and applications with extremely low latency requirements. 
	One feasible solution is to reduce the volume of data to decrease the latency in subsequent communication and processing stages. This is why  we need to perceive the significance, meaning, and  effectiveness-related information of the source data in the SAGSIN. Capturing significance can be achieved through significance-oriented sampling, while capturing the other two can be accomplished through intelligent semantic extraction, which is typically facilitated by DL, knowledge graphs (KGs), and so on. The significance-oriented sampling and semantic extraction are collectively referred to as  perception in this survey.
	
	Perception plays a significant role in various application scenarios. For example, in the emergency response and rescue applications for maritime incidents, ocean buoys deployed in the maritime area are used to collect oceanographic source  data such as sea currents, wind speeds, and wave conditions. UAVs deployed above the sea surface  capture image data with cameras. During non-urgent situations, the sensors operate at a relatively low sampling frequency to conserve energy and reduce data processing burden. However, in the emergency events  such as maritime shipwrecks, oil spills, fires, or drowning, the sensors immediately increase their sampling frequency to keep data freshness and let the server estimate the source status more accurately. The source data are further processed to extract semantic information, such as the situation of stranded crew members and ocean current information. The extracted semantic information is then transmitted  to the emergency rescue center in real-time to facilitate rescue operations.

	\subsection{Computing}
	In the semantic-empowered SAGSIN, computing refers to the AI-related complex operations, such as the feature extraction, semantic coding, and adaptive decision. 
	Therefore, it is imperative to use  AI algorithms to extract semantic information and make intelligent decisions, which will release data traffic and thus decrease processing latency. However, AI algorithms rely heavily on sufficient computing resources, and it is unlikely to achieve satisfactory performance when these algorithms are deployed on end devices such as ocean buoys and UAVs due to limitations in computing  and caching resources. The SAGSIN encompasses  various types of devices, resulting in a large volume of data that usually require real-time processing and analysis. 
	Edge computing technology has emerged to address this issue. 
	
	In recent years, there have been significant academic and industrial interests in on-orbit edge computing based on space-based networks\cite{Proposal35,Proposal36}, as well as edge storage technologies \cite{gushushi,Proposal38}. A large number of cloud servers, edge servers, and other computing units are distributed in the SAGSIN,  providing supports for perception, communication, and actuation techniques.
	For instance, in the context of satellite-based edge computing, efficient interconnection is achieved among space-based, and ground-based networks by leveraging LEO satellites as the core. An edge computing platform is built on the satellites, empowering the entire network with on-orbit capabilities for intelligent data collection, processing, and caching\cite{Proposal39}. The authors assume a geographically dispersed remote team with diverse members. Through augmented reality (AR) and virtual reality (VR) technologies, remote team members can share a virtual workspace, enabling collaborative working and real-time communications. In this scenario, the on-orbit edge servers are equipped with a semantic encoding network driven by DL models. Team members wear head-mounted displays integrated with semantic decoding networks, accessing the shared virtual workspace through VR technology. In this example, the satellite-based edge computing platform provides robust support for the combination of VR technology and semantic communications.
	
	\subsection{Actuation}
	The receiver utilizes the received messages, which are relevant to task performance,  to accomplish the task decision. Based on the decision results, the task execution process will exert an influence on the physical world. In this survey, the task decision and task execution comprise the definition of actuation. Taking remote control services as example, the intelligent decision-maker generates and transmits task decision commands based on the received messages, and then the terminal receives the decision commands and executes specific tasks.
	
	With the development of 5G/6G, the variety of task-critical applications in the SAGSIN continues to enrich. For instance, the stability control of UAVs can be modeled as a classical CartPole problem solution. In this scenario, sensors collects  environmental data, which will be then processed on the edge server using DL-based autoencoders to extract task-related semantic features. These features are transmitted to the remote control center on the ground to generate control commands, enabling remote control of the UAVs \cite{arxiv76}. 
	
	Another example is found in remote surgeries, which is a typical use case of haptic Internet. Sensors located in remote areas (e.g., mountainous areas and remote seas) collect crucial patient data, which are transmitted to the medical center with the aid of satellites. Surgeons receive real-time auditory, visual, and haptic feedback through a human-machine interface. Visual feedback is provided using streaming technologies, such as holographic-type communications, depending on whether surgeons interact with holograms or wear head-mounted devices. Subsequently, surgeons operate haptic devices through the human-machine interface and perform surgical actions based on real-time visual feedback and haptic information transmitted to the robot\cite{ITU-T}.
	
	\subsection{PCCAIP}
	According to the interpretation of perception, computation, actuation, we introduce the perception-communication-computing-actuation-integrated framework under the guidance of PCCAIP. This framework aims to maximize the effectiveness of tasks by establishing closed information flow loop, as shown in Fig. \ref{fig2}.  We consider an emergency detection and rescue scenario in the SAGSIN as an example. It is worth noting that, in the current industrial infrastructure, the emergency detection task and emergency rescue task can both be independently implemented in real world. However, the  collaborative design and coordination of detection \& rescue operations still rely on the comprehensive SAGSIN which provides heterogeneous connections as well as computing resources. 
	
	Specifically, various sensors (e.g., cameras deployed on the UAVs) continuously observe the environment that contains abnormal statuses (e.g., sinking ships or drowning persons), and decide to sample data representing the current statuses when the data are significant enough (e.g. untimeliness, status change,  or environment change). After data sampling, these data will be further processed by the computing-intensive semantic extractor deployed on UAVs and then transmitted to remote edge servers (e.g. servers deployed on maritime base station or space base station), by which multi-level data perceptions are completed by distributed computing resources. 
	
	Based on the received status data, the edge servers conduct real-time decision making and generate commands, which  can guide the terminals (e.g., rescue ships) in responding to the emergencies if necessary.  Subsequently,  the decision commands are transmitted to remote terrestrial cloud server via LEO/GEO satellite relays.  After the cloud server receives the decision commands, it will undertake task execution (e.g., rescuing the ships/persons, or doing nothing) based on the commands, by which task actuation is completed. Moreover, in cases where rescue capabilities are constrained, the cloud servers will firstly aggregate the received commands from all the edge servers and secondly decide which emergencies should be rescued according to various factors (e.g., significance, environment statuses, and effectiveness aspects). After the decision making process, the task execution will be completed by the terminal.
	
	Task execution will further determine the statuses and thus the decisions at the next time interval (for example, if a sinking ship gets rescued by the rescue ship, there will be no emergency in the next period of time, and thus the rescue ship does not need to rescue in the next), by which closed loops for the emergency rescue tasks are formed by the PCCAIP framework.

	\subsection{Lessons Learned from This Section}
	This section elaborates on how the perception, computing, and actuation techniques help realize semantic communications. Overall, perception techniques help select the data with more semantics, computing techniques facilitate intelligent communications in various aspects, and actuation techniques aid in the interaction between system and environment to complete all-inclusive tasks. PCCAIP integrates all these techniques which can realize a closed-loop communication and a joint design to achieve high effectiveness of tasks. Moreover, maritime rescue tasks are adopted as an example to showcase how PCCAIP guides semantic communication system design.
	
	The proposed PCCAIP is essentially a novel paradigm which significantly deviates from traditional integrated sensing and communication (ISAC) paradigm\cite{isac} along with integrated sensing, communication, and computing (ISCC) paradigm\cite{iscc} in two aspects. Firstly, the ISAC and ISCC both focus on the integration of radar sensing (as well as traditional data capturing) and communication (computing), which belong to the category of traditional bit-level communication, while PCCAIP aims at high-level semantic communication facilitated by intelligent significance-oriented sampling and meaning extraction. Secondly, ISAC and ISCC both address the issue of only communication processes, while PCCAIP can facilitate more comprehensive task-critical applications by jointly designing both communication processes and the ultimate actuation processes. Therefore, it is a promising idea to implement semantic-communication-based task-critical applications in the future network via PCCAIP.
	
	As typical semantic communication frameworks facilitated by PCCAIP, we will introduce significance-oriented, meaning-oriented, and effectiveness/task-oriented communications systems respectively in the following sections. In Section \ref{sec3},  we  design a significance-oriented semantic communication system which belongs to a perception-communication-integrated semantic communications. In Section \ref{sec4}, we study a meaning-oriented semantic communication system which falls under the category of a perception-communication-computing-integrated semantic communication system. In Section \ref{sec5}, we  introduce a task-oriented communication system which achieves perception-communication-computing-actuation integration.
	\section{Perception-Communication-Integrated Semantic Communications in the SAGSIN: A Data Significance Perspective} \label{sec3}
	In this section, we will review the research on semantic communications from the data significance perspective. Here, the term ``semantics" is explained as its etymological meaning ``significance", which implies a significance-oriented semantic communication system. 
	
\begin{figure*}[!t]
		\centering
		\includegraphics[width = 0.96\textwidth]{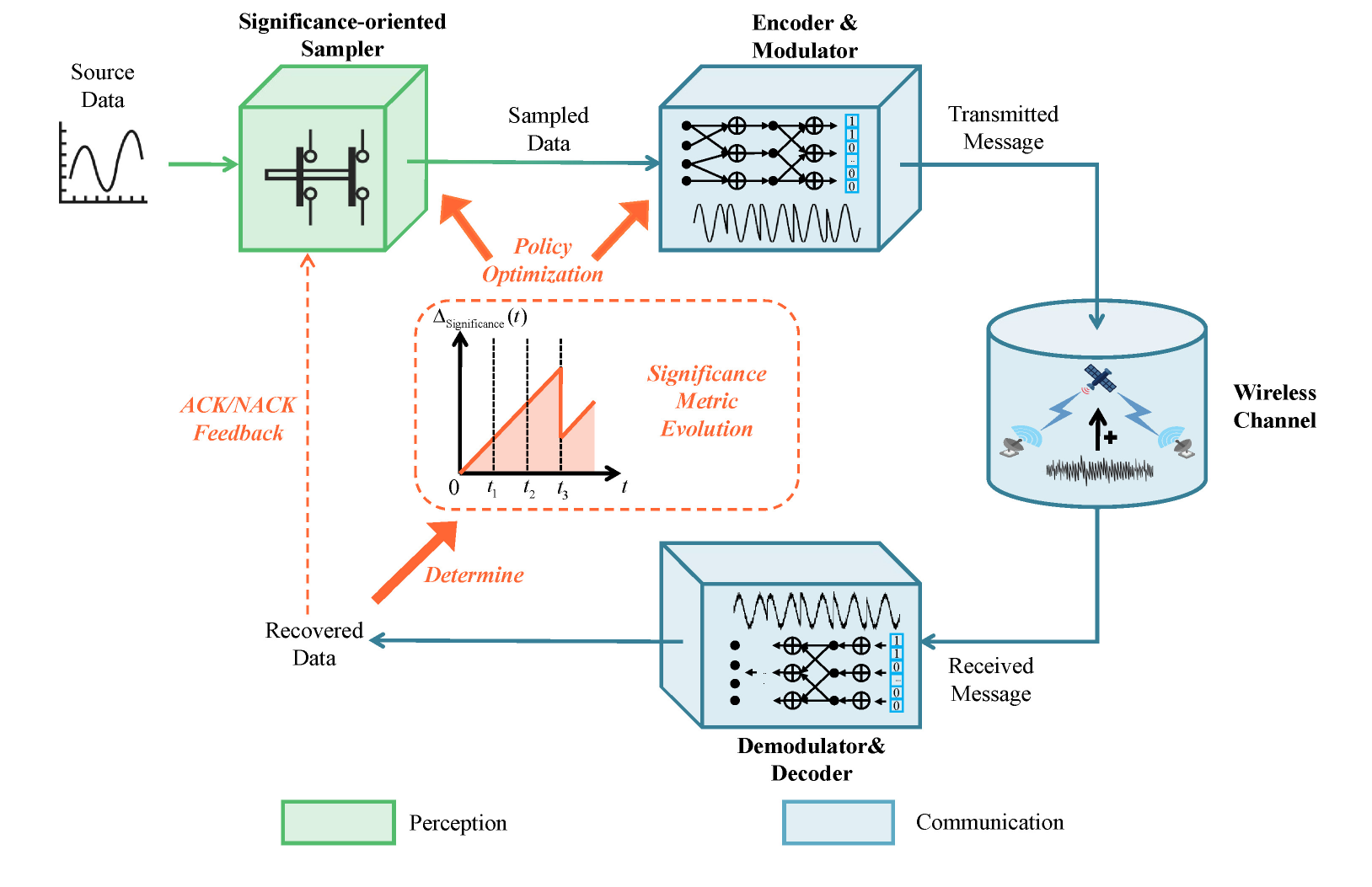}
		\caption{The framework of perception-communication-integrated semantic communications in the SAGSIN.}
		\label{fig3}
	\end{figure*}

	In significance-oriented semantic communications in the SAGSIN, a joint perception-communication framework is adopted as shown in Fig. \ref{fig3}. In Fig. \ref{fig3}, the source data can be of various types, such as given data and status updates. Given data are usually collected from fixed files provided by users, such as texts, images, and audios, while status updates are collected from a time-variant physical process from the real world. Aiming at studying the significance evolution and optimization of a physical process, all of the works reviewed in this section assume that the source data are status updates. Specifically, source data are firstly sampled by significance-oriented sampler in order to release the heave data burden, and the sampled data will be coded and modulated for transmission over wireless channel.  The receiver demodulates and decodes the received message to recover sampled data. According to the recovery results, the significance of source data varies to affect the evolution of significance metric; sequentially, the variation of significance metric will affect the further decision of sampler and coder/modulator at the transmitter via feedback from the receiver. It is worth noting that both coding and modulation techniques mentioned in this section are traditional technologies based on Shannon information theory.
	
	Considering the fundamental limits of the SAGSIN and the time-sensitivity of various tasks facilitated by SAGSIN, the data freshness/timeliness is naturally significant for semantic communication system due to large communication distance. 
	Therefore, timeliness metrics, including age of information (AoI) and its variants, are of vital importance in describing the performance of the tasks.
	Thus, we will first introduce AoI and AoI variants as significance metrics in subsection \ref{sec3A}, and then review the literature on techniques towards significance-protection  in subsection \ref{sec3B}. 
	
	\subsection{Metrics: Towards Significance Representation}\label{sec3A}
	
	\begin{table*}[!t]\normalsize
		\renewcommand{\arraystretch}{1} 
		\caption{SUMMARY OF SIGNIFICANCE METRICS}\label{III_A}
		\begin{tabular}[!t]{|c|c|c|c|c|c|c|} 
			\hline 
			$\textbf{Metrics}$ & $\textbf{Significance Aspect}$ & $\textbf{Properties}$ & \begin{tabular}[c]{@{}c@{}} $\textbf{Application}$ \\ $\textbf{Scenarios}$  \end{tabular} & $\textbf{Evolution Example}$ & $\textbf{Formula}$ & \renewcommand{\arraystretch}{1.2} \begin{tabular}[c]{@{}c@{}} $\textbf{Initial}$ \\ $\textbf{References}$  \end{tabular}  \\ 
			\hline
			
			$\textbf{AoI}$ 
			& \begin{tabular}[c]{@{}c@{}}The staleness of\\ source data\end{tabular} & \begin{tabular}[c]{@{}c@{}}Content-\\ agnostic\end{tabular} & \begin{tabular}[c]{@{}c@{}}{Require for}\\{
					short delay}\end{tabular}
			& \begin{minipage}[c]{0.45\columnwidth}
				\centering
				\vspace{0.5mm} {\includegraphics[width=1\textwidth]{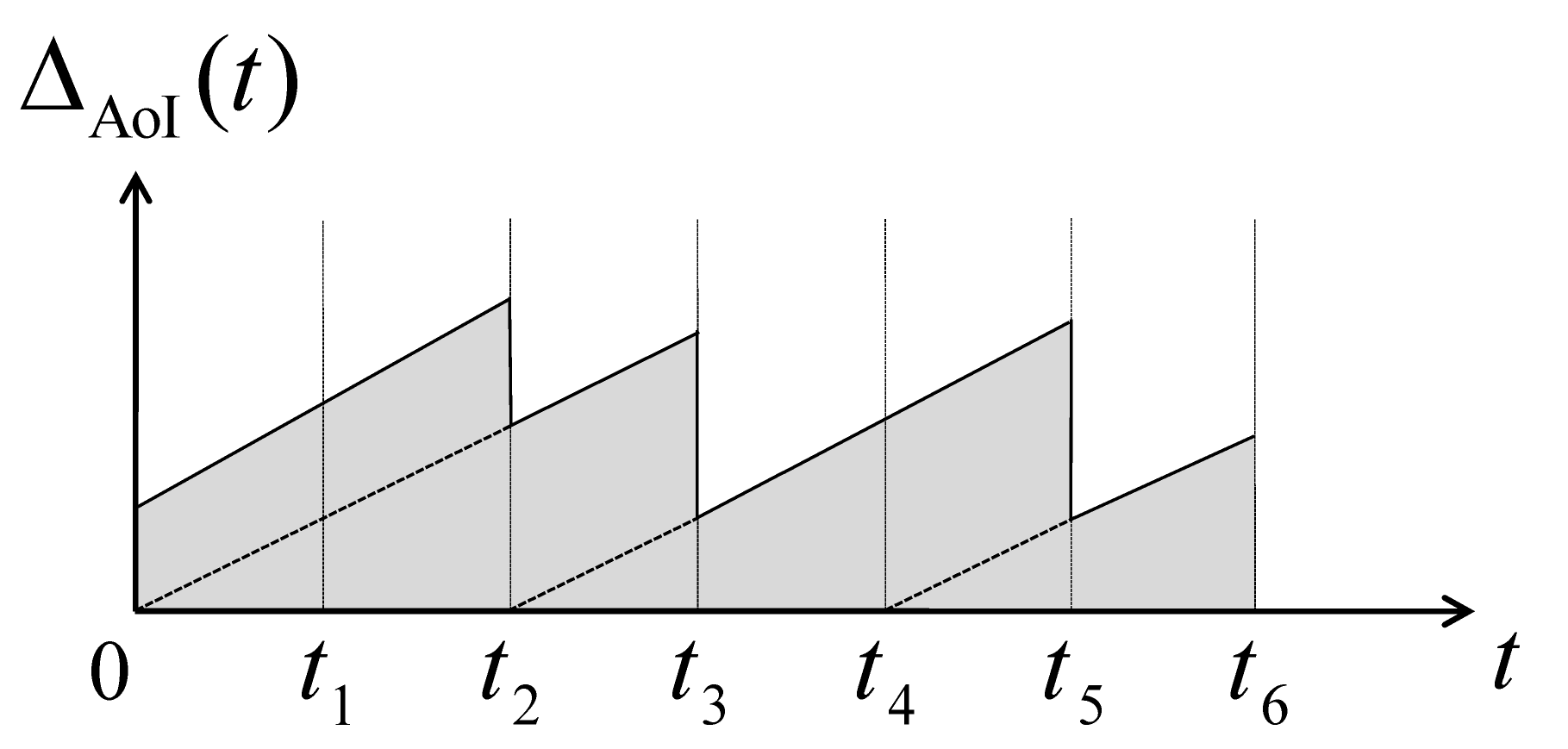}}  
			\end{minipage}
			& (\ref{AoI}) &  \cite{AoI1,AoI2}  \\ \hline
			
			\begin{tabular}[c]{@{}c@{}}$\textbf{Nonlinear}$\\ $\textbf{AoI}$\end{tabular}
			& \begin{tabular}[c]{@{}c@{}}The nonlinear cost  \\ caused by staleness \\ \end{tabular} & \begin{tabular}[c]{@{}c@{}}Content-\\ agnostic\end{tabular}&\begin{tabular}[c]{@{}c@{}}{Require} \\ {for extremely}\\ {low latency} \end{tabular}
			& \begin{minipage}[c]{0.45\columnwidth}
				\centering
				\vspace{0.5mm} {\includegraphics[width=1\textwidth]{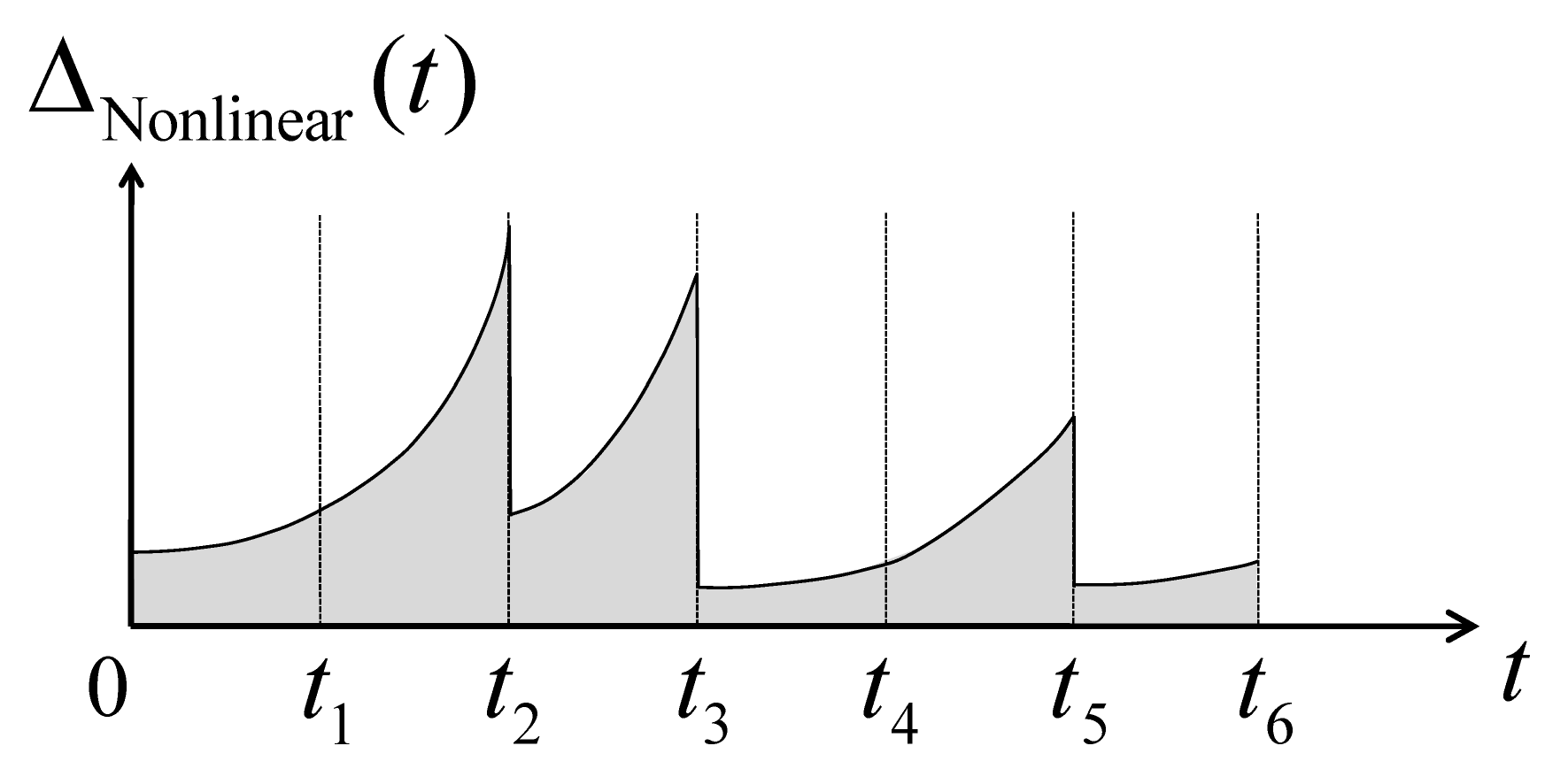}} 
			\end{minipage} 
			& (\ref{nonlinear}) &\cite{voi,AoI6}   \\ \hline
			
			$\textbf{AoS}$
			& \begin{tabular}[c]{@{}c@{}}The cost caused by \\ asynchronization\\ duration  \end{tabular} & \begin{tabular}[c]{@{}c@{}}Content-\\ aware\end{tabular}&\begin{tabular}[c]{@{}c@{}}{Require for}\\ {real statuses}\\{of the system}\end{tabular}
			& \begin{minipage}[c]{0.45\columnwidth}
				\centering
				\vspace{0.5mm} {\includegraphics[width=1\textwidth]{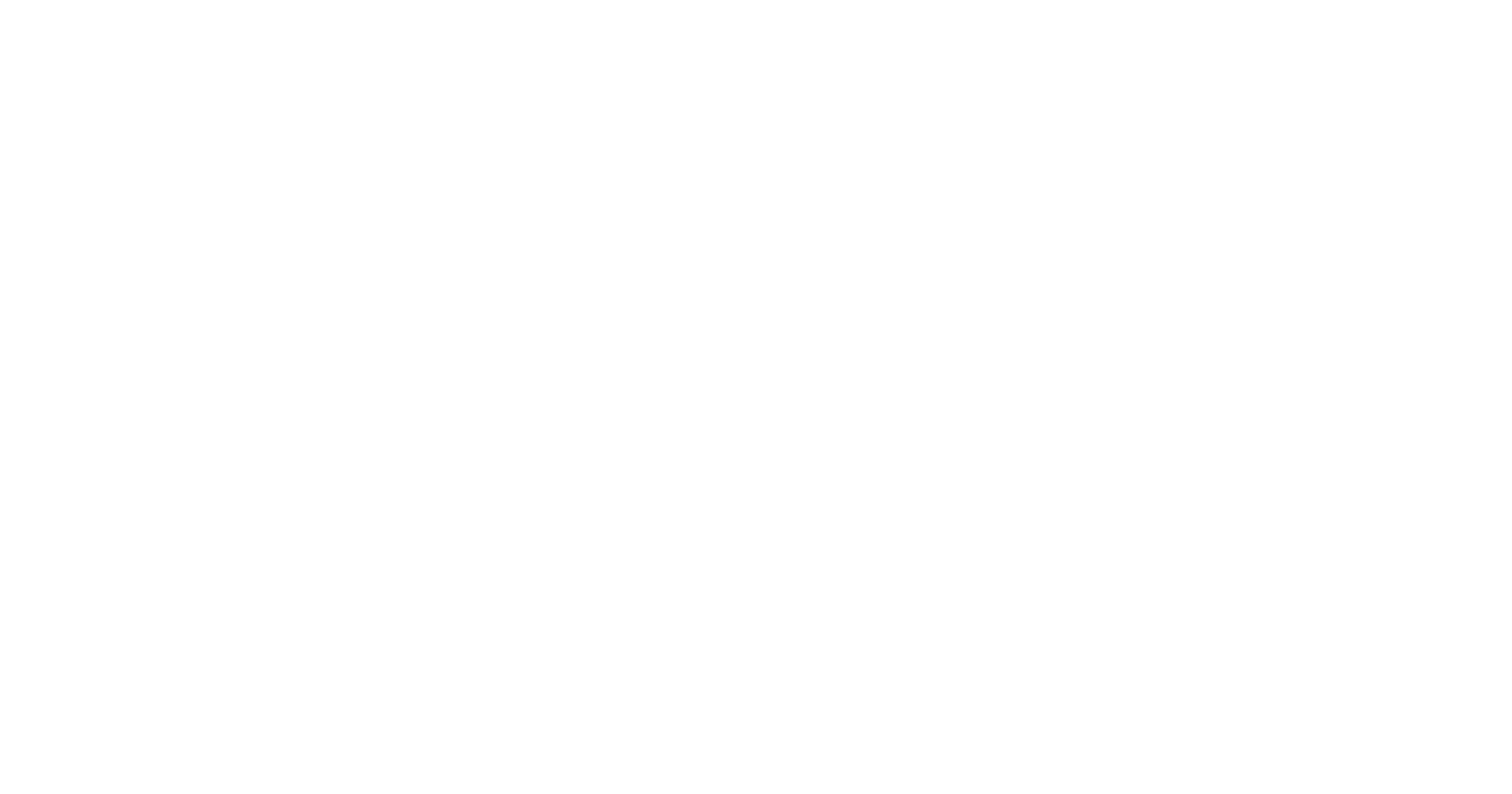}} 
			\end{minipage} 
			& (\ref{AoS}) & \cite{aos}  \\ \hline
			
			$\textbf{AoII}$
			& \begin{tabular}[c]{@{}c@{}}The  cost caused by \\ asynchronization \\ duration and\\ mismatch extent\end{tabular} & \begin{tabular}[c]{@{}c@{}}Content-\\ aware\end{tabular}&\begin{tabular}[c]{@{}c@{}}{Require for}\\{ real statuses}\\{of the system}\end{tabular}
			& \begin{minipage}[c]{0.45\columnwidth}
				\centering
				\vspace{0.5mm} {\includegraphics[width=1\textwidth]{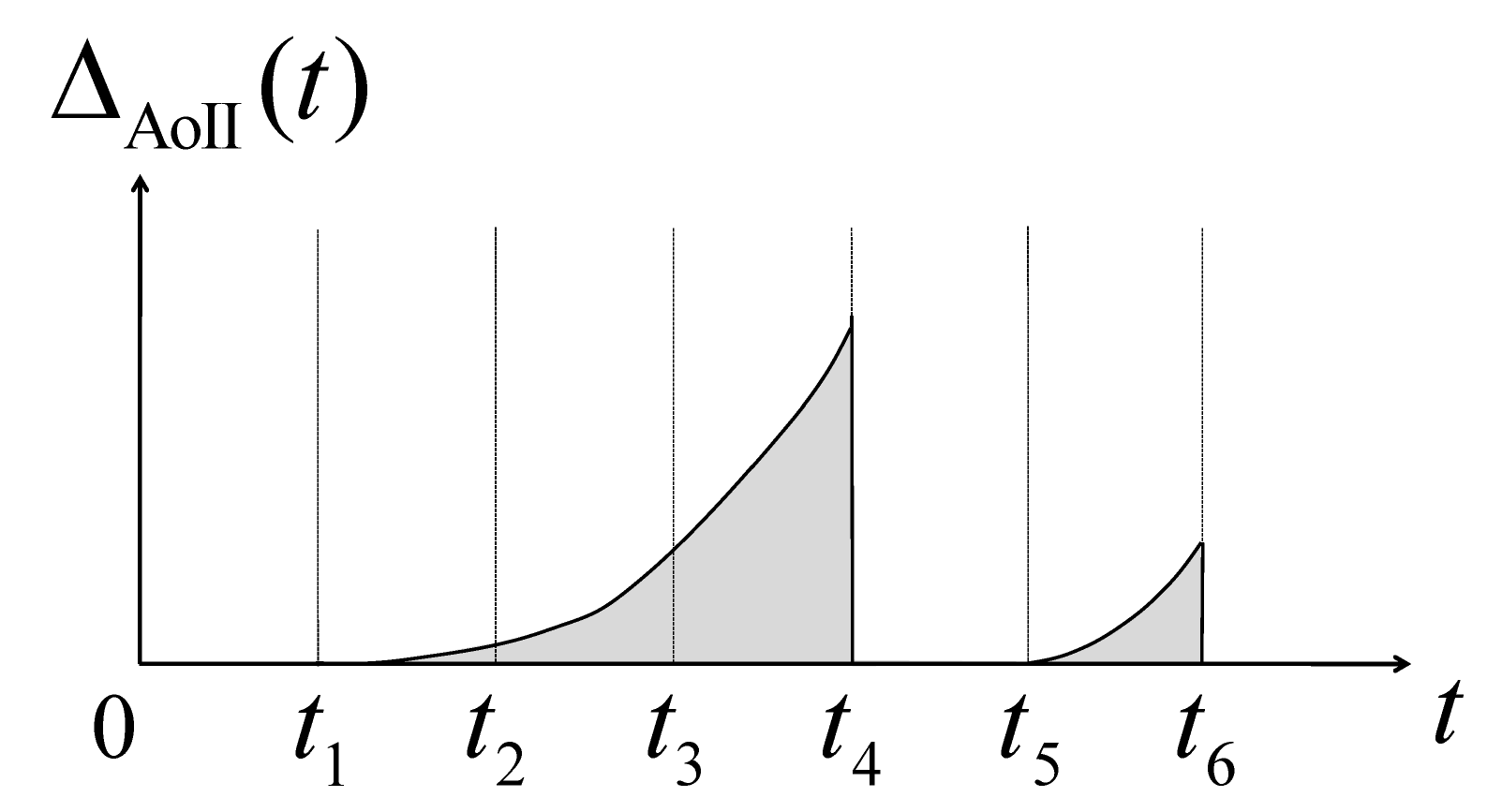}} 
			\end{minipage} 
			& (\ref{AoII}) & \cite{AoII2}  \\ \hline
			
			$\textbf{UoI}$
			& \begin{tabular}[c]{@{}c@{}}Time-variant cost   \\ depending on  \\ external environment \end{tabular} & \begin{tabular}[c]{@{}c@{}}Content-\\ aware\end{tabular}&\begin{tabular}[c]{@{}c@{}}{When}\\{environments}\\ {change rapidly}\end{tabular}
			& \begin{minipage}[c]{0.45\columnwidth}
				\centering
				\vspace{0.5mm} {\includegraphics[width=1\textwidth]{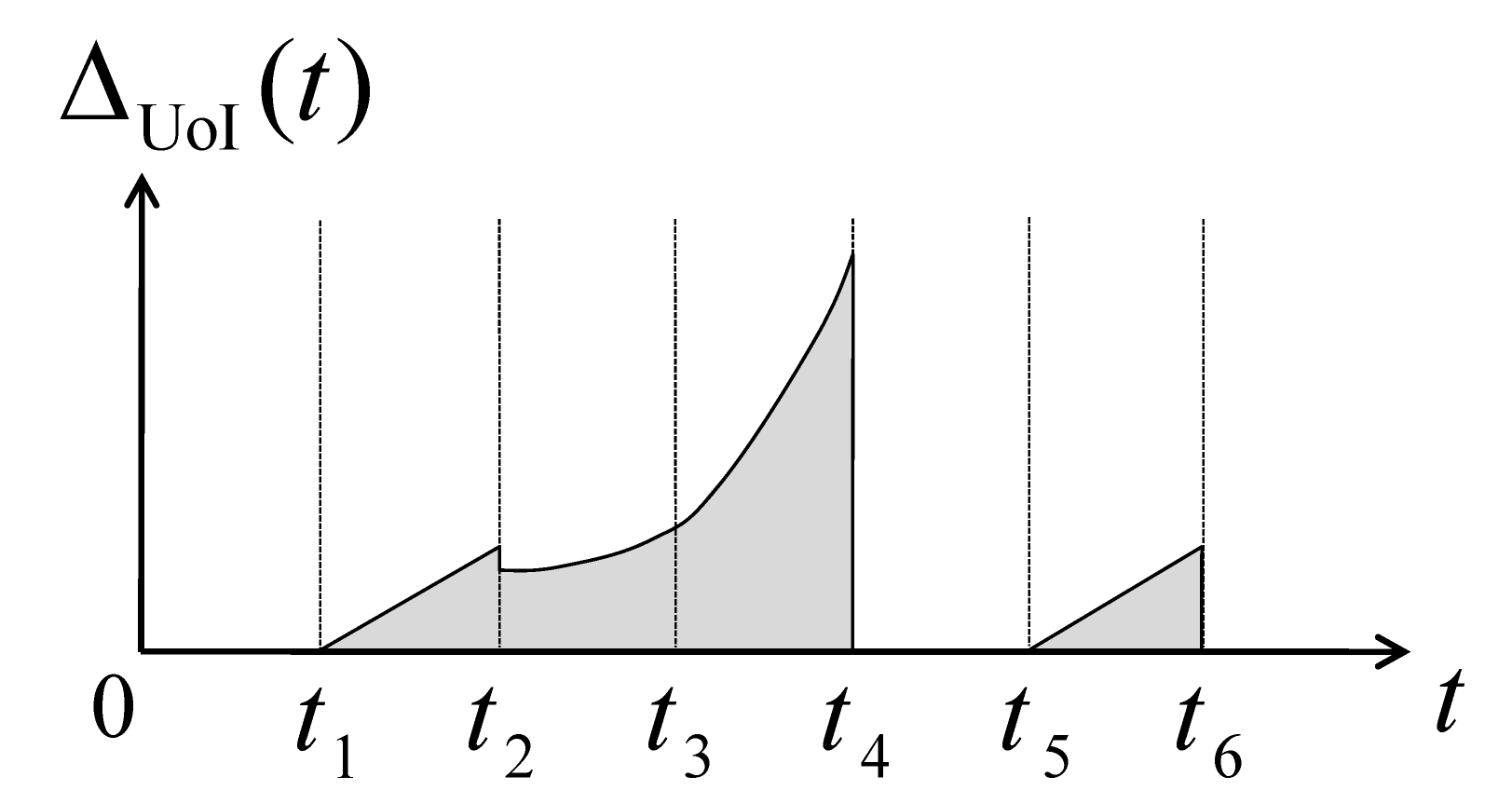}} 
			\end{minipage} 
			&  (\ref{UoI})& \cite{UoI1}  \\ \hline

			$\textbf{GoT}$
			& \begin{tabular}[c]{@{}c@{}} Customized cost  \\  depending on\\ environment, \\ source data, and\\ received data\end{tabular} & Unified &\begin{tabular}[c]{@{}c@{}}{Approximately}\\ 
				{any cases}\end{tabular}
			& \begin{minipage}[c]{0.45\columnwidth}
				\centering
				\vspace{0.5mm} {\includegraphics[width=1\textwidth]{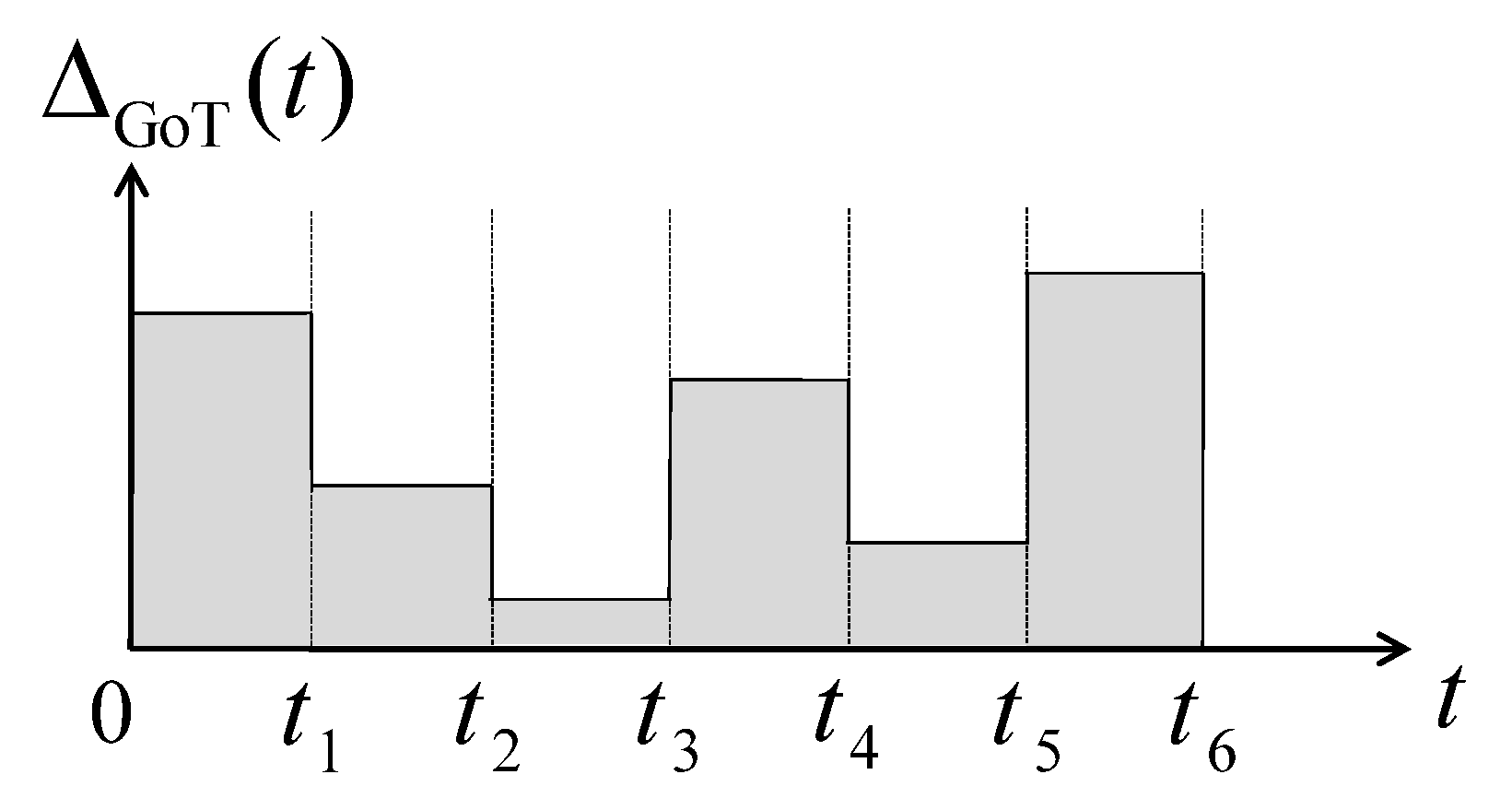}}
			\end{minipage} 
			&  (\ref{GoT})&\cite{beyondAoI} \small  \\ \hline
		\end{tabular}\centering
	\end{table*}
	
	We start our review on significance metrics by AoI, and then we introduce its variants, including nonlinear AoI,  age of synchronization (AoS), age of incorrect information (AoII), urgency of information (UoI) and the recently proposed goal-oriented tensor (GoT), to describe multi-dimensional significance (e.g., timeliness, status synchronization cost, and environment variation cost). The characteristics of significance metrics reviewed in this survey are listed in Table \ref{III_A}.

	AoI is firstly proposed in \cite{AoI1}, where the authors use this metric to find the optimal information transmission rate in vehicular networks. In \cite{AoI2}, the authors further use AoI to evaluate the performance of status update systems. Specifically, AoI is defined as the time elapsed since the latest successfully received status update was generated.  The instantaneous AoI expression for the system is
	\begin{equation}\label{AoI}
		\Delta_{\rm AoI}(t)=t-U(t),
	\end{equation}
	where $t$ represents the current time and $U(t)$ represents the generation time of the currently received information. The evolution of AoI is rather simple, that is, only when a status update is successfully received does AoI update as the age of the newly received update, and otherwise AoI continuously increases in a linear manner.
	
	In the SAGSIN, status update messages are exchanged among UAVs, satellites, ships, and ground base stations through wireless channels.  High AoI implies long delay and frequent errors of the status update transmission, thereby affecting the significance (or timeliness) of data. 
	Usually, the average AoI or peak AoI is used to measure long-term significance of data. Some research also focuses on the AoI at specific query instants to propose query age of information (QAoI)\cite{qaoi} as an AoI variant.
	
	However, the linearly increasing cost of AoI is insufficient in describing the nonlinear significance variation incurred by untimely information. For applications that are more sensitive to time, such as autonomous driving and remote surgery, the performance degradation caused by information aging may not be a linear function of time. For instance, in the state estimation problem of Gaussian linear time-invariant systems, if the system is stable, the state estimation error (which directly determines the significance of source data) is a sublinear function of $\Delta_{\rm AoI}$; if the system is unstable, the state estimation error increases exponentially with $\Delta_{\rm AoI}$\cite{AoI5}. Therefore, it is necessary to evaluate the significance of data which is varying nonlinearly in age. Some scholars have proposed  value of information (VoI)\cite{voi} (also called nonlinear AoI\cite{AoI6}). Nonlinear AoI is a nonlinear function of AoI, that is
	\begin{equation}\label{nonlinear}
		\Delta_{\rm nonlinear AoI}(t)=f(t-U(t)),
	\end{equation}
	where $f(\cdot )$ is a nonlinear penalty function for linear information age. For example, if the penalty function is chosen as exponential function, a possible evolution of nonlinear AoI is shown in Table \ref{III_A}. The choice of penalty function is dependent on how the significance varies in data aging. If the significance increases slowly in time, we adopt functions with decreasing derivatives in time, such as logarithmic function; in contrast, if the significance increases rapidly, we adopt functions with increasing derivatives in time, such as exponential function. 
	
	However, both AoI and nonlinear AoI only reflect the time elapsed from the generation of the current information to its successful reception, without considering whether the receiver correctly estimates the information from the source. That is, they only focus on the timeliness of the currently transmitting information and ignore the real-time status synchronization between the recovered information and the current source status information. This characteristic of AoI (and nonlinear AoI)  is also called content-agnostic. However, synchronization of statuses between transceivers also affects the significance of data, since wrong estimation will lead to biased decision on current status and cause severe consequences. For instance, missed detection of the fire will cause considerably large financial loss due to untimely rescue. Therefore, to reflect whether the receiver fully understands the source information, researchers have proposed a series of content-aware significance metrics including AoS, AoII, and UoI. 
	
	In \cite{aos} the authors define a novel metric called  AoS, which extends the freshness of information to the synchronization duration of information. Unlike AoI, AoS is calculated as the duration since the latest time of perfect inference of the current process status, that is
	\begin{equation}\label{AoS}
		\Delta_{\rm AoS}(t)=t-W(t),
	\end{equation}
	where $W(t)$ represents the latest time slot that the statuses between transceivers are synchronized. A possible AoS evolution shown in Table \ref{III_A} demonstrates that when statuses are synchronized,  AoS is equal to zero, and otherwise AoS linearly increases in asynchronized status duration.
	
	As a general case of AoS, AoII is defined as a  function of AoS. Specifically, in  \cite{AoII2}, AoII is described as a (non)linear function with regard to time duration multiplying a function reflecting the difference between source status and estimated status, that is
	\begin{equation}\label{AoII}
		\Delta_{\rm AoII}(t)=f(t)\times g(X(t),\hat{X}(t)),
	\end{equation}
	where $X(t)$ is the source status at time $t$, $\hat{X}(t)$ is the estimated result of $X(t)$ predicted by the receiver, $f(t)$ is an increasing time penalty function, and $g(X(t),\hat{X}(t))$ is an information penalty function that reflects the difference between the predicted result of the receiver and the actual status. Since the difference of two identical statuses is zero, the information penalty function $g(X(t),\hat{X}(t))$ should be equal to zero when $X(t)=\hat{X}(t)$. The AoII evolution is similar to that of AoS, that is, if the statuses between transceivers are synchronized, AoII is zero; if they are asynchronized, then AoII increases with the time duration of the asynchronized state in a (non)linear manner.
	
	Furthermore, \cite{UoI1} proposes a new metric called UoI, which is used to measure the nonlinear time-varying significance of state information by considering the impact of external environment. Specifically, UoI is defined as
	\begin{equation}\label{UoI}
		{\Delta _{{\text{UoI}}}}(t) = f(t,{\mathbf{\Phi}}(t)) \cdot g(X(t),\hat X(t)),
	\end{equation}
	where ${\mathbf{\Phi}}(t)$ represents the environment states at time $t$. That is, the time penalty function $f(t)$ is time-variant according to the current environment state ${\mathbf{\Phi}}(t)$. As shown in Table \ref{III_A}, the UoI evolution is rather complex due to time-variant environment. Even the current statuses between transceivers keep asynchronized during time interval $[t_1,t_4]$, the manners of UoI increase are different in $[t_1,t_2]$ (linear) and $[t_2,t_4]$ (exponential), for the data in the latter time interval are more significant due to more urgent environment.

\begin{figure*}[t]
		\centering
		\includegraphics[width = 0.96\textwidth]{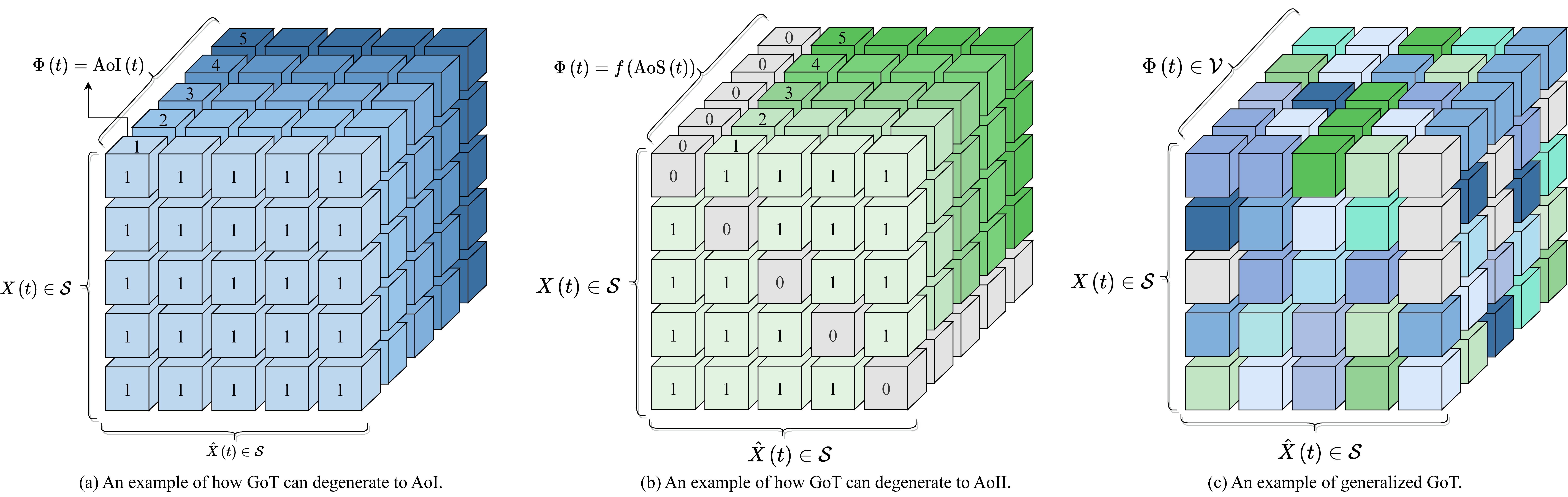}
		\caption{Examples of how GoT can degenerate to existing metrics. By setting the environment states as related to current message ages, the GoT can characterize AoI shown in subfigure (a). By setting the matrices composed of source status dimension and estimated status dimension as symmetric with values on the main diagonal as zeros, the GoT can describe AoII shown in subfigure (b). Naturally, by relaxing these constraints, a generalized metric called GoT can be constructed as shown in subfigure (c).}
		\label{figGoT}
	\end{figure*}

	The recent works enrich the concept of significance to propose comprehensive tensor-based metric which unifies the above significance metrics. Specifically, the authors in \cite{beyondAoI} investigate a novel performance metric called GoT to directly quantify the significance of data considering the impacts of environments, status synchronization, and inherent costs, which is defined as
	\begin{equation}\label{GoT}
		{\Delta _{{\text{GoT}}}}(t) = {\text{GoT}}(X(t),\hat X(t),{\mathbf{\Phi}}(t)),
	\end{equation}
	where $\text{GoT}(\cdot)$ is a tensor with three dimensions, namely source status, estimated status, and environment states. Each element of GoT reflects the significance value (i.e., inherent cost) depending on the difference of source and estimated statuses between transceivers and current environment state. As shown in Table \ref{III_A}, the significance values in different time intervals are rather distinct. The authors prove that the GoT metric can reduce to existing metrics such as AoI, AoS, AoII, and UoI under certain conditions, as illustrated in Fig. \ref{figGoT}. 
	
	\begin{remark}
		{ The preference of significance metric depends on our demands on system performance. Specifically, content-agnostic metrics can describe the delay aspect (or staleness), while content-aware metrics captures the fidelity aspect (or mismatch). Therefore, we should firstly explore the demand of specific services and secondly choose corresponding metrics. For instance, in a remote control scenario under time-varying environment, UoI may demonstrate the significance of information better than AoII, since remote control services necessitate precise source status information calling for content-aware metrics, and UoI is more suitable for describing time-varying costs. Also, since GoT is a unified significance metric, it is preferable  in most services as long as the number of statuses $X(t)$ is sufficiently large. In table \ref{III_A}, we have briefly compared the application scenarios of each metric.}
	\end{remark}

	\subsection{Techniques: Towards Significance Protection}\label{sec3B}
	This subsection discusses the design of a significance-oriented semantic communication system within SAGSIN. The main focus is on ensuring appropriate sampling and accurate transmission of information with high significance, referred to as ``significance protection". Taking into account the characteristics of SAGSIN, such as large-scale scenarios, highly dynamic channels, and limited device capabilities, we categorize existing research into the design of sampling, coding, and modulation policies, which we term as SPTs. These techniques aim to optimize the significance of transmitted data throughout the signal processing and transmission stages, which are listed in Table \ref{III-B}.

	\begin{table*}[!t]\normalsize 
		\renewcommand{\arraystretch}{1.4} 
		\caption{SUMMARY OF SIGNIFICANCE-PROTECTION TECHNIQUES}\label{III-B}
		\begin{tabular}[b]{|c|c|c|cc|c|} 
			\hline 
			$\textbf{Perspective}$ & \renewcommand{\arraystretch}{1.2} \begin{tabular}[c]{@{}c@{}}$\textbf{\normalsize SAGSIN}$\\ $\textbf{\normalsize Issues}$ \\ $\textbf{\normalsize Addressment}$ \end{tabular} & \renewcommand{\arraystretch}{1.2}  \begin{tabular}[c]{@{}c@{}}$\textbf{Related}$\\$\textbf{Metrics}$\end{tabular}& \multicolumn{2}{c|}{$\textbf{Technical Details}$} & \renewcommand{\arraystretch}{1.2}  \begin{tabular}[c]{@{}c@{}}$\textbf{Related}$\\$\textbf{References}$\end{tabular} \\ 
			\hline
			
			\multirow{8.2}{*}{$\textbf{Sampling}$} 
			& \multirow{8.2}{*}{\begin{tabular}[c]{@{}c@{}}\normalsize Limited device\\ \normalsize capabilities\end{tabular}} & \multirow{8.2}{*}{\begin{tabular}[c]{@{}c@{}} AoI, \\ AoS, \\AoII, \\ Nonlinear\\ AoI \end{tabular}} & \multicolumn{1}{c|}{\multirow{4.2}{*}{\begin{tabular}[c]{@{}c@{}} Content- \\ agnostic \end{tabular}}} & \begin{tabular}[c]{@{}c@{}} AoI-optimal threshold-based sampling \\ with (non)linear age penalty \end{tabular} & \small \cite{updateorwait, sampleiot, AoI6} \\ \cline{5-6}								
			& & &\multicolumn{1}{c|}{}  & \begin{tabular}[c]{@{}c@{}} AoI-optimal threshold-based  sampling \\ for Wiener or Ornstein-Uhlenbeck sources \end{tabular} & \small \cite{wiener, OUprocess} \\ \cline{5-6}
			& & &\multicolumn{1}{c|}{}  & Multi source joint sampling &  \small \cite{multisourcesample}\\  \cline{4-6}		
			& &	& \multicolumn{1}{c|}{\multirow{4}{*}{\begin{tabular}[c]{@{}c@{}} Content- \\ aware \end{tabular}}} & AoS-optimal threshold-based sampling & \small \cite{aos}  \\ \cline{5-6}	
			& &	& \multicolumn{1}{c|}{} & AoII-optimal threshold-based sampling & \small \cite{AoII2,aoiipower,semanticaoii} \\ \cline{5-6}	
			& &	& \multicolumn{1}{c|}{} & AoII-optimal sampling under random delay  & \small \cite{AoIIdelay} \\ \cline{5-6}	
			& &	& \multicolumn{1}{c|}{} & AoII-optimal sampling with retransmission & \small \cite{AoIIHARQ} \\ \cline{5-6}	
			\hline
			
			\multirow{7}{*}{$\textbf{Coding}$} 
			& \multirow{7}{*}{\begin{tabular}[c]{@{}c@{}}\normalsize Highly dynamic\\ \normalsize channels, \\ large-scale \\ scenarios \end{tabular}} & \multirow{7}{*}{AoI} & \multicolumn{1}{c|}{\multirow{3}{*}{\begin{tabular}[c]{@{}c@{}} Neglecting \\ propagation \\ delay \end{tabular}}} & \multicolumn{1}{c|}{HARQ-IR} & \small \cite{RateAlloc8, RateAlloc9}  \\ \cline{5-6}	
			& & & \multicolumn{1}{c|}{} & \multicolumn{1}{c|}{Truncated HARQ-CC for LDPC code} & \small \cite{RateAlloc7} \\ \cline{5-6}
			& & & \multicolumn{1}{c|}{} & \multicolumn{1}{c|}{HARQ-based Polar coded system} & \small \cite{RateAlloc1} \\ \cline{4-6}
			& & & \multicolumn{1}{c|}{\multirow{4}{*}{\begin{tabular}[c]{@{}c@{}} Considering \\ propagation \\ delay \end{tabular}}} & \multicolumn{1}{c|}{Codeblock assignment} & \small \cite{RateAlloc3} \\ \cline{5-6}
			& & & \multicolumn{1}{c|}{} & \multicolumn{1}{c|}{HARQ-IR } & \small \cite{RateAlloc4, RateAlloc5, RateAlloc6} \\ 	\cline{5-6}
			& & & \multicolumn{1}{c|}{} & \multicolumn{1}{c|}{HARQ-based Spinal coded system } & \small \cite{RateAlloc2} \\ 	\cline{5-6}
			& & & \multicolumn{1}{c|}{} & \multicolumn{1}{c|}{HARQ-based Polar coded system } & \small \cite{RateAlloc10} \\ 	
			
			\hline
			
			\multirow{3}{*}{$\textbf{Modulation}$} 
			& \multirow{3}{*}{\begin{tabular}[c]{@{}c@{}}\normalsize Limited device\\ \normalsize capabilities\end{tabular}} & \multirow{3}{*}{AoI} & \multicolumn{2}{c|}{NOMA/OMA} & \small \cite{PowerAlloc4, PowerAlloc1}  \\ \cline{4-6}	
			& & & \multicolumn{2}{c|}{HARQ-CC aided NOMA} & \small \cite{PowerAlloc2} \\ \cline{4-6}
			& & & \multicolumn{2}{c|}{NOMA in S-IoT network} & \small \cite{PowerAlloc5, PowerAlloc6} \\ 	 								
			\hline
			
		\end{tabular}\centering
	\end{table*}
	
	\subsubsection{Significance-oriented sampling policy design}	
\begin{figure}[t]
		\centering
		\includegraphics[width = 0.48\textwidth]{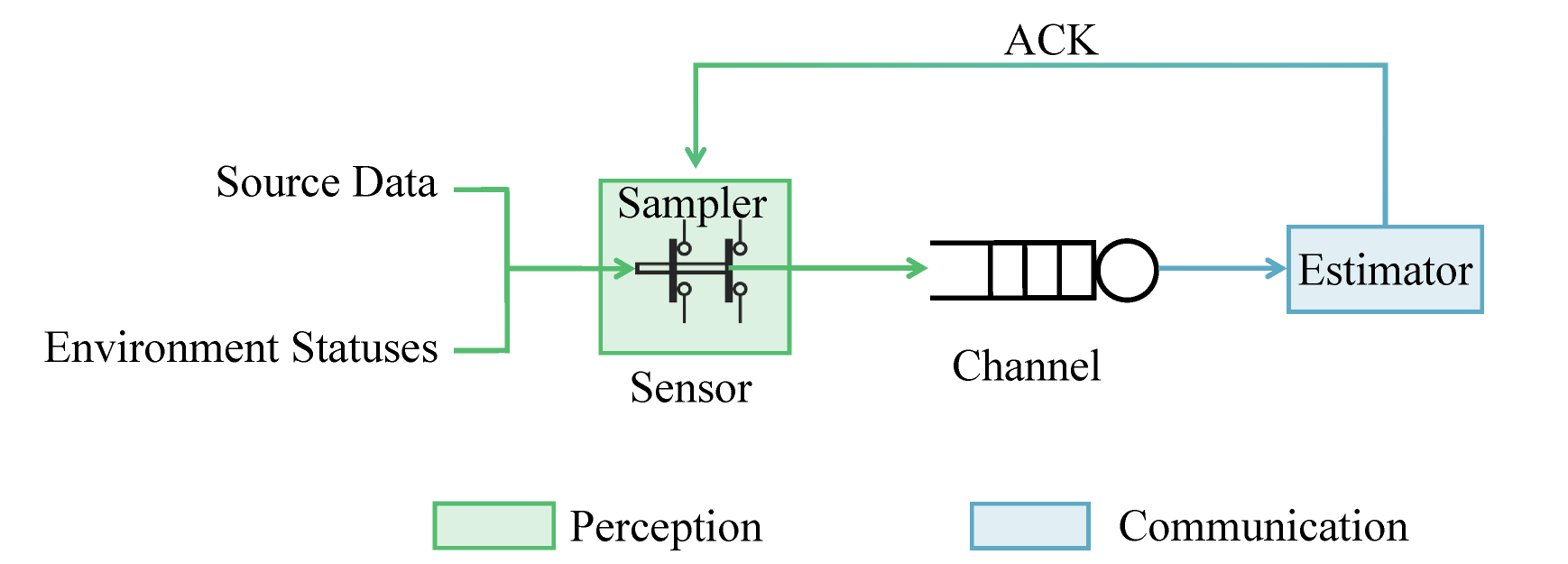}
		\caption{A general system model of status update system facilitated by significance-oriented sampling. It is assumed that source data, environment statuses, and channel are all time-variant, affecting the significance of source data and thus determining whether to sample the current update. Moreover, the channel is usually assumed to be a queuing system with/without preemption, which will also affect the optimal sampling policy.}
		\label{figsample}
	\end{figure}

	Given the limited device capabilities of the SAGSIN, significance-oriented sampling policy is designed to alleviate the burden on data transmission. Generally, in a status update system shown in Fig. \ref{figsample}, a status update will be sampled only when the significance metric of the system reaches a certain threshold, implying that the current status update is significant enough that  the receiver should acquire the current source status. By such significance-oriented sampling, most of data with less importance are filtered by the sampler, with only data with more semantics (i.e., significance) being sampled and transmitted to the receiver. Thus, the pre-processing burden of transmitter gets initially released, saving considerably large perception and communication resources which is severely limited the SAGSIN. 
	
	\textit{Content-agnostic sampling policy design. }Since AoI is a typical timeliness metric, the design of AoI-related sampling policy has been studied widely in the literature. Intuitively, a zero-wait sampling policy, where a new update is sampled just when an acknowledgment (ACK) feedback is received by transmitter, will be naturally AoI-optimal since the newest sampled status update is the freshest and contains the most semantics. However, the authors in  \cite{updateorwait} point out that zero-wait policy is not always AoI-optimal, especially when the status varies slowly, since most of the status updates are repetitive and thus useless. To solve the AoI-optimal sampling problem for such cases, \cite{updateorwait} adopts a penalty function to measure the dissatisfaction level on data staleness, and formulates the age penalty minimization problem as a semi Markov decision process (MDP), which is solved by a divide-and-conquer approach. The zero-wait policy is proven to be optimal when  the variance of log-normal distribution for service time is small; in contrast, the policy of waiting is optimal when the variance is large (i.e., a heavy tail distribution) or the penalty function increases rapidly with age growing. Similarly, the AoI-optimal sampling policy under energy constraint is solved in  \cite{sampleiot}. The choice of nonlinear age penalty and its impact on optimal sampling policy can also be seen in \cite{AoI6}.
	
	With regard to AoI-optimal sampling policy for specific distributions of source data,  \cite{wiener} discusses the optimal sampling policies for remote estimation of a Wiener process for different metrics including AoI and MSE. Specifically, the AoI-optimal sampling policy follows a threshold structure, that is, only when AoI reaches a threshold does the sampler generate a new status update for transmission. Meanwhile, the MSE-optimal sampling policy is equivalent to AoI-optimal one only when sampling times  are independent with Wiener process; however, when the sampling times are determined by the knowledge of the Wiener process, the AoI-optimal sampling policy will no longer achieve the minimum estimation error as compared to MSE-optimal one. Another similar work is  \cite{OUprocess}, where the optimal sampling policy of a Ornstein-Uhlenbeck source is studied serving as a general case of the study in \cite{wiener}. The sampling optimization problem is formulated as a time-continuous MDP, and the solution is also proven to be a threshold structure related to instantaneous estimation error (which is related to MSE). 
	
	The above research focuses on sampling design of single source. For the case of multiple sources,  \cite{multisourcesample} studies the total-peak-AoI-optimal and total-average-AoI-optimal sampling strategies, respectively. To simplify the solution of overall scheduling-sampling policy for multi-source case, the authors firstly prove that a maximum age first  scheduling policy provides the best age performance. Next,  the optimal sampling policy for each source can be solved easily by a dynamic programming algorithm based on the maximum age first policy. The total-peak-AoI-optimal sampling policy is proven to be zero-wait,  while the total-average-AoI-optimal sampling policy can be solved by water filling, which holds a threshold structure similar to \cite{updateorwait,wiener,OUprocess}.
	
	\textit{Content-aware sampling policy design. }Although AoI-oriented sampling policy does select more significant data for transmission, this design is inherently content-agnostic, since the sampler selects only fresh status updates without knowing the content of data. In order to design a content-aware significance-oriented sampling policy, we should resort to  AoI variants such as AoS and AoII, etc.

	A major content-aware significance metric is AoII, and thus we review AoII-related data sampling optimization. In  \cite{aos}, the AoS-optimal and AoI-optimal sampling rates are respectively solved under a multi-source scenario. As a general case of AoS, the authors in  \cite{AoII2} discuss AoII-optimal sampling policy with and without power constraints respectively. Under no power constraint, the optimal policy is ``always update", which also achieves the minimum AoI and MSE simultaneously; however, the optimal policy is complex under a power constraint, which is solved by constraint MDP.   \cite{aoiipower} further studies AoII-optimal sampling policy under power constraint over an unreliable channel, where a threshold-based policy is proven to be optimal and adopted to reduce the complexity of global optimization problem. The authors in  \cite{semanticaoii} point out the semantic (i.e., significance) characteristics of AoII, and showcase the superiority of significance-oriented AoII-optimal policy as compared to MSE-optimal and AoI-optimal ones. Considering the non-trivial transmission delay which is a fundamental characteristic in the SAGSIN,  \cite{AoIIdelay} solves AoII-optimal sampling policy when the status updates experience random delay, which also holds a threshold structure related to the maximum transmission delay. Also, a recent work\cite{AoIIHARQ} introduces limited retransmission with resource constraint to enhance the AoII performance.
	
	\subsubsection{Significance-oriented coding policy design}
\begin{figure*}[ht]
		\centering
		\subfigure[ARQ.]{\label{arq} \includegraphics[angle=0,width=0.49\textwidth]{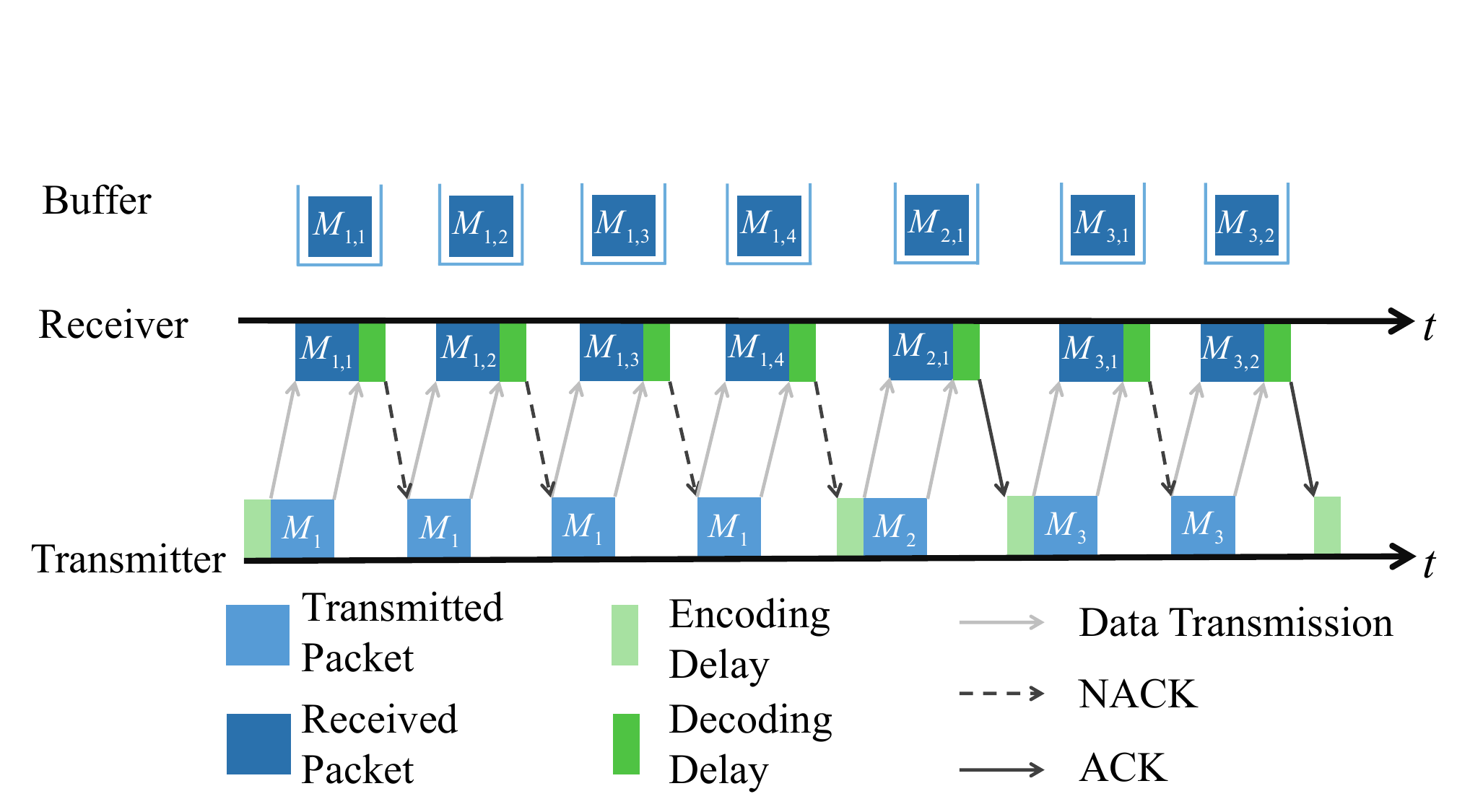}}
		\subfigure[HARQ-CC.]{\label{harqcc}	\includegraphics[angle=0,width=0.49\textwidth]{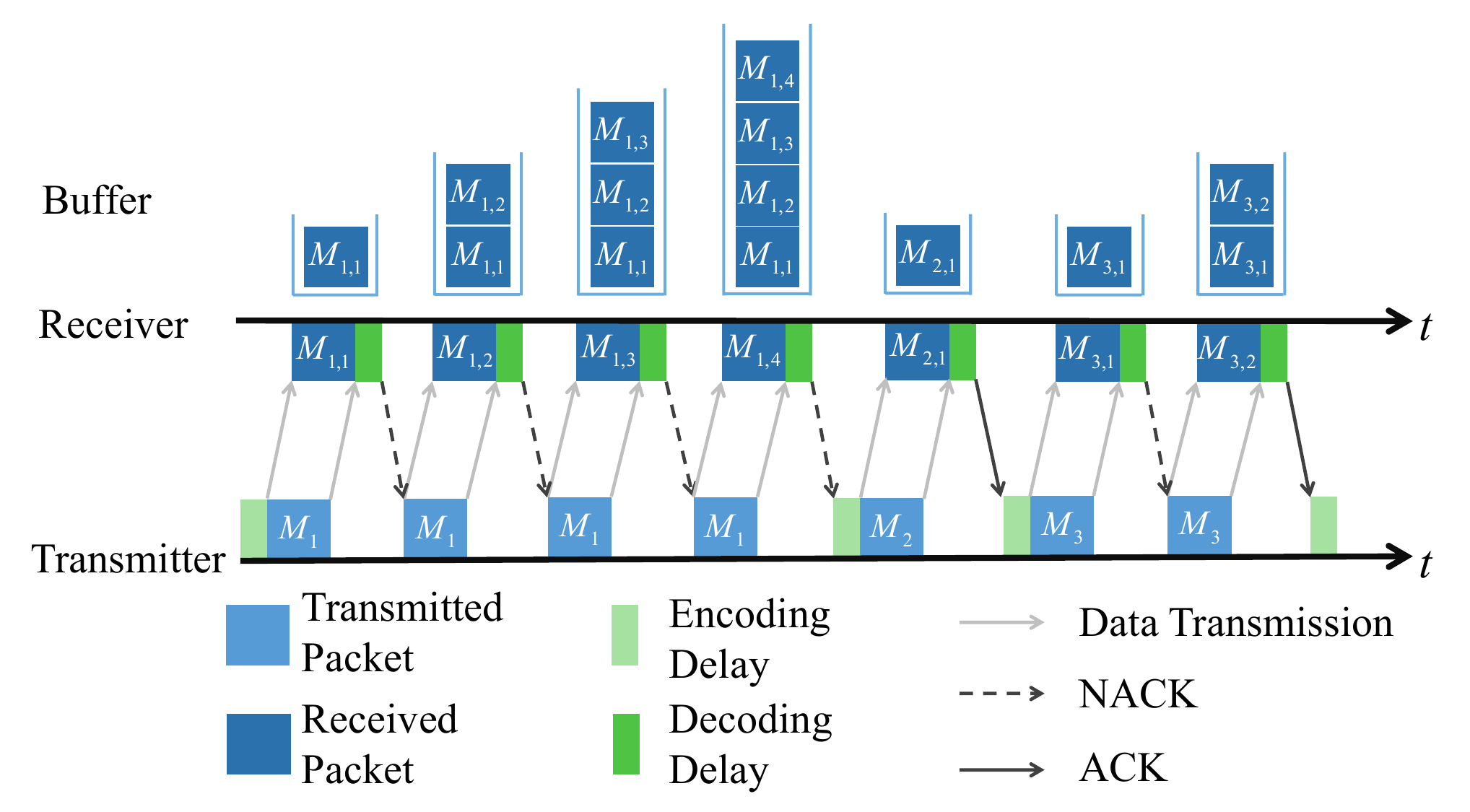}}
		\subfigure[HARQ-IR.]{\label{harqir}	\includegraphics[angle=0,width=0.49\textwidth]{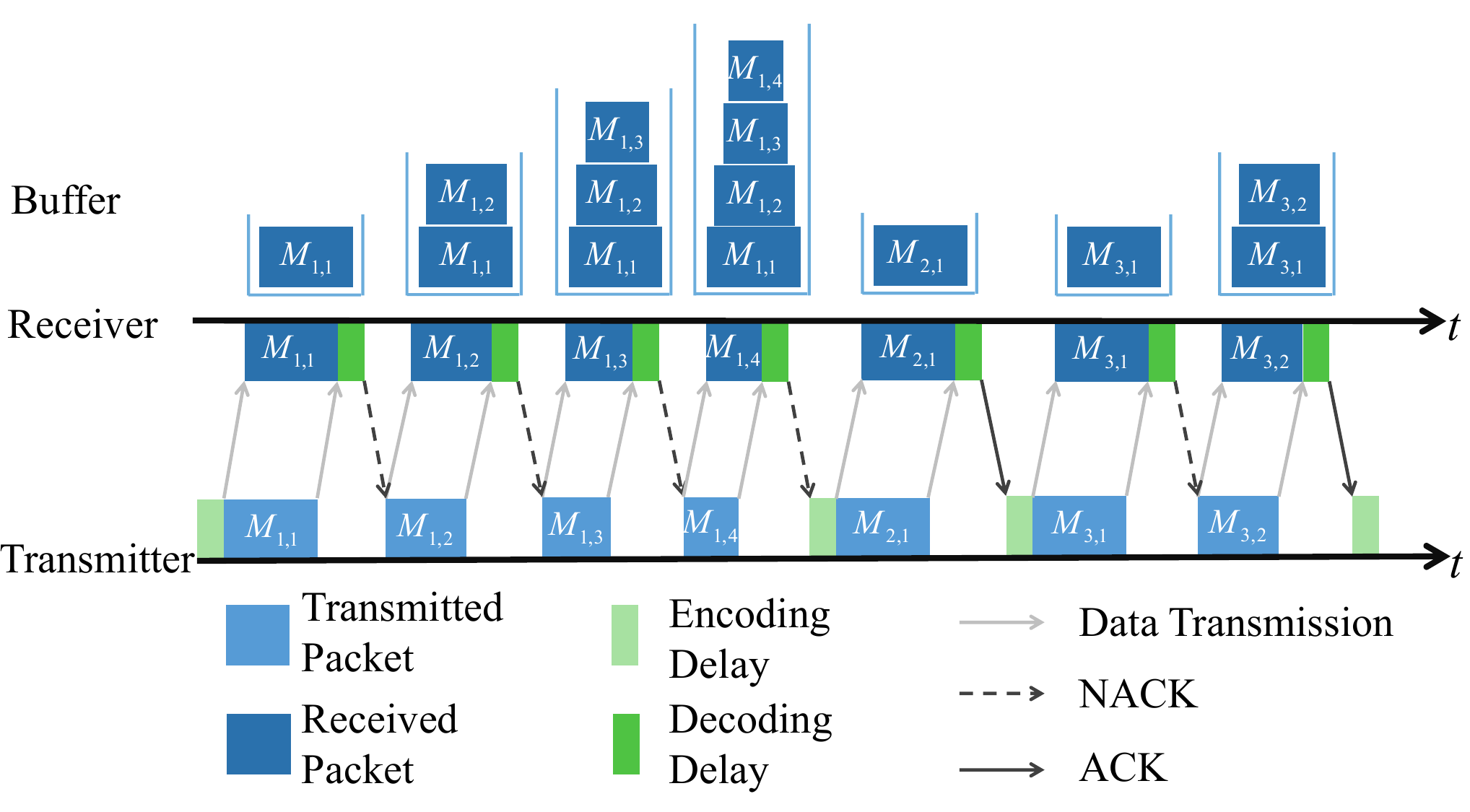}}
		\subfigure[Proactive HARQ-CC.]{\label{proactive}	\includegraphics[angle=0,width=0.49\textwidth]{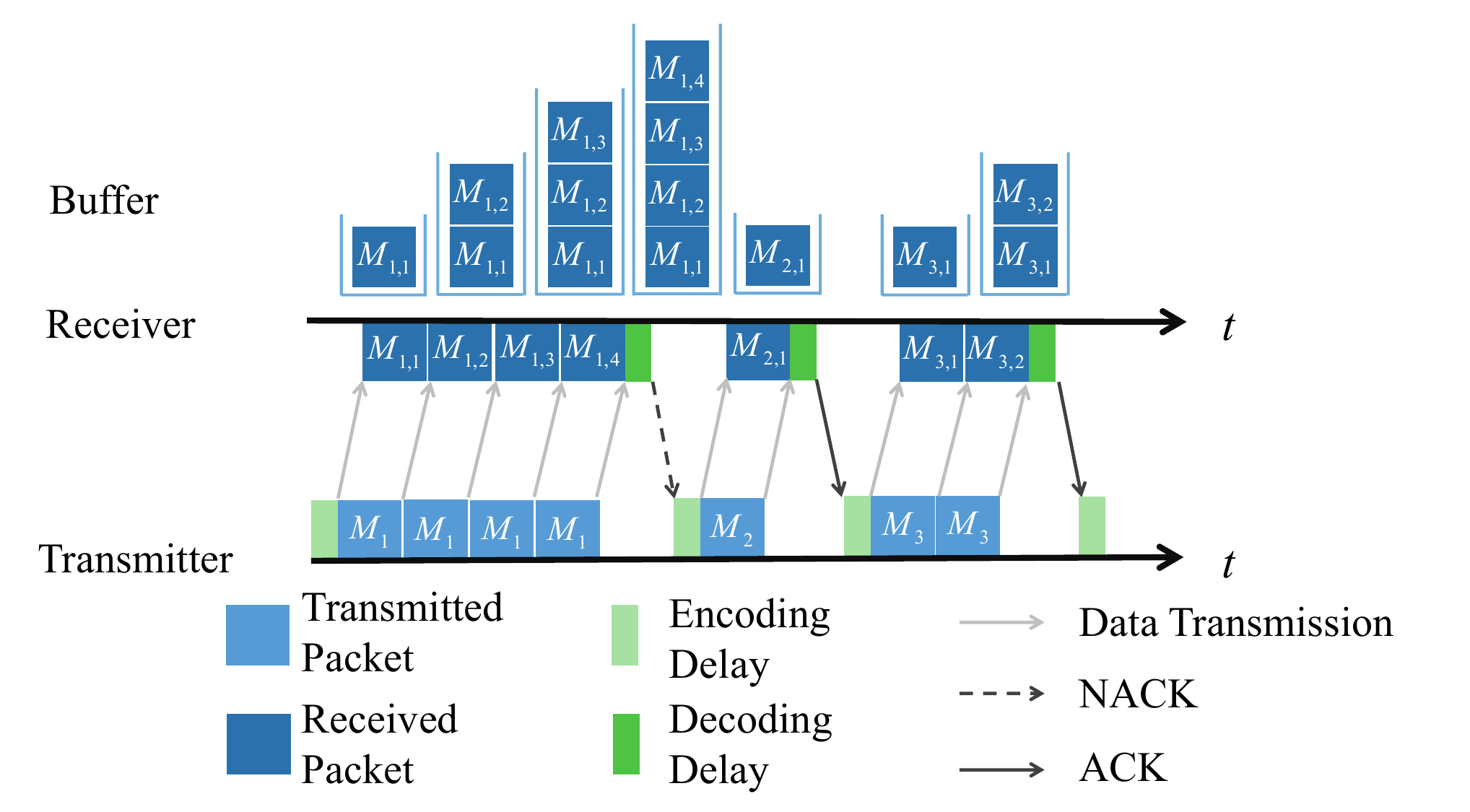}}
		\caption{The comparison of ARQ, HARQ-CC, and HARQ-IR. Specifically, the main difference between ARQ and HARQ is whether receiver uses the previous packets (or sub-blocks) stored in the buffer in the current decoding. Moreover, the packets transmitted in HARQ-CC are the same in the retransmission rounds, and the receiver simply combine the received packets using maximum ratio combining method. However, the packets transmitted in HARQ-IR (also called sub-blocks) are different in the retransmission rounds (usually with different code lengths), and the receiver combine the received sub-blocks into a whole vector for decoding. Hence, HARQ-IR usually achieves high coding gain than HARQ-CC, while the protocol design of the former is naturally more complex. Note that the subfigures (b) and (c) both show  reactive HARQ protocols where a retransmission occurs only if a negative acknowledgment (NACK) feedback is received by transmitter, while subfigure (d) shows a proactive HARQ protocol with chase combining. In proactive HARQ, the transmitter will continuously retransmit the packets no matter whether the message is successfully received or not. Unless otherwise specified, the mentioned HARQ protocols in this paper are all reactive HARQ.
		}\label{figHARQ}
	\end{figure*}

	Considering the large-scale scenarios and highly dynamic channels of the SAGSIN, the implementation of significance-oriented coding policy necessitates the allocation of data rates,   striking a trade-off between system efficiency and reliability. In particular, in scenarios involving bad channel conditions or a large volume of significant data, introducing more redundancy can improve the successful transmission probability. However, in contrast, when the channel conditions are favorable or there are less important data, using fewer codewords can still yield satisfactory results. A widely used rate allocation method is introducing retransmission protocols, such as automatic repeat request (ARQ) and hybrid ARQ (HARQ) including HARQ with chase combining (HARQ-CC) and HARQ with incremental redundancy (HARQ-IR). ARQ and HARQ protocols accomplish code rate allocation by dynamically adapting the number of retransmissions or/and the redundancy length in each retransmission, taking into account the factors such as channel conditions, delay constraints, and data significance. The differences among simple ARQ protocol, HARQ-CC protocol, and HARQ-IR protocol are illustrated in Fig. \ref{figHARQ}.
	
	\textit{Coding policy design without considering propagation delay.} Most of the works in this field have only considered the transmission delay and neglected the propagation delay, i.e., the only delay elements that affect AoI evolution is the code length. For example,  \cite{RateAlloc8} considers a scenario  where the sensor transmits the collected data to a central entity over an unreliable link, and coding policies with low latency are designed to protect significant data. HARQ is used for the transmission of updates and the relationship between average timeliness  and physical layer  decisions (i.e., whether to retransmit the old data or send a new data) is analyzed. The analysis reveals a trade-off between average feedback rate and average timeliness. Specifically, for a constrained set of HARQ code word lengths, refining the code length can improve the average timeliness at the receiver. The work formulates the optimization problem with average age as the objective function and finds out the block allocation vector that minimizes the average age under the constraint of the average feedback rate. The results show that HARQ can greatly outperform ARQ protocol and fixed-length coding schemes with no retransmission when the size of the incremental redundancy sub-block is properly chosen. The HARQ protocol considered in \cite{RateAlloc8} is called HARQ-IR (illustrated in Fig. \ref{harqir}).  Different from the ARQ protocol where  the same packet is retransmitted after a failed decoding until ACK reception, in the HARQ-IR protocol, a different sub-block from the previous packet will be transmitted in each retransmission, and the receiver combines previous  sub-blocks with the current sub-block in the decoding process to enhance successful decoding  probability\cite{compareHARQ}.
	
	In \cite{RateAlloc9}, the authors utilize the HARQ-IR protocol  in the status update system over a noisy channel. In case of decoding failure at the receiver, IR bits are transmitted to increase the successful decoding probability in the future. If decoding remains unsuccessful after a specified duration, a new status update is transmitted as a replacement. Conversely, when decoding is successful, the transmitter enters an idle state for a certain period after successful transmission, prior to sending a new update. The research  focuses on optimizing the code word and IR length for each update, along with the waiting time, with the objective of minimizing the long-term average AoI over the binary symmetric channel (BSC).
	
	In recent years, there has been an increase in research focusing on analyzing and comparing AoI based on retransmission protocols for specific state-of-the-art channel coding techniques. The authors in \cite{RateAlloc7} firstly investigate the AoI performance of specific coding schemes. Specifically, the authors analyze the average AoI and energy cost in LDPC coded status update system with and without ARQ using a fixed redundancy scheme. Different coding policies, including the non-ARQ, classical ARQ, truncated ARQ, and truncated HARQ-CC protocols, are analyzed and compared. Among them, truncated ARQ (illustrated in Fig. \ref{arq}) involves repeated transmission of the current update until a maximum number of transmissions or a successful reception is achieved. Moreover, truncated HARQ-CC (illustrated in Fig. \ref{harqcc}) combines previous packets with currently received packet for decoding using maximum ratio combining method. In HARQ-CC, additional encoding schemes are unnecessary as the packets are the same for each retransmission. Additionally, the accumulation effect of signal-to-noise ratio (SNR) in each HARQ retransmission improves the successful decoding probability. These inherent characteristics empower truncated HARQ-CC to strike a balance between AoI and energy consumption, achieving the best average age and moderate energy cost among the considered protocols. 
	
	As for AoI performance of status update system coded by capacity-achieving codes, the average AoI performance of Polar coded status updates over the additive white Gaussian noise (AWGN) channel is firstly investigated in \cite{RateAlloc1}. Similar to the conclusion drawn by \cite{RateAlloc7}, compared to non-ARQ protocols, the HARQ-CC protocol can effectively reduce the average AoI, especially in low SNR regions. To further optimize the AoI of the system, two methods are proposed to optimize the design SNR and puncture length, both of which are important construction parameters of Polar codes. 
	
	\textit{Coding policy design considering propagation delay. }Instead of only concentrating on transmission delay, some research further considers other types of delays (including propagation delay) and specific application scenarios, thereby providing a more precise depiction of the transmission process in the SAGSIN \cite{RateAlloc3,RateAlloc4,RateAlloc5,RateAlloc6}. The authors in \cite{RateAlloc3} conduct a comprehensive analysis of the AoI for two types of HARQ protocols, namely reactive HARQ and proactive HARQ (the latter protocol is illustrated in Fig. \ref{proactive}). Considering various types of delay, including coding delay, transmission delay, propagation delay, decoding delay, and feedback delay, \cite{RateAlloc3} derives unified closed-form expressions for the average AoI and average peak AoI for the two protocols. Based on the derived explicit expressions, an optimization problem is formulated to minimize the AoI by investigating the optimal block assignment strategy in the finite block-length (FBL) regime. The numerical results  show that proactive HARQ offers advantages in terms of both age performance and system robustness. 
	
	In \cite{RateAlloc4}, the authors focus on the scenarios characterized by non-trivial propagation delays, such as space communications and satellite communications. The authors study AoI performance in a communication system with non-trivial propagation delays, where status updates  are transmitted to the receiver through a binary erasure channel (BEC). In order to mitigate the effect of erasures on timeliness performance, an HARQ-IR protocol with a predetermined maximum number of retransmissions is employed. It is shown that a critical  upper threshold related to propagation delay exists, within which retransmissions benefits the  AoI. In cases where the propagation delay is longer than the threshold, allocating all the symbols   to the first transmission is the optimal coding policy. Following \cite{RateAlloc4}, in \cite{RateAlloc5}, the author considers a satellite-based IoT system, where IoT devices observe physical processes and transmit status updates to a monitor node through an error-prone channel with significant propagation delay. \cite{RateAlloc5} applies  HARQ-IR protocol   to the satellite-based IoT system and investigates the age-optimal redundancy allocation problem under reliability constraints. The results also demonstrate the superiority of HARQ-IR in minimizing AoI below a certain propagation delay threshold. In SAGSIN scenarios, \cite{RateAlloc6} presents a novel fast HARQ-IR protocol  which omits the  decoding and feedback operations of the standard HARQ-IR in the first few rounds. Under the constraint of finite block lengths, it is also demonstrated that the fast HARQ-IR protocol outperforms in the age-optimal rate allocation problem when the propagation delay is shorter than a certain threshold.
	
	Taking into account specific channel coding schemes, the authors in \cite{RateAlloc2} firstly consider a Spinal coded timely status update system based on HARQ with all practical delay elements. An upper bound of average AoI for Spinal codes is derived. Moreover, the HARQ transmission protocol is optimized in two steps to minimize the AoI. Firstly, the authors optimize the puncture pattern of Spinal codes and propose a transmission scheme based on incremental tail transmission puncturing pattern. Secondly, an optimal HARQ protocol is solved based on coarse-grained incremental tail transmission puncturing pattern, where the number of code symbols (i.e. the sub-block length) in each round is refined. 
	
	Based on the previous work \cite{RateAlloc1}, the authors in \cite{RateAlloc10} investigate a Polar coded status update system that accounts for encoding, transmission, propagation, decoding, and feedback delays. The average AoI of the proposed system with different transmission protocols is analyzed. Based on the analysis, the authors focus on the optimization for Polar-coded HARQ design. Specifically, an effective algorithm is devised to optimize the design SNR in code construction and the code length in HARQ-CC transmission. Additionally, a greedy algorithm is utilized to optimize the code lengths for each transmission and the number of transmissions in Polar-coded HARQ-IR. 
	
	\subsubsection{Significance-oriented modulation policy design}
	Modulation techniques is crucial in the wireless communication system, for they can enhance the reliability, spectral efficiency, and data transmission rate. However, the modulation technique design encounters  challenges in efficiently utilizing limited spectrum resources to ensure timeliness in the SAGSIN. To address this, NOMA has emerged as a promising solution, which enables multiple users to share time-frequency resources by using power domain multiplexing. By adopting power allocation (i.e., assigning varying power levels to different users) which is a modulation policy, NOMA facilitates simultaneous message transmission of multiple users. Allocating more power to the data of higher significance not only ensures reliable transmission of significant data but also enhances the timeliness of the system by directly influencing data rates and system capacity.
	
	Various techniques are employed to enhance the timeliness performance of NOMA systems. \cite{PowerAlloc4} considers the problem of energy-efficient scheduling for minimizing the AoI in an opportunistic NOMA/orthogonal-multiple-access (NOMA/OMA) downlink broadcast wireless network. The initial step involves formulating a resource allocation problem to minimize the average AoI in the network. To address energy-efficiency considerations, the work takes into account both a long-term average power constraint and a maximum power constraint. Then the Lyapunov framework is employed to approximate the original problem as a queue stability problem. The obtained single time-slot optimization problem is further decomposed into power allocation and user scheduling subproblems. An efficient piece-wise linear approximation is utilized to solve the non-convex power allocation subproblem, while linear programming algorithm is adopted to solve the user scheduling subproblem. Unlike the approach discussed in \cite{PowerAlloc4} that focuses solely on NOMA and OMA transmission schemes, \cite{PowerAlloc1} proposes an adaptive NOMA/OMA/cooperative-SWIPT-NOMA transmission scheme. In the NOMA scheme, the base station is responsible for allocating power to each user. However, in the cooperative-SWIPT-NOMA scheme, the base station not only allocates power to each user but also determines the power splitting coefficient (which determines the amount of harvested energy). The authors focus on a short packet communication system within a wireless network, where a base station transmits timely status updates to the users. The objective is to minimize the weighted average AoI  by  selecting appropriate multiple access techniques and corresponding power allocations. The resource allocation problem for the adaptive transmission scheme is formulated as an MDP and solved by iteration algorithms. 
	Similar to \cite{PowerAlloc1}, the authors in \cite{PowerAlloc2} investigate the AoI performance in a downlink wireless communication system, while HARQ-CC aided NOMA technology is employed in \cite{PowerAlloc2}.  \cite{PowerAlloc2} considers a more flexible scenario where both power allocation decisions and retransmission decisions are made adaptively to minimize the system average AoI in each time slot. An AoI minimization problem is formulated, and the optimal policy is obtained by transforming the problem to MDP. Furthermore, the authors notice the issue of fairness between users, and propose a greedy algorithm to minimize the users' maximal expected AoI, ensuring fairness by preventing the far user (with weak channel condition) from being deprived of timely service. 
	
	The aforementioned works primarily focus on ground-based wireless communication scenarios, while \cite{PowerAlloc5,PowerAlloc6} consider the satellite-based IoT (S-IoT) network. In particular, the authors in \cite{PowerAlloc5} propose an AoI-minimal resource allocation scheme in NOMA-based S-IoT downlink network. In this system, the satellite sends timely status updates to multiple user equipment  devices. To ensure the freshness of the status updates in the network, an AoI optimization problem is formulated, taking into account long-term average power,  peak power, and throughput constraints. The problem is then transformed into a series of online power allocation problems using Lyapunov optimization tools. Considering the limited computing resources of satellites, the particle swarm optimization (PSO) algorithm is employed to solve the non-convex optimization problem with linear computational complexity, named NOMA-AoI scheme. In the PSO algorithm, each particle means a possible solution which is the power allocation factors. By iteratively updating the velocities and positions of particles based on their historical information and the globally best information, the PSO algorithm explores the search space to find the optimal power allocation strategy. Based on \cite{PowerAlloc5}, the authors in  \cite{PowerAlloc6} further consider the network stability constraint in AoI optimization problem under NOMA-based S-IoT network. The problem is then transformed into three queue stability problems using the Lyapunov framework, which also converts the optimization problem into a series of single time slot deterministic optimization problems. The weights for the queue backlog and channel conditions are derived using the ListNet algorithm, a machine-learning-based approach, to obtain an optimized power allocation order with linear complexity. Similar to \cite{PowerAlloc5}, the proposed NOMA long-term AoI minimization power allocation problem is also addressed using the PSO algorithm.

	\subsection{Lessons Learned from This Section}
	This section focuses on significance-oriented sampling, coding, and modulation policies to release heavy burden caused by massive source data and ensure that only significant data are sampled and  transmitted for further processing and actuation. In fact, in the SAGSIN scenarios, although data generated by distributed edge devices are massive and multi-modal, not all of the data are significant enough, especially when the source data vary slowly as compared to data transmission. Therefore, as the first step of semantic signal processing, significance-oriented sampling filters most of unimportant source data, saving communication resources for sequential coding and modulation. 
	
	Moreover, even the sampled ``significant enough" data may have different levels of importance according to the varying environment. Therefore, significance-oriented coding and modulation policy designs provide unequal protection for sampled data with different levels of importance. Specifically, more significant data (e.g., status updates with lower age) are given more coding redundancy or more power allocation to ensure their successful acknowledgment at the receiver. By significance-oriented semantic communication system, implementation of semantic communication systems in the SAGSIN (which will be elaborated in the following two sections) will be more energy-efficient.

	Finally, it is also a promising idea to enrich the concept of ``significance" metrics such that they are meaning-aware or effectiveness-aware, and study meaning- (effectiveness-) aware communication policy optimization. In fact, there have been pioneering works on refining AoI metric by considering spacial-temporal information correlation\cite{caaoi}. In SAGSIN, such information correlations are usually meaning-related, because the adjacent (in space or time) statuses in real world have similar meanings (such as similar fire statuses among the same area or during a period of time). Also, some preliminary efforts have sought to refine the AoII metric by introducing meaning-oriented metrics (which will be elaborated in the next section) as information penalty function\cite{aoiixr}. By such means, the refined AoII can capture the meaning similarity of the statuses between transceivers. Furthermore, if we introduce the cost of decision/actuation as penalty function, effectiveness-aware significance metrics can also be proposed. Note that in cases when we utilize such meaning- (effectiveness-) aware metrics to guide SPT design, we should firstly acquire the meaning (or effectiveness-related information) before sampling/coding/modulation processes. The methods on acquiring meaning and effectiveness-related information will be elaborated on Section \ref{sec4} and Section \ref{sec5}, respectively.

	\section{Perception-Communication-Computing-Integrated Semantic Communications in the SAGSIN: A Meaning Extraction and Reconstruction Perspective}\label{sec4}

\begin{figure*}[!t]
		\centering
		\includegraphics[width = 0.96\textwidth]{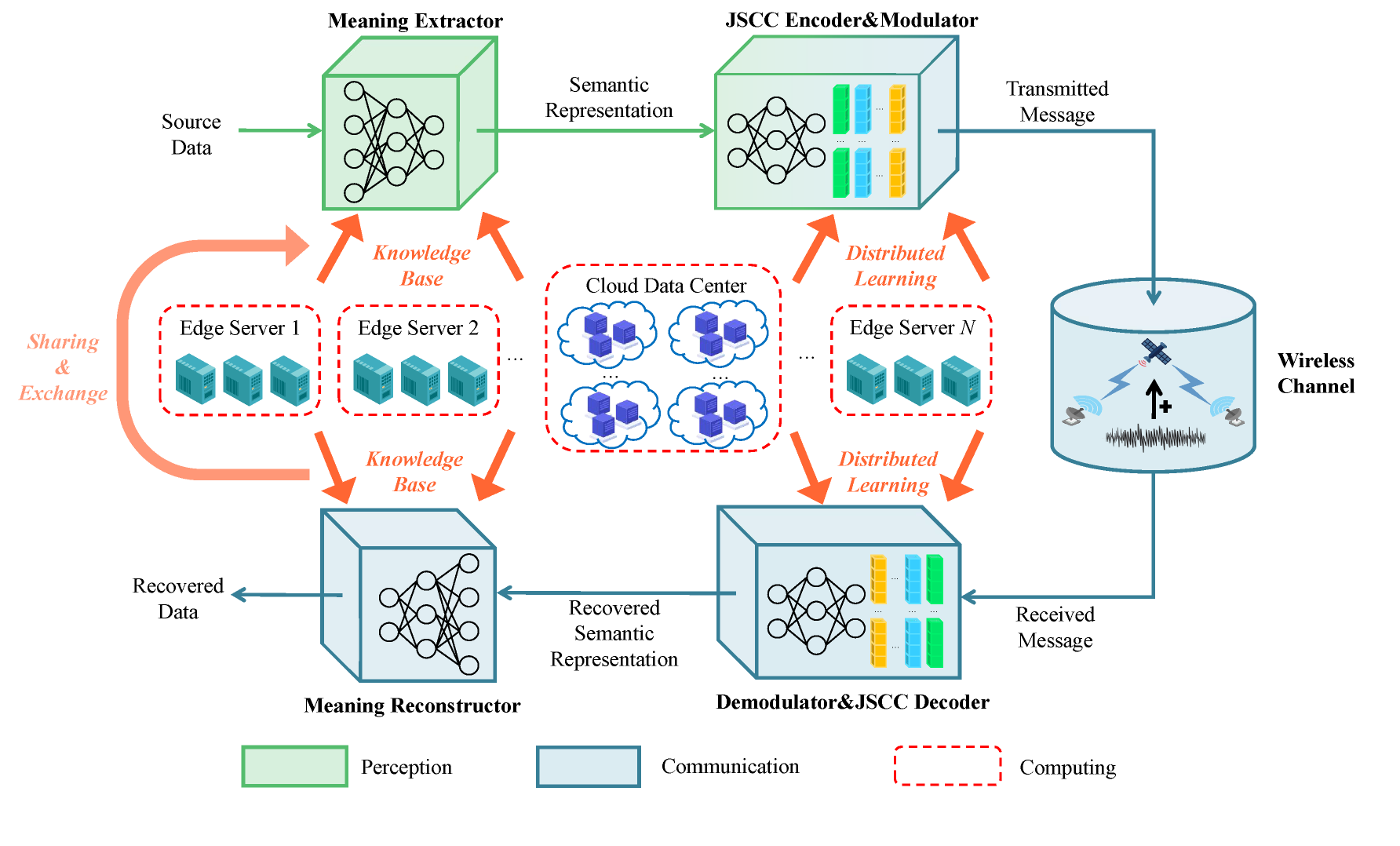}
		\caption{The framework of joint perception-communication-computing semantic communications in the SAGSIN.}
		\label{fig4}
	\end{figure*}

	In this section, we will review another major semantic communication system, called meaning-oriented semantic communication system. Since the primary purpose of such meaning-oriented system design is to achieve perfect recovery of the original data, this meaning-oriented semantic communication system is characterized by a strict symmetrical structure as shown in Fig. \ref{fig4}.
	
	In the SAGSIN, even after the semantic-aware sampling mentioned in Section \ref{sec3}, there are still a large volume of multi-modal source data (such as texts, images, audios, and videos) collected by numerous perception devices. To reduce the amount of transmitting data, it is necessary to perform semantic extraction from the source data. Here, the term ``semantics" is interpreted as ``meaning". However, the meaning of data is quite implicit to the extent that traditional codes cannot easily capture it. Hopefully, because of the similarity, we may associate meaning extraction with the process of human language learning. Humans gradually grasp semantics through extensive ``training" such as teacher instruction, reading books, conversations with others, and so on. Also, a fact is that the original intention of the design for artificial neural networks (ANNs) is to mimic the human brain. Taking the deep neural network (DNN) as an example, through training on large amounts of data, it can learn something from the data to some extent just like humans. Thus, using DNNs to construct the system codec has become the most popular alternative nowadays to “learn” the semantics from original data. 
	
	Based on above, the meaning-oriented semantic communication system adopting a joint perception-communication-computing framework is established, which is illustrated in Fig. \ref{fig4}. Aiming at conducting intelligent semantic extraction, coding, and modulation, we assume that the type of source data are given data in this section and the next section, unless otherwise specified. Specifically, DL-based meaning extractor captures the semantic information of source data to generate semantic representation, which serves as input of a JSCC encoder and modulator to generate transmitted messages for transmission. At the receiver, after demodulation and JSCC decoding, the semantic representation is recovered, which will be processed by DL-based meaning reconstructor to recover the source data. During the training of such large-scale neural networks, edge servers and cloud data centers must be utilized to provide knowledge bases and facilitate distributed learning, since end devices are usually limited in computing resources in the SAGSIN scenarios. Also, the knowledge bases distributed among edge/cloud servers can be synchronized via data sharing between transceivers. 
	
	It is worth noting that in distributed learning scenario, the meaning extractor/reconstructor and JSCC encoder/decoder \& modulator/demodulator will be firstly pre-trained by various data sets on the edge/cloud servers before being deployed. Secondly, the pre-trained models will be downloaded to the corresponding modules (or devices). Next, according to whether the module parameters are updated during communication processes, different communication modules will adopt either offline learning policy (without parameter updating) or online learning policy (with parameter updating). In online learning policy, module parameters will be also updated based on the distance between real data and the output data, similar to the gradient descent methods in training processes. In the following, unless otherwise specified, the mentioned works on meaning-oriented semantic communications all adopt  offline learning methods. 
	
	In the following subsections, we will firstly elaborate on the metrics for representing the system performance in terms of the ``meaning similarity" in Section \ref{sec41}. Next, in Section \ref{sec42}, we will provide a detailed description of the DNNs used to construct the system codec, which strongly intersects with the field of computer science. Finally, we discuss the related techniques for meaning enhancement in Section \ref{sec43}.
	\subsection{Metrics: Towards Meaning Similarity}\label{sec41}
	In this subsection, we review the metrics for the meaning-oriented semantic communication system. A perfect metric can maximally represent the ``meaning similarity" between the recovered data at the receiver and the original one at the transmitter. We classify these metrics according to the source type and acquisition method. For the former, we consider text, image/video, and speech sources. As for the latter, based on using whether explicit formulas or neural networks (NNs) to obtain metric values, we analyze objective metrics and learning-based metrics. Please note that the NNs for obtaining metrics are different from those for building semantic communication systems. The comparison of all the involved metrics is presented in Table \ref{IV_A}, and these metrics play crucial guiding roles for the system design in Section \ref{sec43}. As will be elaborated later, learning-based metrics are usually more effective in describing the meaning similarity than objective metrics. The reason behind is that the NNs that evaluate the similarity are trained from various aspects, while objective metrics usually demonstrate the similarity from only one of the aspects.

	\begin{table*}[!t]\normalsize 
		\renewcommand{\arraystretch}{1.3} 
		\caption{SUMMARY OF METRICS BASED ON MEANING SIMILARITY}\label{IV_A}
		\begin{tabular}[b]{|c|cc|c|c|c|} 
			\hline 
			$\textbf{Sources}$ & \multicolumn{2}{c|}{$\textbf{Metrics}$} & $\textbf{Explanation and Application Scenarios}$ & $\textbf{Formula}$ & \renewcommand{\arraystretch}{1.2}  \begin{tabular}[c]{@{}c@{}}$\textbf{Initial}$\\$\textbf{Reference}$\end{tabular} \\ 
			\hline
			
			\multirow{6}{*}{$\textbf{Text}$} 
			& \multicolumn{1}{c|}{\multirow{4}{*}{Objective}} & {WER} & \small Focusing on individual words & \small (\ref{WER}) & /  \\ \cline{3-6} 
			& \multicolumn{1}{c|}{}                           & {BLEU} &\small Focusing on consecutive words & \small (\ref{BLEU-1}) & \small \cite{BLEU}     \\ \cline{3-6} 
			& \multicolumn{1}{c|}{}                           & {CIDEr} &\small Reflecting the consistency and quality & \small ($\ast$) & \small \cite{CIDEr}   \\ \cline{3-6} 
			& \multicolumn{1}{c|}{}                           & {MSS} &\small Measuring the similarity and correctness & \small ($\ast$) & \small \cite{arxiv64}  \\ \cline{2-6} 
			& \multicolumn{1}{c|}{\multirow{2}{*}{\begin{tabular}[c]{@{}c@{}}Learning-\\ based\end{tabular}}} 
			& {BERT score} &\small A precision score by contextual understanding & \small (\ref{BERT-1}) & \small \cite{BERT_score} \\ \cline{3-6} 
			& \multicolumn{1}{c|}{} & {Sentence similarity} &\small Focusing on complete sentences & \small (\ref{sentence_similarity}) & \small \cite{SentenceSimilarity}  \\ 
			\hline
			
			\multirow{7}{*}{\begin{tabular}[c]{@{}c@{}}$\textbf{Image/}$ \\ $\textbf{Video}$\end{tabular}}  
			& \multicolumn{1}{c|}{\multirow{3}{*}{Objective}} & {PSNR} &\small Focusing on the pixel grayscale values & \small (\ref{PSNR}) & /  \\ \cline{3-6} 
			& \multicolumn{1}{c|}{}                           & {SSIM} &\small Considering the luminance, contrast and structure & \small (\ref{SSIM-1})  & \small \cite{SSIMandMSSSIM}  \\ \cline{3-6} 
			& \multicolumn{1}{c|}{}                           & {MS-SSIM} &\small Analyzing SSIMs at different scales & \small (\ref{MS-SSIM}) & \small \cite{SSIMandMSSSIM}  \\ \cline{2-6} 
			& \multicolumn{1}{c|}{\multirow{4}{*}{\begin{tabular}[c]{@{}c@{}}Learning-\\ based\end{tabular}}} 
			& {LPIPS} &\small Starting from human perception & \small (\ref{LPIPS}) & \small \cite{LPIPS}  \\ \cline{3-6} 
			& \multicolumn{1}{c|}{}  & {FID} &\small A biased estimator & \small (\ref{FID}) & \small \cite{FID}  \\ \cline{3-6} 
			& \multicolumn{1}{c|}{}  & {KID} &\small An unbiased estimator & \small (\ref{KID}) & \small \cite{KID}  \\ \cline{3-6}
			& \multicolumn{1}{c|}{}  & {AKD} &\small Analyzing facial keypoints & \small ($\ast$) & \small \cite{arxiv111}  \\ 
			\hline
			
			\multirow{7}{*}{$\textbf{Speech}$} 
			& \multicolumn{1}{c|}{\multirow{3}{*}{Objective}}  & {SDR}   &\small Comparing signal quality and distortion  & \small (\ref{SDR}) & \small \cite{SDR}    \\ \cline{3-6} 
			& \multicolumn{1}{c|}{}                            & {PESQ}  &\small Calculating speech quality scores & \small ($\ast$)  & \small \cite{PESQ}   \\ \cline{3-6} 
			& \multicolumn{1}{c|}{}                            & {MCD}   &\small Measuring spectral distortion & \small ($\ast$)  & \small \cite{MCD}  \\ \cline{2-6} 
			& \multicolumn{1}{c|}{\multirow{4}{*}{\begin{tabular}[c]{@{}c@{}}Learning-\\ based\end{tabular}}}
			& {FDSD}  &\small Extension based on FID  & \small (\ref{FDSD}) & \small \cite{FDSDKDSD}  \\ \cline{3-6} 
			& \multicolumn{1}{c|}{}     & {KDSD}  &\small Extension based on KID & \small (\ref{KDSD}) & \small \cite{FDSDKDSD}  \\ \cline{3-6} 
			& \multicolumn{1}{c|}{}     & {cFDSD} &\small FDSD under specific feature distributions & \small ($\ast$)     & \small \cite{FDSDKDSD}  \\ \cline{3-6} 
			& \multicolumn{1}{c|}{}     & {cKDSD} &\small KDSD under specific feature distributions & \small ($\ast$)     & \small \cite{FDSDKDSD}  \\ 
			\hline
		\end{tabular}\centering
	\end{table*}
	
	\subsubsection{Text sources}
	Here, the ``meaning similarity" can be interpreted as the similar degree of the meaning contained in the texts (which is composed of words and sentences) of the transmitter and receiver.
	
	\textit{Objective metrics.}
	Similar to the bit error rate (BER) in traditional communication systems, the word error rate (WER) is a fundamental metric to measure the text differences. It is calculated by
	\begin{equation}\label{WER}
		\text{WER} = \frac{{S + D + I}}{N},
	\end{equation}
	where ${S}$, ${D}$, ${I}$ denote the numbers of word substations, deletions, and insertions respectively, and ${N}$ is the word number of the original text. Unlike BER, WER can be greater than 1 due to a large number of insertions.

	WER focuses on the similarity of individual words, while the bilingual evaluation understudy (BLEU) score\cite{BLEU} simultaneously considers the differences between $n$-length combinations of consecutive words (named $n$-grams) in two texts. As a typical method for measuring translation quality in machine translation, BLEU is now commonly used to describe the recovery performance in semantic communication systems. It is commonly used in logarithmic form as
	\begin{equation}\label{BLEU-1}
		\log{\text{BLEU}} = \min (1 - \frac{{{l_{\hat s}}}}{{{l_s}}},0) + \sum\limits_{n = 1}^N {{w_n}\log {p_n}}, 
	\end{equation}
	where ${l_{\hat s}}$ and ${l_s}$ are the lengths of recovered texts ${\hat s}$ and original texts ${s}$, ${w_n}$ is the weight of $n$-grams, $N$ is the total considering number of grams, ${p_n}$ is the $n$-grams score, which is defined as
	\begin{equation}\label{BLEU-2}
		{p_n} = \frac{{\sum\limits_{k = 1}^{{K_n}} {\min ({C_k}(\hat s),{C_k}(s))} }}{{\sum\limits_{k = 1}^{{K_n}} {\min ({C_k}(\hat s))} }},
	\end{equation}	
	where ${K_n}$ is the number of elements in the ${n}$-th gram, and ${C_k(\cdot)}$
	is the frequency count function for the ${k}$-th element in the ${n}$-th gram. The BLEU score falls within the range of 0 to 1, and a higher score indicates greater text similarity.
	
	Besides, consensus-based image description evaluation (CIDEr)\cite{CIDEr} is used to assess the consistency and quality between the reference and generated image descriptions, and thus it can be seen as a metric for evaluating the similarity between texts. Moreover, the metric of semantic similarity (MSS)\cite{arxiv64} can measure both the meaning similarity between texts and the correctness of the recovered text (i.e., whether it is a fluent statement), and the contribution weights of the two aspects for the metric can be adjusted.
	
	The aforementioned metrics are all based on the text words, and their exploration for text semantics is far from sufficient. Taking the BLEU as an example, sentences ``I am available on the weekend." and ``I am not busy during the weekend." have the same semantics, but the BLEU score for them is not 1. To explore deep semantics of sentences (such as synonyms and expressions with similar meanings), we should resort to learning-based metrics.
	
	\textit{Learning-based metrics.}
	The bidirectional encoder representations from Transformers (BERT) model\cite{BERT_model} is a pre-trained language model that learns rich language representations from a vast amount of text data through large-scale unsupervised training. 
	
	The BERT score\cite{BERT_score} is a metric built upon the BERT model to evaluate text similarity, utilizing the contextual understanding learned by the BERT neural network. \cite{BERT_score} mentions three types of BERT scores, which are recall, precision, and F1 scores\footnote{For the definitions of these scores, readers can go to Section \ref{sec5A} for more details.}. Considering the need to measure the meaning similarity, the BERT score here specifically refers to the precision score. We assume that the representation vector generated by the embedding model for the original text $\left\langle {{x_1},{x_2}, \ldots ,{x_k}} \right\rangle$ is denoted as $\left\langle {{\bm{x}_1},{\bm{x}_2}, \ldots ,{\bm{x}_k}} \right\rangle$, and the representation vector for the recovered text $\left\langle {{{\hat x}_1},{{\hat x}_2}, \ldots ,{{\hat x}_m}} \right\rangle$ is denoted as $\left\langle {{\bm{\hat x}_1},{\bm{\hat x}_2}, \ldots ,{\bm{\hat x}_m}} \right\rangle$. The BERT score is given by
	\begin{equation}\label{BERT-1}
		{\text{BERT}} = \frac{{\sum\limits_{{{\hat x}_j} \in \hat x} {{\text{idf}}({{\hat x}_j})\mathop {\max }\limits_{{x_i} \in x} \bm{x}_i^{\rm T}{\bm{\hat x}_j}} }}{{\sum\limits_{{{\hat x}_j} \in \hat x} {{\text{idf}}({{\hat x}_j})} }},
	\end{equation}	
	where $\text{idf}(\cdot)$ is the importance weighting function. Given $M$ sentences from the test corpus $\{ {x^{(i)}}\} _{i = 1}^M$, $\text{idf}(\cdot)$ is defined as
	\begin{equation}\label{BERT-2}
		{\text{idf}}(w) =  - \log \frac{1}{M}\sum\limits_{i = 1}^M {{\text I}[w \in {x^{(i)}}]},
	\end{equation}	
	where ${\text I}[\cdot]$ is an indicator function. To increase the score readability, we adjust the scale of the BERT score in relation to its empirical lower bound $b$. The rescaled value $\hat{\text{BERT}}$ of $\text{BERT}$ is
	\begin{equation}\label{BERT-3}
		{\hat{\text{BERT}} = \frac{{\text{BERT} - b}}{{1 - b}}}.
	\end{equation}	
	
	$\hat{\text{BERT}}$ is between 0 and 1, and a higher score implies the better text similarity.
	
	Sentence similarity\cite{SentenceSimilarity} is another metric based on the BERT model that effectively evaluates the meaning similarity between two sentences. It is given by
	\begin{equation}\label{sentence_similarity}
		{\text{match}}(\hat s,s) = \frac{{\bm{B_\Phi }(s) \cdot \bm{B_\Phi }{{(\hat s)}^{\rm T}}}}{{\left\| {\bm{B_\Phi }(s)} \right\|\left\| {\bm{B_\Phi }(\hat s)} \right\|}},
	\end{equation}		
	where ${\bm{B_\Phi }(\cdot)}$ is the BERT model to map the sentence to its semantic vector space. Like the rescaled BERT score, sentence similarity also ranges from 0 to 1. When the similarity between two sentences is maximum, the sentence similarity value is equal to 1. \footnoteONLYtext{In table \ref{IV_A}, formulas for some metrics are omitted by * due to space limits in the main text.
		Readers can go to corresponding initial references for details.}
	
	The above two metrics are both based on the BERT model, which leverages the learning from billions of sentences to comprehend certain semantic information within the sentences. Therefore, they can provide a more accurate evaluation of the meaning similarity between texts than objective metrics.
	
	\subsubsection{Image (Video) sources}
	A digital image is composed of a finite number of pixels, while a video is composed of a continuous sequence of images, with each image referred to as a ``frame". Therefore, when discussing metrics, images and videos can be grouped together for analysis, and we take image sources as an example next. The semantics of images heavily rely on context, making it challenging to directly define the ``meaning similarity". For instance, without providing contextual information, we cannot determine whether ``a yellow triangle" is more similar to ``a green triangle" or ``a yellow square".
	
	\textit{Objective metrics.}
	Assume that the original and recovered images are respectively $\bm x$ and $\hat {\bm x}$. The matrix dimensions depend on the image properties, such as two dimensions for grayscale images, and three dimensions for colorful images. 
	
	The peak signal-to-noise ratio (PSNR), comparing each corresponding pixel of two images, is one of the typical metrics to measure the differences between images. The PSNR of a grayscale image is defined as
	\begin{equation}\label{PSNR}
		\text {PSNR}(\bm x,\hat {\bm x}) = 20\lg \frac{\text {MAX}}{{\sqrt {\text {MSE}(\bm x,\hat {\bm x})} }},
	\end{equation}
	where $\text {MAX}$ is the maximum grayscale value (typically 255 for an 8-bit image), and $\text {MSE}(\bm x,\hat {\bm x})$ (mean squared error) is given by
	\begin{equation}\label{MSE}
		{\text {MSE}(\bm x,\hat {\bm x}) = \left\| {\bm x - \hat {\bm x}} \right\|_2^2}.
	\end{equation}
	
	As for colorful images, one common approach is to calculate the MSE for each color channel separately and then take the average, which is used in \eqref{PSNR} to get the PSNR. Obviously, PSNR cannot provide an accurate similarity measurement of colorful images.
	
	The structural similarity index (SSIM)\cite{SSIMandMSSSIM}, considering the similarity of the luminance, the contrast and the structure, is effective in evaluating the colorful image similarity. It is defined as
	\begin{equation}\label{SSIM-1}
		\text {SSIM}(\bm x,\bm{\hat x}) = {\left[ {l(\bm x, \bm{\hat x})} \right]^\alpha } \cdot {\left[ {c(\bm x, \bm{\hat x})} \right]^\beta } \cdot {\left[ {s(\bm x, \bm{\hat x})} \right]^\gamma },
	\end{equation}
	where ${l(\bm x, \bm{\hat x})}$, ${c(\bm x, \bm{\hat x})}$, and ${s(\bm x, \bm{\hat x})}$ represent the similarity of the luminance, contrast, and structure comparisons, respectively. They are given by
	\begin{equation}\label{SSIM-2}
		{l(\bm x,\bm {\hat x}) = \frac{{2{\mu _{\bm x}}{\mu _{\bm {\hat x}}} + {C_1}}}{{\mu _{\bm{x}}^{2} + \mu _{\bm {\hat x}}^{2} + {C_1}}}},
	\end{equation}
	\begin{equation}\label{SSIM-3}
		{c(\bm x,\bm {\hat x}) = \frac{{2{\sigma _{\bm x}}{\sigma _{\bm {\hat x}}} + {C_2}}}{{\sigma _{\bm{x}}^{2} + \sigma _{\bm {\hat x}}^{2} + {C_2}}}},
	\end{equation}
	\begin{equation}\label{SSIM-4}
		s(\bm x,\bm {\hat x}) = \frac{{{\sigma _{\bm x}{_{\bm{\hat x}}}} + {C_3}}}{{{\sigma _{\bm x}}{\sigma _{\bm{\hat x}}} + {C_3}}},
	\end{equation}
	where $\mu _{\bm x}$, $\sigma _{\bm{x}}^{2}$ and $\sigma _{{{_{\bm x}}{_{\bm{\hat x}}}}}^2$ are the mean of $\bm x$, variance of $\bm x$, and the covariance between $\bm x$ and $\bm {\hat x}$, respectively, and $C_1$, $C_2$, and $C_3$ are the constants to avoid instability. If we set $\alpha=\beta=\gamma=1$ and $C_3=C_2/2$, we obtain a simplified expression, which is given by 
	\begin{equation}\label{SSIM-5}
		\text {SSIM}(\bm x,\bm{\hat x}) = \frac{{(2{\mu _{\bm x}}{\mu _{\bm{\hat x}}} + {C_1})(2{\sigma _{\bm x}}_{\bm{\hat x}} + {C_2})}}{{(\mu _{\bm x}^2 + \mu _{\bm{\hat x}}^2 + {C_1})(\sigma _{\bm x}^2 + \sigma _{\bm{\hat x}}^2 + {C_2})}}.
	\end{equation}	
	
	SSIM is a value between 0 and 1, and a larger value indicates smaller differences between images.
	
	The multi-scale structural similarity index (MS-SSIM)\cite{SSIMandMSSSIM} is an extension of SSIM. It applies multiple downsampling operations on the original image to obtain images at different scales. Then, the SSIMs for these images are calculated separately, and finally  a weighted summation operation of these SSIMs is performed, that is
	\begin{equation}\label{MS-SSIM}\begin{split}
			\text {MS-SSIM}(\bm x,{\bm{\hat x}})=& {\left[ {{l_M}(\bm x,\bm {\hat x})} \right]^{{\alpha _M}}} \\
			&\cdot \prod\limits_{j = 1}^M {{{\left[ {{c_j}(\bm x,\bm {\hat x})} \right]}^{{\beta _j}}} \cdot {{\left[ {{s_j}(\bm x,\bm{\hat x})} \right]}^{{\gamma _j}}}},
	\end{split}\end{equation}	
	where $\alpha_M$, $\beta_j$, and $\gamma_j$ are used to adjust the relative importance for different components. MS-SSIM takes into account local details and global structures of images, thereby generally providing a better evaluation of the image quality compared to SSIM.
	
	Considering the need for the accurate measurement of meaning similarity between images, the aforementioned three metrics are simple and shallow, and thus cannot capture the subtle differences perceived by humans.
	
	\textit{Learning-based metrics.}
	The learned perceptual image patch similarity (LPIPS)\cite{LPIPS} learns image semantics from the human perception perspective through DNNs. It evaluates image similarity by calculating the distance between the feature representations of images extracted from pretrained DNNs such as the SqueezeNet, AlexNet and VGG. LPIPS is defined as
	\begin{equation}\label{LPIPS}
		\text {LPIPS}= \sum\limits_l {\frac{1}{{{H_l}{W_l}}}\sum\limits_{h,w} {\left\| {{w_l} \odot (f_{hw}^l - \hat f_{hw}^l)} \right\|_2^2} },
	\end{equation}	
	where $H_l$ and $W_l$ denote the height and width of layer $l$, ${f^l}$ and ${{\hat f}^l}$ are the normalized latent feature maps by layer $l$ of the specific neural network for two images, $h$ and $w$ represent the $(h, w)$-th element of the feature map, and $\odot$ denotes the scale operation with scale vector ${w_l}$. By comparing the feature representations, LPIPS (lower values being better) can capture higher-level visual information.

	Another category of metrics measures the meaning similarity between the generating image (created by the generative adversarial network (GAN) based on latent features from the real image distribution) and the real image, including the Fréchet Inception distance (FID)\cite{FID} and the kernel Inception distance (KID)\cite{KID}, which utilize the pre-trained model from the Inception network for feature representations. FID is defined as
	\begin{equation}\label{FID}\begin{split}
			\text{FID} &= {d^2}\left( ({\bm m,\bm C),({\bm m_w},{\bm C_w})} \right)\\
			& = \left\| {\bm m - {\bm m_w}} \right\|_2^2 + {\rm{Tr}}\left( {\bm C + {\bm C_w} - 2{{(\bm C{\bm C_w})}^{\frac{1}{2}}}} \right),
	\end{split}\end{equation}		
	where $\bm m$ and $\bm C$ are the mean vector and covariance matrix in the feature space of the generating image distribution, $\bm m_w$ and $\bm C_w$ are the mean vector and covariance matrix of the real image distribution, and $\rm{Tr}(\cdot)$ indicates the trace of a square matrix. During the derivation of FID, \cite{FID} assumes that both the real and generated images follow the normal distribution (which is an ideal assumption), and the final FID is a biased estimator. As an improvement, \cite{KID} proposed KID, which is an unbiased estimator without the assumption of a normal distribution. It is given by
	\begin{equation}\label{KID}\begin{split}
			\text{KID} = & \frac{1}{{m(m - 1)}}\sum\limits_{i \ne j}^m {k({\bm x_i},{\bm x_j})}  \\
			&+ \frac{1}{{n(n - 1)}}\sum\limits_{i \ne j}^n {k({\bm y_i},{\bm y_j})} 
			- \frac{2}{{mn}}\sum\limits_{i = 1}^m {\sum\limits_{j = 1}^n {k({\bm x_i},{\bm y_j})} },
	\end{split}\end{equation}
	where $m$ and $n$ are respectively the number of samples in the generating and real images, $\bm x_i$, $\bm x_j$ are the 2048-dimensional vectors obtained from the Inception network for the generating image,  $\bm y_i$, $\bm y_j$ are the ones for the real image, and $k(\cdot)$ is the polynomial kernel defined as 
	\begin{equation}\label{kernel}
		k(\bm x,\bm y) = {\left( {\frac{1}{d}{\bm x^{\rm{T}}}\bm y + 1} \right)^3},
	\end{equation}	
	where $d$ is the representation dimension, which equals to 2048 here. 
	
	Smaller values of FID and KID indicate greater meaning similarity between images. They are consistent with human perception and can reveal semantic differences between images at a deeper level.
	
	In addition, considering the video conferencing scenario, \cite{arxiv111} proposes the average keypoint distance (AKD) based on the keypoints (containing certain semantic information) extracted by a pretrained facial landmark detector to evaluate the differences between images.
	
	The learning-based metrics above go beyond the pixel level and instead extract deeper semantic information from images using different pre-trained models. Thus, they can effectively measure the meaning similarity between images. 
	\subsubsection{Speech sources}	
	The semantics in speeches, which encompass not only the vocabulary and grammar but also the intonation, timbre, and speed of speech, are significantly complex and subjective. The same sentence with different intonations can convey different meanings. There are two evaluation methods, respectively evaluating the meaning similarity of the texts contained in the speeches or the speeches themselves. For the former, we can use all the aforementioned metrics for text sources. For the latter, we focus on the ``meaning similarity" of speeches, which can be interpreted as the similarity degree of the inferred meanings derived from the speeches at the transmitter and receiver. The metrics related to the second method are introduced below.
	
	\textit{Objective metrics.}
	Assume that the original and recovered speeches are respectively $\bm s$ and $\hat {\bm s}$. Similar to the SNR, the signal-to-distortion ratio (SDR) \cite{SDR} was initially proposed to calculate the ratio between the signal quality and distortion. As an extension, the SDR can be used to measure the similarity between two speeches\cite{arxiv117}. 
	\begin{equation}\label{SDR}
		\text {SDR}(\bm s,\hat{\bm s}) = 10\lg \frac{{{{\left\| \bm s \right\|}^2}}}{{{{\left\| {\bm s - \hat{\bm s}} \right\|}^2}}},
	\end{equation}	
	where ${\left\| \bm s \right\|}^2 = \left\langle {\bm s,\bm s} \right\rangle $ represents the energy of $\bm s$. A larger SDR indicates a higher similarity between speeches.
	
	Another metric called perceptual evaluation of speech quality (PESQ)\cite{PESQ} calculates the speech quality score by comparing the differences between speeches, which was established as Recommendation ITU-T P.862 in 2001. The specific process of calculating the PESQ score for two speeches mainly includes two stages: the alignment routine and the perceptual model. The former consists of the computation of the overall system gain, the intelligent reflecting surface (IRS) filtering, the time alignment (including the envelope-based alignment, the fine time alignment, the utterance splitting, and the perceptual realignment), and so on. The latter includes the IRS-receive filtering, the short-term fast Fourier transform (FFT), the calculation of the pitch power densities, the realignment of bad intervals, the computation of the PESQ score, and so on. The range of the PESQ score is -0.5 to 4.5, and the larger value reveals the higher similarity between speeches. The calculation process of the PESQ is quite complex, and due to limited space, we have provided only a brief description here. For more details, see, e.g., \cite{PESQ} and the references therein.
	
	Furthermore, the Mel cepstral distortion (MCD)\cite{MCD}, which measures the spectral distortion between synthesized and target speeches, can be regarded as a metric to evaluate the similarity between speeches. The smaller the MCD is, the closer the spectra between speeches are.
	
	The aforementioned metrics do not sufficiently represent the semantic information of speeches, making it difficult to accurately evaluate the similarity between speeches at the semantic level. 
	
	\textit{Learning-based metrics.}
	Based on the FID and KID, the unconditional Fréchet DeepSpeech distance (FDSD) and unconditional kernel DeepSpeech distance (KDSD) are proposed in \cite{FDSDKDSD} to measure the meaning similarity between speeches. By utilizing the pre-trained DeepSpeech2 model from the NVIDIA OpenSeq2Seq library, the KDSD and FDSD are able to capture the semantic information in the high-level speech feature space. The computational differences between the FDSD and FID are only reflected in the source and network types. According to \eqref{FID}, the FDSD is defined as
	\begin{equation}\label{FDSD}\begin{split}
			\text{FDSD} = &{d^2}\left( {({\bm \mu _{\bm s}},{\bm \Sigma _{\bm s}}),({\bm \mu _{\hat {\bm s}}},{\bm \Sigma _{\hat {\bm s}}})} \right)\\
			=& \left\| {{\bm \mu _{\bm s}} - {\bm \mu _{\hat {\bm s}}}} \right\|_2^2 
			 + {\rm{Tr}}\left( {{\bm \Sigma _{\bm s}} + {\bm \Sigma _{\hat {\bm s}}} - 2({\bm \Sigma _{\bm s}}{\bm \Sigma _{\hat {\bm s}}}){)^{\frac{1}{2}}}} \right),
	\end{split}\end{equation}		
	where $\bm \mu _{\bm s}$, $\bm \Sigma _{\bm s}$ are the mean vector and covariance matrix in the feature space of the original speech distribution, $\bm \mu _{\hat {\bm s}}$, $\bm \Sigma _{\hat {\bm s}}$ are the ones of the recovered speech distribution, and $\rm{Tr}(\cdot)$ indicates the trace of a square matrix. Moreover, the KDSD is given by
	\begin{equation}\label{KDSD}\begin{split}
			\text{KDSD} = &\frac{1}{{K(K - 1)}}\sum\limits_{i \ne j}^K {k({\bm s_i},{\bm s_j})} \\ &+\frac{1}{{\hat K(\hat K - 1)}}\sum\limits_{i \ne j}^{\hat K} {k({{\hat {\bm s}}_i},{{\hat {\bm s}}_j})} 
			+ \sum\limits_{i = 1}^K {\sum\limits_{j = 1}^{\hat K} {k({\bm s_i},{{\hat {\bm s}}_j})} },
	\end{split}\end{equation}
	where $K$, $\hat K$ are respectively the number of samples in the original and recovered speeches, $\bm s_i$, $\bm s_j$ are the vectors obtained from the DeepSpeech2 model for the original speech,  $\hat {\bm s}_i$, $\hat {\bm s}_j$ are the ones for the recovered speech, and $k(\cdot)$ is the polynomial kernel defined as \eqref{kernel}. Additionally, \cite{FDSDKDSD} also proposes the conditional Fréchet DeepSpeech distance (cFDSD) and conditional kernel DeepSpeech distance (cKDSD) to evaluate the meaning similarity between speeches under the specific linguistic feature distribution condition.
	
	Compared to objective metrics, learning-based metrics utilize NNs to assist in capturing the semantic information, which can better represent the meaning similarity of speeches.
	\begin{remark}\label{4-1-1}
		``Perfect" metrics are believed to completely represent the ``meaning similarity". However, there is not a unified definition for ``semantics". Hopefully, with the development of semantic information theory, the metrics may also evolve accordingly.
	\end{remark}
	\begin{remark}\label{4-1-2}

		The aforementioned objective metrics are not differentiable. Also, the learning-based metrics cannot be expressed in an explicit form, making it impossible to determine the differentiability as well. Therefore, they cannot be used to design the loss function in the training of semantic communication systems. The inconsistency between the objective (loss) function in the training phase and the evaluation metrics in the testing phase restricts the performance of meaning-oriented semantic communication systems. One obvious solution is to focus on researching differentiable evaluation metrics. However, these metrics should represent the ``meaning similarity" as accurately as possible at the same time. One possible solution is to leverage the reinforcement learning (RL), where non-differentiable metrics can be used to guide learning processes and train high-performing systems by maximizing long-term rewards\cite{arxiv69, arxiv71}.
	\end{remark}
	
	\subsection{Methods: Based on DNNs}\label{sec42}
	
	Shannon’s separation theorem\cite{Shannon} proves that under the condition of infinite block lengths, the separate source-channel coding (SSCC) scheme can achieve the same performance as the JSCC scheme. Due to its modularity and ease of implementation, the SSCC scheme is widely applied in traditional communication systems, with source codes and channel codes being proposed in parallel on separate tracks.
	
	However, affected by the typical characteristics of large-scale scenarios, highly dynamic channels, and limited device capabilities in the SAGSIN, the communication systems have extremely stringent requirements on latency, while the computing resources of IoT devices are limited. Moreover, in the semantic communication systems, redundant information is removed through  semantic extraction before encoding. All these facts lead to the prevalence of finite block lengths (even short ones) in the SAGSIN. Thus, the SSCC scheme cannot achieve the Shannon limit, and the JSCC one regains attention.
	
	The mainstream approach in constructing semantic communication systems currently involves utilizing DNNs as source (semantic) and channel codecs, and training them jointly, called deep learning based JSCC (DeepJSCC), which is firstly applied in \cite{DeepJSCC} to implement semantic communications. Taking the transmitter as an example here, DNNs sequentially perform operations such as semantic extraction and encoding on the source data. The output of the network can be directly transmitted over the channel. Therefore, it is necessary to review typical DNNs which can be used in JSCC.
	
	In semantic communication systems, commonly used DNN-related models primarily include four  categories: convolutional neural Networks (CNNs), recurrent neural networks (RNNs), Transformer and GANs. These networks, which are listed in Table \ref{IV_B}, become the base of the system in Section \ref{sec43}.
	\begin{table*}[!t]\normalsize 
		\renewcommand{\arraystretch}{1.3} 
		\caption{SUMMARY OF NEURAL NETWORKS USED IN JSCC}\label{IV_B}
		\begin{tabular}[b]{|c|c|c|c|} 
			\hline 
			$\textbf{Types}$ & {$\textbf{Neural Networks}$} & $\textbf{Explanation and Application Scenarios}$  & \renewcommand{\arraystretch}{1.2}  \begin{tabular}[c]{@{}c@{}}$\textbf{Initial}$\\$\textbf{Reference}$\end{tabular} \\ 
			\hline
			
			\multirow{6}{*}{$\textbf{CNN-Related}$} 
			& {LeNet-5} & \small First CNN model, laying the foundation for CNN structure  & \cite{LeNet5}  \\ \cline{2-4} 
			& {AlexNet} &\small Deeper CNN, having improvements compared to LeNet-5  & \small \cite{AlexNet}     \\ \cline{2-4} 
			& {FCN} &\small Consisting of only convolutional layers for semantic segmentation & \small \cite{FCN}  \\ \cline{2-4} 
			
			& {ResNet} &\small A residual block via ``shortcut connection"  & \small \cite{ResNet} \\ \cline{2-4} 
			& {Others} &\small \begin{tabular}{c} Including CNN variants such as ZFNet, VGGNet,\\ GoogLeNet, DenseNet, PSPNet, and PatchGAN\end{tabular}& \small \begin{tabular}{c} \cite{ZFNet,VGGNet,GoogLeNet},\\\cite{DenseNet,PSPNet,PatchGAN} \end{tabular} \\ 
			\hline
			
			\multirow{5}{*}{$\textbf{RNN-Related}$}  
			& {Hopfield/Jordan network} &\small Initial exploration on standard RNNs  & \small \cite{Hopfield,Jordan}  \\ \cline{2-4} 
			& {Elman network} &\small Standard RNN that can utilize historical information   & \small \cite{Elman}  \\ \cline{2-4} 
			& {LSTM network} &\small Addressing long-term dependency issue by memory units  & \small \cite{LSTM}  \\ \cline{2-4} 
			
			& {BiRNN/BiLSTM network} &\small Two independent networks to capture more context information  & \small \cite{BiRNN,BiLSTM,DeepBiLSTM}  \\ \cline{2-4} 
			& {GRU network} &\small Simpler than LSTM model  & \small \cite{GRU}  \\ \cline{2-4} 
			
			\hline
			
			\multirow{3}{*}{$\textbf{Transformer}$} 
			& {Classic Transformer}   &\small Initially for NLP tasks, extracting long-term dependency   & \small \cite{Transformer}    \\ \cline{2-4} 
			& {Others}   &\small \begin{tabular}{c}Including ViT for image classification, DETR for object\\ detection, and swin Transformer \end{tabular} & \small \cite{ViT,DETR,swin}  \\ \cline{2-4} 
			
			\hline
			\multirow{8}{*}{$\textbf{GAN}$} 
			& {Classic GAN}   &\small Generator-discriminator unsupervised adversarial training  & \small \cite{GAN}    \\ \cline{2-4} 
			& {CGAN}  &\small Introducing supervised learning to avoid unstable training  & \small \cite{CGAN}   \\ \cline{2-4} 
			& {DCGAN}  &\small Lighter GAN model implementation by CNN  & \small \cite{DCGAN}   \\ \cline{2-4}
			& {WGAN}   &\small Introducing Wasserstein distance as loss function  & \small \cite{WGAN}  \\ \cline{2-4} 
			
			& {SAGAN}  &\small Attention-based GAN   & \small \cite{SAGAN}  \\ \cline{2-4} 
			& {BigGAN}  &\small Large scale training for better image quality & \small \cite{BigGAN}  \\ \cline{2-4} 
			& {Others} &\small \begin{tabular}{c}
				Including StyleGAN for customized images and\\ HiFi-GAN for audio sources
			\end{tabular}    & \small \cite{StyleGAN,HiFiGAN}  \\ \cline{2-4} 
			
			\hline
		\end{tabular}\centering
	\end{table*}
	\subsubsection{CNN-related models}	
	CNN is primarily used for space relevant sources, particularly image ones. The development of CNN can be traced back to 1962 when Hubel and Wiesel introduced the concept of Receptive fields \cite{1962}. In 1980, \cite{1980} proposed a neural network including convolutional and pooling layers. The first formal CNN model, called LeNet-5\cite{LeNet5}, was proposed in 1998. It laid the foundation for the basic structure of CNNs: convolutional, pooling, and fully connected layers. However, due to limited computing resources and the vanishing gradient caused by sigmoidal activation functions, LeNet-5 did not receive the deserving attention.
	
	AlexNet \cite{AlexNet} marks the triumphant return of CNN, which consists of eight layers, including five convolutional ones and three fully connected ones. Compared to LeNet-5, AlexNet has the following four main improvements: First, a deeper network architecture is used to learn more intricate semantic features. Second, the sigmoidal activation function is replaced by the rectified linear unit (ReLU) one, alleviating the vanishing gradient. Third, researchers utilize two data augmentation methods and dropout technique to reduce overfitting. Fourth, overlapping pooling further enhances the richness of feature maps and extracts more detailed semantic information. Later, ZFNet\cite{ZFNet} (a fine-tuned version of AlexNet), VGGNet\cite{VGGNet} (a ``very deep" version of AlexNet), and GoogLeNet \cite{GoogLeNet} (also known as Inception-V1) are successively proposed. 
	
	The aforementioned CNNs are primarily used for image classification tasks, and thus the network output is the classification result. However, for image semantic segmentation tasks, the network should output an image of the same size as the input one and classify each pixel of the image into semantic categories (e.g., person, car, and tree). Based on this purpose, \cite{FCN} proposes the fully convolutional network (FCN), which consists entirely of convolutional layers.

	The development history of CNNs implies a belief: increasing the network depth does not decrease the model performance. A straightforward example is that adding a layer with the ``identity mapping" does not change the model performance, but makes the network deeper. However, ``Is learning better networks as easy as stacking more layers?\cite{ResNet}" Researchers discovered the ``degradation" phenomenon through experiments, revealing a negative answer to the question. In order to make the network fully learn the data semantic features, activation functions are set to be nonlinear, to the extent that deep layers cannot simply perform the linear transformation (e.g., ``identity mapping"). One solution is to build a ``shortcut connection" between certain layers to strike better balance between nonlinear and linear transformations, forming a residual block. This is the ResNet\cite{ResNet}. 
	
	Recently, some new variants of CNNs have been also proposed consecutively, including DenseNet\cite{DenseNet} (employing dense connection structures), PSPNet\cite{PSPNet} (utilizing pyramid pooling), and PatchGAN\cite{PatchGAN} (commonly used as a  discriminator in GANs).
	\subsubsection{RNN-related models}	
	RNN is mainly used for time relevant sources, particularly text and speech ones. In 1982, a Hopfield network was introduced by \cite{Hopfield} as a precursor to RNNs. Later, a Jordan network was proposed in \cite{Jordan}. In \cite{Elman}, a Jordan network was improved, resulting in a Elman network, which is also known as the standard RNN. It consists of an input layer, a hidden layer, and an output layer. The neurons in the hidden layer are recurrently connected over time, allowing the output from the previous time step to be fed as input to the current one, thus enabling the network to remember and utilize historical information. This is why RNNs can process sequential data and learn semantic features from the context.
	
	When the input sequence is too long, the time steps of the standard RNN become large, which may lead to the vanishing/exploding gradients during training. The solution is to utilize techniques like gradient clipping or more stable RNN variants such as long short-term memory (LSTM) networks. LSTM\cite{LSTM} effectively addresses the issue of long-term dependencies in standard RNNs by introducing memory units and gate mechanisms (an input gate, a forget gate and an output gate). It can efficiently propagate and express information in long sequences without neglecting/forgetting useful information from earlier time steps. In fact, only a portion of the input sequence is crucial (i.e., having more semantic information), and it needs to be remembered for a long time, while the remaining one can be appropriately forgotten. Therefore, the concept of ``long short-term memory" was proposed in \cite{LSTM}.
	
	Because standard RNNs only consider past information of the input sequence, \cite{BiRNN} introduced two independent RNN modules (forward and backward processing) simultaneously to capture both past and future context information, resulting in the bidirectional RNN (BiRNN). Afterwards, \cite{BiLSTM} combined the essence of LSTM and BiRNN to create bidirectional LSTM (BiLSTM) networks. Furthermore, \cite{DeepBiLSTM} proposed deep BiLSTM networks to achieve better extraction and representation of semantic features. Gated recurrent unit (GRU) networks\cite{GRU}, a variant of LSTM, controls the flow of information by using an update gate and a reset gate. It is simpler than the LSTM model and can achieve comparable performance to LSTM with higher training efficiency.
	
	\subsubsection{Transformer-related models}	
	
	The Transformer \cite{Transformer} was initially proposed in the field of natural language processing (NLP) (which is for text sources). Its primary purpose is to address the limitations of RNNs in handling sequential data, such as long-term dependencies and computational efficiency. However, due to its powerful capabilities, the Transformer has been successfully applied to other domains, including computer vision (image sources) and speech processing (speech sources).

	The Transformer primarily comprises an encoder and a decoder. The encoder utilizes multiple (typically 6) encoding layers to convert input sequences into semantic space representations. Each encoding layer consists of a multi-head self-attention sub-layer and a position-wise fully connected feed-forward network sub-layer. The former is employed to establish global dependencies, while the latter performs nonlinear transformations. Residual connections are applied around each of the two sub-layers, followed by layer normalization. In comparison to the encoder, the decoder introduces an extra masked multi-head self-attention sub-layer at each decoding layer. By setting the attention weights of future positions to negative infinity, the masking operation prevents leakage of future information.

	The Transformer is powerful due to the following aspects: Firstly, the multi-head self-attention mechanism divides the model into multiple heads to form subspaces and allows the model to automatically learn the relationships between different positions in the input sequence, capturing contextual information more effectively. Secondly, the self-attention mechanism allows for ``parallel" computation of the encoder and decoder, effectively harnessing the parallel computing capabilities of modern hardware accelerators such as graphics processing units (GPUs). Thirdly, the Transformer effectively models long-term dependencies in sequences, alleviating the issue of gradient vanishing/exploding in RNNs when processing long sequences. Fourthly, the Transformer structure is simple, highly modular, and easy to extend and modify.
	
	In addition, there are several variants of the Transformer proposed recently, such as the vision Transformer (ViT) \cite{ViT} (for image classification tasks), the detection Transformer (DETR)\cite{DETR} (for object detection tasks), the shifted window (swin) Transformer\cite{swin} (reducing computational complexities by the window self-attention mechanism), and so on.
	
	\subsubsection{GAN-related models}	
	In practical applications, the GAN can be applied to various types of sources. The classic GAN was first proposed in \cite{GAN}, which consists of a generator and a discriminator. The purpose of the generator is to generate instances that appear natural and realistic, closely resembling the original data. And the role of the discriminator is to determine whether a given instance is real or fake. Clearly, they have an ``adversarial" relationship. The generator and discriminator are trained alternately, progressing together in the ``game" and ideally reaching a dynamic equilibrium (i.e., Nash equilibrium). At this point, the generated instances from the generator are indistinguishable from real ones. There are two key points to note: Firstly, the ultimate goal is to obtain a well-trained generator, while the discriminator is an additional benefit. Secondly, there are no specific constraints on DNNs used for the generator and discriminator, and thus they can be CNNs, RNNs, or even Transformers.
	
	The initiall GAN is an unsupervised learning model, and its training process does not require labeled data. Conditional GANs (CGANs)\cite{CGAN} bring GANs back into the realm of supervised learning, alleviating the problem of unstable training in GANs. Deep convolutional GANs (DCGANs)\cite{DCGAN} are the first successful implementation of GANs that use CNNs instead of fully connected layers. This significantly reduces the number of network parameters while greatly improving the quality of generated data. Wasserstein GANs (WGANs)\cite{WGAN} improve the stability of model training by introducing the Wasserstein distance as a loss function. Self-attention GANs (SAGANs)\cite{SAGAN} introduce an attention mechanism that enables the network to better understand the semantic information. The introduction of BigGANs\cite{BigGAN} reveals that GAN network training can also benefit from large-scale data and computing resources. BigGANs achieve unprecedented levels of image generation quality. Recently, some new variants of GANs have been proposed, such as StyleGANs\cite{StyleGAN} (controlling the style and features of generated images), HiFi-GAN \cite{HiFiGAN} (focusing on generating high-fidelity audio signals), and so on.

	\begin{remark}\label{4-2-1}
		There are also other neural networks used in JSCC schemes. Graph convolutional networks (GCNs)\cite{GCN}, which are commonly utilized to extract semantic features from non-Euclidean structured images (e.g., social networks and knowledge graphs), are used in \cite{arxiv13} to construct joint source-channel codecs. \cite{arxiv57, arxiv58, arxiv69, arxiv71, arxiv102} employ reinforcement learning networks (RLNs) to construct semantic communication systems. 
		
	\end{remark}
	
	\begin{remark}\label{4-2-2}
		Semantic communication systems can be also constructed by DL-based SSCC (DeepSSCC) schemes, which consist of two categories. The first one uses traditional codecs for channel coding, and only the source (semantic) codec are trained\cite{arxiv50, arxiv93, arxiv99, dytext4}. The source and channel codecs of the second one are both DNNs, but trained separately \cite{arxiv34, arxiv91}.
	\end{remark}
	
	\subsection{Techniques: Towards Meaning Enhancement}\label{sec43}
	
	\begin{table*}[!t]\small
		\renewcommand{\arraystretch}{1.4} 
		\caption{SUMMARY OF MEANING-ENHANCEMENT TECHNIQUES}\label{IV_C}
		\begin{tabular}[b]{|c|c|cc|c|} 
			\hline 
			$\textbf{\normalsize Techniques}$ & \renewcommand{\arraystretch}{1.3} \begin{tabular}[c]{@{}c@{}}$\textbf{\normalsize SAGSIN}$\\ $\textbf{\normalsize Issue}$ \\ $\textbf{\normalsize Addressment}$ \end{tabular} & \multicolumn{2}{c|}{$\textbf{\normalsize Technical Details}$} & \renewcommand{\arraystretch}{1.2} \begin{tabular}[c]{@{}c@{}}$\textbf{\normalsize Related}$\\ $\textbf{\normalsize References}$\end{tabular}  \\ 
			\hline
			
			\multirow{11}{*}{\begin{tabular}[c]{@{}c@{}}$\textbf{\normalsize Channel-}$\\ $\textbf{\normalsize related METs}$\end{tabular}} 
			& \multirow{11}{*}{\begin{tabular}[c]{@{}c@{}}\normalsize Highly dynamic\\\normalsize channels\end{tabular}} & \multicolumn{1}{c|}{\multirow{2}{*}{\normalsize Extracting channel semantics}} & Role as a new research direction & \small \cite{arxiv36, arxiv37} \\ \cline{4-5}
			& 							     & \multicolumn{1}{c|}{}             & Concept of ``channel semantics" & \small \cite{arxiv1} \\ \cline{3-5}
			&                                & \multicolumn{1}{c|}{\multirow{3}{*}{\begin{tabular}[c]{@{}c@{}}\normalsize Designing noise-\\ \normalsize resistant mechanisms\end{tabular}}} & Joint semantic-noise coding & \small \cite{arxiv71} \\ \cline{4-5}
			& 							     & \multicolumn{1}{c|}{}             & \begin{tabular}[c]{@{}c@{}}Literal semantic noise and\\ adversarial semantic noise\end{tabular} & \small \cite{arxiv2} \\ \cline{3-5}
			&                                & \multicolumn{1}{c|}{\multirow{6}{*}{\normalsize Capturing and utilizing CSI}} & Channel ModNet & \small \cite{arxiv12, arxiv114} \\ \cline{4-5}
			&                                & \multicolumn{1}{c|}{}             & Response network & \small \cite{arxiv47} \\ \cline{4-5}
			&                                & \multicolumn{1}{c|}{}             & Attention feature module & \small \cite{arxiv25} \\ \cline{4-5}
			&                                & \multicolumn{1}{c|}{}             & \begin{tabular}[c]{@{}c@{}}Encoding and modulation\\ match with CSI\end{tabular} & \small \cite{arxiv54} \\ \cline{4-5}
			&                                & \multicolumn{1}{c|}{}             & Training with CSI & \small \cite{dytext3} \\ 
			\hline
			
			\multirow{9}{*}{\begin{tabular}[c]{@{}c@{}}$\textbf{\normalsize Rate-control-}$\\ $\textbf{\normalsize related METs}$\end{tabular}} 
			& \multirow{9}{*}{\begin{tabular}[c]{@{}c@{}}\normalsize Highly dynamic\\\normalsize channels\end{tabular}} & \multicolumn{1}{c|}{\multirow{3}{*}{\normalsize Retransmission-based}} & SC-RS-HARQ, SCHARQ & \small \cite{dytext4} \\ \cline{4-5}
			& 							     & \multicolumn{1}{c|}{}             & SVC-HARQ & \small \cite{arxiv111} \\ \cline{4-5}
			& 							     & \multicolumn{1}{c|}{}             & IK-HARQ & \small \cite{arxiv11} \\ \cline{3-5}
			&                                & \multicolumn{1}{c|}{\multirow{6}{*}{\normalsize Non-retransmission-based}} & Channel feedback & \small \cite{dyimage4} \\ \cline{4-5}
			& 							     & \multicolumn{1}{c|}{}             & Policy network & \small \cite{dyimage6} \\  \cline{4-5}
			& 							     & \multicolumn{1}{c|}{}             & Rate adaptive transmission & \small \cite{dyimage7, IEEE1} \\ \cline{4-5}
			& 							     & \multicolumn{1}{c|}{}             & Rate allocation network & \small \cite{arxiv86} \\ \cline{4-5}
			& 							     & \multicolumn{1}{c|}{}             & Adaptive density learning module & \small \cite{arxiv50} \\ \cline{4-5}
			& 							     & \multicolumn{1}{c|}{}             & Rate controller & \small \cite{arxiv102} \\ 
			\hline
			
			\multirow{9}{*}{\begin{tabular}[c]{@{}c@{}}$\textbf{\normalsize Interpretability-}$\\ $\textbf{\normalsize related METs}$\end{tabular}} 
			& \multirow{9}{*}{\begin{tabular}[c]{@{}c@{}}\normalsize Limited device\\ \normalsize capabilities\end{tabular}} & \multicolumn{1}{c|}{\multirow{3.2}{*}{\normalsize KG}} & \begin{tabular}[c]{@{}c@{}}Interpretable semantic information\\ detection algorithm\end{tabular} & \small \cite{arxiv19} \\ \cline{4-5}
			& 							     & \multicolumn{1}{c|}{}            & Transformer-based knowledge extractor & \small \cite{arxiv51} \\ \cline{4-5}
			& 							     & \multicolumn{1}{c|}{}            & Semantic representation based-on KG & \small \cite{arxiv64} \\ \cline{3-5}
			& 							     & \multicolumn{1}{c|}{\multirow{2.2}{*}{\normalsize ProbLog}}  & \begin{tabular}[c]{@{}c@{}}Conversion from NPM\\ to a symbolic graph\end{tabular} & \small \cite{arxiv101} \\ \cline{4-5}
			& 							     & \multicolumn{1}{c|}{}             & ProbLog-based KB & \small \cite{arxiv7} \\ \cline{3-5}
			& 							     & \multicolumn{1}{c|}{\multirow{1.2}{*}{\normalsize Conceptual space}} & \begin{tabular}[c]{@{}c@{}}Mapping semantic information into \\ coordinates in geometric space\end{tabular} & \small \cite{arxiv80} \\ 
			\hline
			
			\multirow{7.2}{*}{\begin{tabular}[c]{@{}c@{}}$\textbf{\normalsize Security-}$\\ $\textbf{\normalsize related METs}$\end{tabular}}
			& \multirow{7.2}{*}{\normalsize Security} & \multicolumn{1}{c|}{\multirow{2.2}{*}{\begin{tabular}[c]{@{}c@{}}\normalsize Understanding attack\\ \normalsize mechanisms\end{tabular}}} & \begin{tabular}[c]{@{}c@{}}Destructive physical layer\\ semantic attacks\end{tabular} & \small \cite{arxiv17,newjsac4} \\ \cline{4-5}
			& 							     & \multicolumn{1}{c|}{}            & Backdoor (Trojan) attacks & \small \cite{arxiv108} \\ \cline{3-5}
			
			& 							     & \multicolumn{1}{c|}{\multirow{3.2}{*}{\begin{tabular}[c]{@{}c@{}}\normalsize Designing defense\\ \normalsize mechanisms\end{tabular}}} & \begin{tabular}[c]{@{}c@{}}Counter-eavesdropping\\ DeepJSCC model\end{tabular} & \small \cite{arxiv24} \\ \cline{4-5}
			& 							     & \multicolumn{1}{c|}{}             & Adversarial encryption training scheme & \small \cite{arxiv34} \\ \cline{4-5}
			& 							     & \multicolumn{1}{c|}{}             & Privacy filter & \small \cite{arxiv42} \\ \cline{3-5}
			& 							     & \multicolumn{1}{c|}{\multirow{1.2}{*}{\normalsize New security-related metrics}} & SOP, DFP & \small \cite{arxiv72} \\ 
			\hline
		\end{tabular}\centering
	\end{table*}

	From a macro perspective, the meaning-oriented semantic communication system proposed in this section integrates the perception, communication, and computing parts. The coordinated collaboration of these three parts contributes to better application of the system in various scenarios within the SAGSIN. Now, let's narrow our focus to some specific details and consider the practical works of system implementation. It is evident that DNNs serve as the core to build semantic codecs. However,  considering the objective realities such as limited device capabilities, highly dynamic channels, and sensitive communication security issues in the SAGSIN, it is necessary to incorporate additional targeted ``designs" for the  components of semantic communication systems. We name these ``designs" as METs and categorize them into four major classes, which  respectively focus on the channel adaptation, rate control, interpretability, and security. In the following, we will review the existing related works for each category, and the summary of these techniques is shown in Table \ref{IV_C}.
	\subsubsection{Channel-related METs}	
\begin{figure}[!t]
		\centering
		\includegraphics[width = 0.48\textwidth]{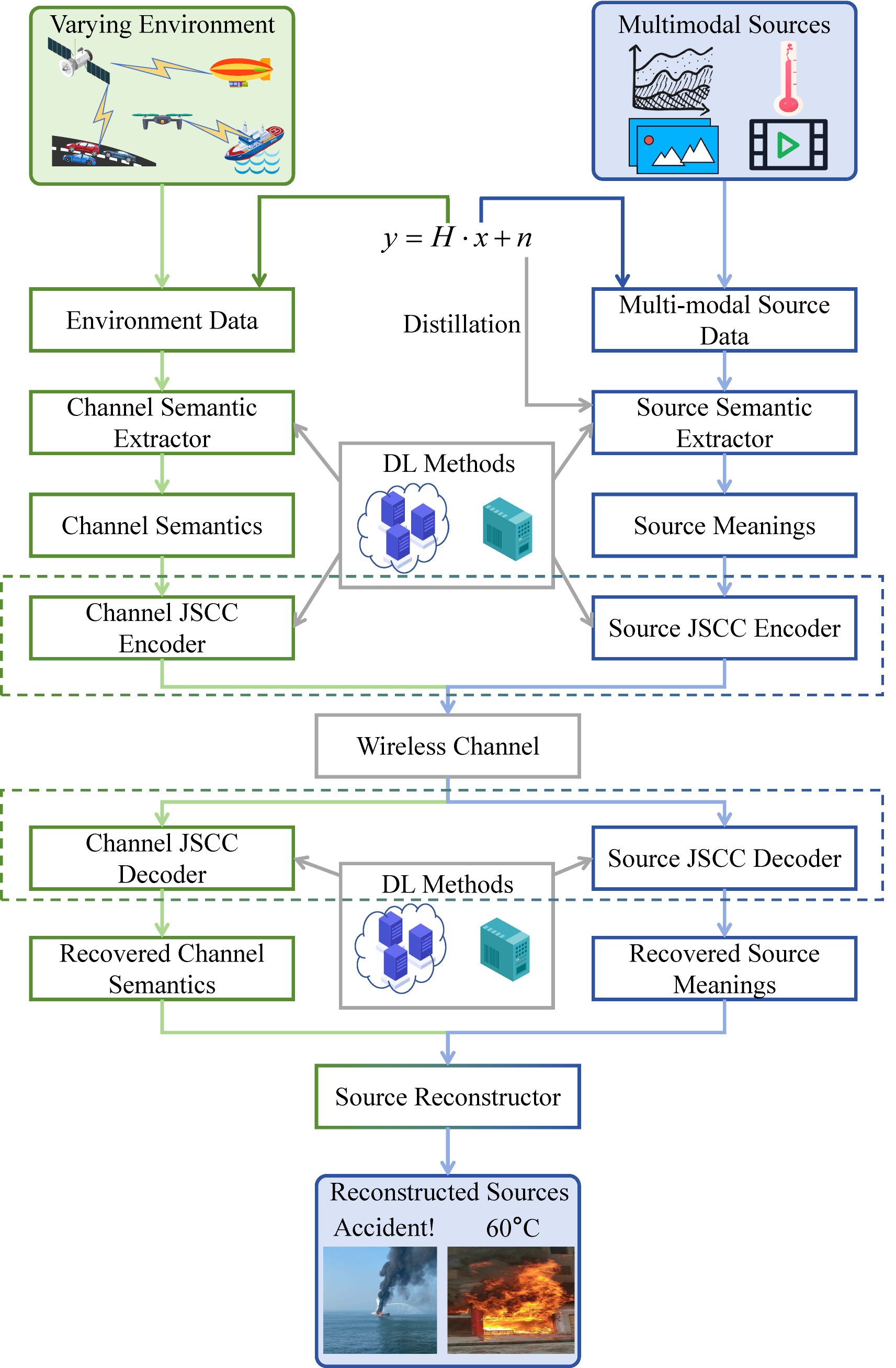}
		\caption{An illustration of typical channel-related METs, including channel semantic extraction (CSI involved) and noise resisting. Here, $x$ denotes the original source data, $H$ denotes the channel gain parameter (which is a part of CSI), $n$ denotes the noise, and $y$ denotes the channel outputs. The ``Environment Data" in the figure include far more than the CSI, and here we use the CSI extraction as an example to illustrate the functions of theses METs. Also, it is worth noting that the source and channel JSCC encoders (decoders) are usually integrated by one module in the literature.}
		\label{figchannel}
	\end{figure}

	Complex and highly dynamic channels in the SAGSIN pose challenges for channel modeling in system design. Meanwhile, in order to ensure high recovery performance, DL-based systems require similar channel conditions during both training and testing phases. Therefore, it is necessary to conduct targeted channel-related analysis and design, such as extraction of channel semantics, noise-resistant mechanisms, and the capturing \& utilization of channel state information (CSI). The main procedures of realizing these channel-related METs are demonstrated in Fig. \ref{figchannel}.

	Analyzing and extracting the channel semantics to assist source information encoding is a new research direction \cite{arxiv36, arxiv37}. In \cite{arxiv1}, the authors introduce the concept of ``channel semantics", which encompasses parameter semantics (e.g., angle of departure, angle of arrival, number of paths, and so on) and environmental semantics (e.g., the layout, shape, and category of objects in the images). The system obtains environmental information in the SAGSIN through devices such as cameras and radars, and extracts channel semantics by DL techniques. The system considering channel semantics exhibits superiority in  terms of meaning similarity metrics such as  WER and FDSD.
	
	To address the impact of noise in complex channels, a confidence-based distillation mechanism is established in\cite{arxiv71} to achieve the ``joint semantic-noise coding (JSNC)". The distillation time in the semantic distillation mechanism of JSNC can be automatically adjusted to accommodate the fluctuating channel conditions. Taking the example of the AWGN channel with 10 dB SNR, JSNC achieves a 24\% lower WER compared to the Transformer baseline. Furthermore, considering the unique ``semantic channel" (the information flow from semantic representation to semantic recovery) of the semantic communication system, the authors in \cite{arxiv2} classify semantic noise for text sources into ``literal semantic noise" and ``adversarial semantic noise", and propose a robust DL-based semantic communication system (R-DeepSC) to resist them. Specifically,  a calibrated self-attention mechanism is utilized to counteract the former, and an adversarial training method is employed to combat the latter. Compared to other baselines considering physical noise only, R-DeepSC achieves superior BLEU and BERT scores at different SNRs.
	
	Effectively capturing and utilizing the rich CSI in the SAGSIN will bring extra recovery performance gains. The authors of \cite{arxiv12} design a plugin-in CSI modulation module, called ``Channel ModNet". \cite{arxiv12} inserts it into the DeepJSCC decoder and proposes an adaptive semantic communication (ASC) based on the overfitting of the source and channel. Different from offline learning method adopted by most works on semantic communications, \cite{arxiv12} firstly introduces online learning policies, where the model parameters will be updated in model inference procedure besides training. This design adapts well to different SNRs in block fading channels. Reaching the same PSNR, ASC saves bandwidth up to 41\% compared to the benchmark. Similarly, \cite{arxiv47} also adopts an online design in order to adapt to dynamic channel conditions. The authors introduce a ``response network" enabled nonlinear transform source-channel coding (NTSCC) framework to explicitly construct response functions and directly embed CSI into the parameters of the DeepJSCC codec. Under the AWGN channel with 10 dB SNR, this architecture achieves higher PSNR than traditional transmission schemes with standardized image codecs across a wide range of channel bandwidth ratios.
	
	Also, in \cite{arxiv114}, the authors propose a wireless image transmission Transformer (WITT) model, utilizing the Swin Transformer and incorporating ``Channel ModNet"\cite{arxiv12} into the codec. Compared to two baselines, namely the CNN-based DeepJSCC and the traditional BPG-LDPC SSCC, WITT exhibits superior PSNR and MS-SSIM at a wide range of (particularly low) SNRs. \cite{arxiv25} proposes an improved version of the classical DeepJSCC, called the adaptive DeepJSCC. Specifically, an ``attention feature module" is utilized to interact with codec DNNs, which can process the input CSI to adapt to different channel conditions. Experimental results show that the adaptive DeepJSCC achieves better PSNR than the classical DeepJSCC under a wide range of SNRs.  Besides, a joint coding-modulation scheme based on binary phase shift keying (BPSK) modulation for digital semantic communications is proposed in \cite{arxiv54}, matching the encoding and modulation processes with CSI better and achieving larger PSNR. Also, a lite distributed DL-based semantic communication system  in \cite{dytext3} is assisted by CSI during training, reducing the impact of fading channels on the transmission and increasing BLEU scores.
	\begin{remark}
		\rm Though introducing online learning method will achieve performance gain owing to adaptability to dynamic channels\cite{arxiv12,arxiv47}, the inherent spacial/temporal computing complexity cannot be neglected, especially for the devices with limited capabilities distributed in SAGSIN. Specifically, the deep networks utilized in semantic extraction \& reconstruction possess huge amount of parameters, and updating such networks will consume caching resources and cause long processing delays. Therefore, striking a trade-off between performance gain and computing complexity in online learning policy will be a promising idea.
	\end{remark}
	
	\subsubsection{Rate-control-related METs}	
	Rate control is another measure to address the issues of highly dynamic channels for the SAGSIN. In the JSCC encoding process, allocating flexible rates to the extracted semantics allows for better adaptation to time-variant channels. Specifically, based on whether the retransmission protocols are adopted to prompt the correct decoding of the current codeword after decoding failure, rate-control techniques can be classified into retransmission-based ones and non-retransmission-based ones.
	
	A typical retransmission-based technique is the HARQ. Recently, a few variants of HARQ have been proposed in semantic communication systems. It is worth noting that the HARQs here are improved versions to adapt to the meaning-oriented semantic communication system, which is the main difference compared to the ones discussed in Section \ref{sec3B}. In \cite{dytext4} and \cite{arxiv111}, HARQ-IR used in traditional communication systems is improved to apply to semantic communication systems.  The authors in \cite{dytext4} firstly propose SC-RS-HARQ, which belongs to the first category of DeepSSCC mentioned in Remark \ref{4-2-2}. The semantic codec is based on Transformers, while the channel codec utilize Reed-Solomon (RS) codes. Because HARQ-IR is only based on RS codes, the authors directly incorporate it into SC-RS-HARQ. To further improve system performance, they improve HARQ-IR and apply it to the DeepJSCC architecture, resulting in SCHARQ. Simulation results show that in a wide range of (especially low) SNRs, SCHARQ achieves lower WER compared to SC-RS-HARQ. \cite{arxiv111} proposes a semantic video conferencing (SVC) network and improves HARQ-IR specifically for SVC, resulting in SVC-HARQ. The semantic error detector is employed to ascertain whether an incremental transmission is necessary for the received frame. Compared to traditional communication systems, SVC-HARQ exhibits lower AKD at high BERs. In \cite{arxiv11}, the authors propose a progressive semantic HARQ scheme based on the incremental knowledge (IK-HARQ). An encoding mechanism with adaptive semantic rate control is employed to dynamically adjust rates based on the contextual semantic complexity and channel conditions. Compared to other benchmarks, IK-HARQ outperforms in terms of BLEU in a wide range of SNRs.
	
	It is worth noting that non-retransmission does not imply non-feedback. In \cite{dyimage4},  no retransmission mechanism is employed, but the authors utilize feedback signals from the current codeword to guide the encoding of subsequent codewords, and introduce an autoencoder-based JSCC scheme, called DeepJSCC-f. They employ layered autoencoders and leverage channel output feedback to achieve variable codeword length transmission. The simulation results demonstrate the benefits of feedback in improving system performance. Compared to the non-feedback DeepJSCC and traditional communication systems, DeepJSCC-f reaches higher PSNR at different SNRs. 
	
	The following works on rate control are all without retransmission or feedback. \cite{dyimage6} proposes an adaptive JSCC for wireless image transmission. The ``policy network" is introduced to activate or freeze visual features by dynamically generating binary masks, thereby achieving rate control. In scenarios with a high SNR and lower information content in the image, this model can learn reasonable strategies with less bandwidth. \cite{dyimage7} and \cite{IEEE1} propose the same ``rate adaptive transmission" mechanism to apply to embedding vectors in latent representation of the encoder side. This enables the model to learn to allocate limited bandwidth resources in order to maximize overall performance. The model in \cite{dyimage7} exhibits superior performance in terms of PSNR and MS-SSIM compared to DeepJSCC and traditional communication systems at various SNRs. In \cite{IEEE1}, the proposed architecture achieves up to 50\% savings in channel bandwidth costs compared to traditional communication systems while maintaining the same PSNR and MS-SSIM performances. Besides, \cite{arxiv86} introduces a variable-length semantic-channel coding  method, where a ``rate allocation network" is utilized to estimate the optimal code length and enhance the PSNR and LPIPS performance. The forward-adaption-based  autoencoder scheme proposed in \cite{arxiv50} improves performance in MS-SSIM and PSNR by introducing an ``adaptive density learning module" at the transmitter to adaptively construct the codebook. \cite{arxiv102} proposes a reinforcement learning based adaptive semantic coding (RL-ASC) model, which uses a ``rate controller" to adaptively adjust the code length. Under different rates, RL-ASC exhibits superior performance to benchmarks in terms of PSNR, SSIM, FID, and KID.
	
	\subsubsection{Interpretability-related METs} 
	One of the widely criticized flaws of DL is lack of interpretability. The black-box nature makes it hard for people to understand the internal decision process of models, hindering model improvement and optimization. Therefore, enhancing the interpretability of semantic communication systems by specific means (e.g., KG and probabilistic logic programming language (ProbLog)) can help further improve communication performance in the SAGSIN. Meanwhile, enhancing interpretability also facilitates the efficient learning of NNs by accelerating the convergence of complex models. This can better address the characteristic of limited device capabilities in the SAGSIN.
	
	KG is a graphical structure used to describe the relationships between entities. It is an extension and development of a knowledge base (KB). KGs contain a series of semantic triplets in the form of ``entity-relation-entity". As an advantage compared to KBs, KGs provide richer semantic associations and contextual information. Therefore, with the backdrop of building the meaning-oriented semantic communication system, KGs gain increasing attention recently. \cite{arxiv19} proposes an interpretable semantic information detection algorithm by using KG triplets as semantic symbols. Also, an efficient semantic correction algorithm is also introduced by extracting inference rules from the KG. The KG-based semantic communication system in \cite{arxiv19} exhibits high sentence similarity in a wide range of (particularly low) SNRs. The authors in \cite{arxiv51} present a KG-enhanced semantic communication framework. In particular, they design a Transformer-based knowledge extractor in the decoder side to extract relevant factual triplets based on the received noisy signal, thereby enhancing the semantic decoding capability. The experimental results demonstrate that regardless of channel types, in low SNRs, the semantic communication system based on knowledge extractor always yields a BLEU improvement of over 5\% for the model. An interpretable semantic communication framework for text data transmission is considered in \cite{arxiv64}. The semantic information is  represented through a collection of semantic triplets based on a KG.  In addition, the authors also define a new metric, namely MSS, which comprehensively measures the accuracy and completeness of the reconstructed data. The framework proposed in \cite{arxiv64}, compared to traditional communication systems, can reduce the amount of transmission data by 41.3\% and improve the overall MSS by two times.
	
	ProbLog is a programming language that combines logic programming with probability theory, extending traditional logic programming. It represents knowledge in the form of logical clauses and allows specifying probability facts and rules within these logical clauses, enabling the representation and inference of knowledge involving uncertainty and probability. ProbLog has been widely applied in various fields, including machine learning (ML), NLP, and so on. In \cite{arxiv101}, the authors propose a semantic protocol model (SPM) to enhance the interpretability of  the NN-based protocol models (NPMs). Specifically, ProbLog is employed to convert the NPM into a symbolic graph. The experimental results confirm that the SPM performs closely to the NPM  while occupying only 0.02\% of the memory. \cite{arxiv7} introduces a KB-based model interpreted by the ProbLog to facilitate semantic information exchange at the semantic communication level.
	
	In addition, \cite{arxiv80} proposes semantic communications based on the conceptual space. Conceptual spaces are similar to KBs, wherein the coder maps the semantic information to the coordinates of a geometric space. By this means, semantic distortion is quantified as the distance between received information and original information in the conceptual space. This approach also enhances the interpretability of the model.
	\subsubsection{Security-related METs} 
	\begin{figure*}[!t]
		\centering
		\includegraphics[width = 0.96\textwidth]{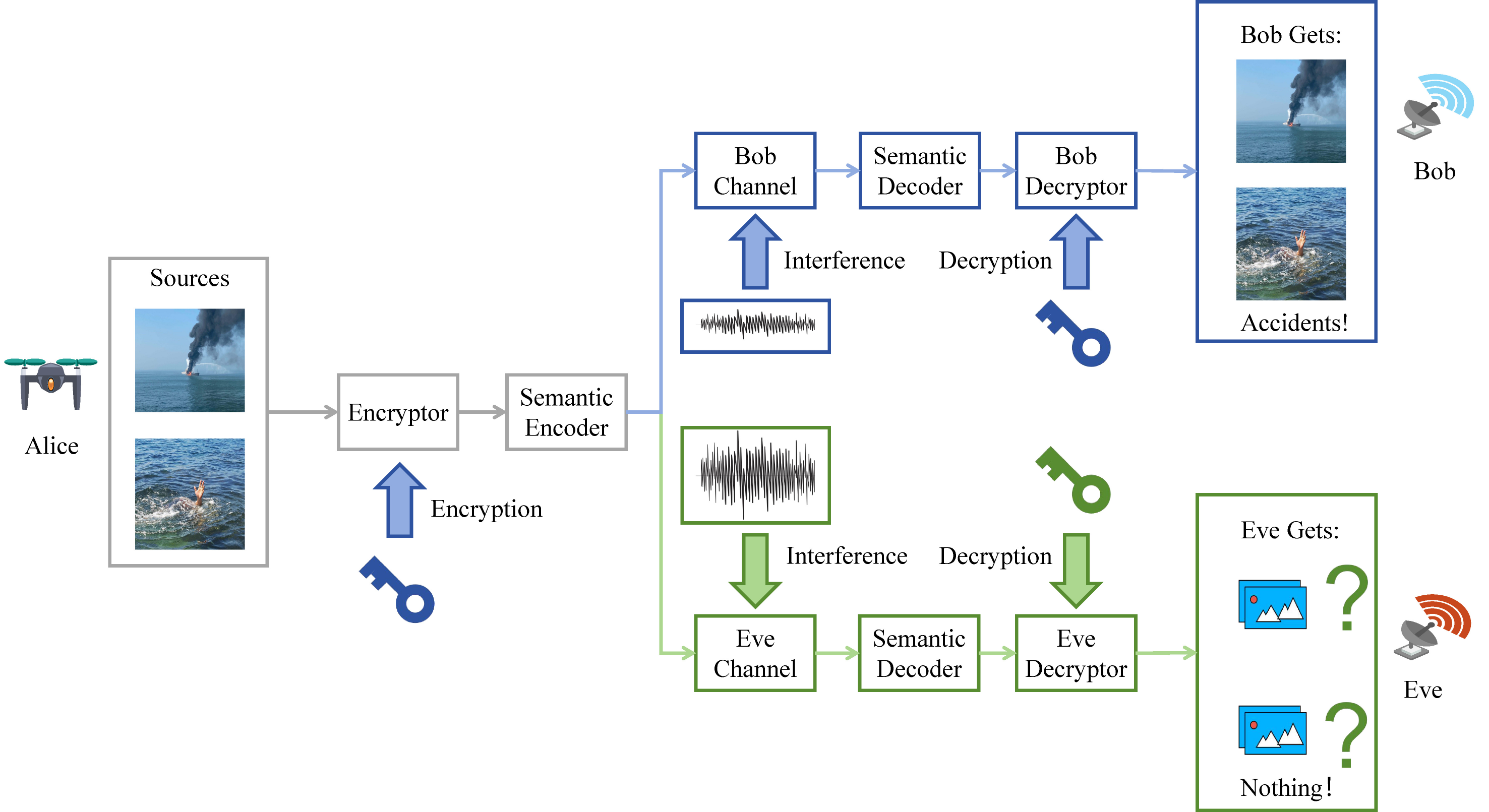}
		\caption{An illustration of typical security communication in the SAGSIN scenario. Alice and Bob share the key such at the source data encrypted by Alice can be decrypted by Bob and cannot be decrypted by those who do not have the key. Eve aims at eavesdropping the transmitted data while suffering from severe interference and not knowing the key. Thus, Eve cannot recover the data sent from Alice.}
		\label{figsecure}
	\end{figure*}
	Communication security, as a prerequisite for accurately transmitting information, has always been an essential concern, especially in the SAGSIN. This comprehensive network consists of numerous nodes, handling massive services at any given moment. When two specific nodes engage in the end-to-end communications, they do not want to be eavesdropped  by other unrelated nodes. Therefore, focusing on the security issues under the background of semantic communications becomes an emerging research direction \cite{arxiv75}. A typical semantic communication system with eavesdropping is shown in Fig. \ref{figsecure}.

	The first step of designing techniques to defend against attacks is understanding the attack mechanisms. A few works validate the impact of certain specific attacks on the performance of the semantic communication system, demonstrating the necessity of researching semantic communication security. \cite{arxiv17} proposes a model called SemBLK, which can learn to generate destructive physical layer semantic attacks for end-to-end semantic communication systems in a black-box setting. Semantic perturbations are generated by introducing a surrogate semantic encoder in the SemBLK. Experimental results demonstrate the destructive and imperceptible nature of black-box attacks in the SemBLK by comparing metrics such as PSNR and SSIM with benchmarks. Similarly, in the recent work \cite{newjsac4}, a semantic perturbation generator called SemAdv is trained for physical-layer attacks that can pollute the images with specific semantics in an imperceptible, controllable, and input-agnostic manner. Besides, the authors in \cite{arxiv108} demonstrate that the meaning-oriented semantic communication system exhibits weak resistance against backdoor (Trojan) attacks. The experimental results indicate that backdoor attacks are stealthy and selective. 
	
	The second step is to design defense mechanisms to enhance the semantic security. \cite{arxiv24} proposes a counter-eavesdropping DeepJSCC model, called DeepJSCEC. It can resist the eavesdroppers' chosen-plaintext attacks without assuming their channel conditions. Compared to the SSCC encryption schemes of traditional communication systems, this model exhibits comparable or even superior PSNR, SSIM, and MS-SSIM performance at different SNRs. An encrypted semantic communication system (ESCS) is proposed in \cite{arxiv34}. Specifically, the authors introduce an adversarial encryption training scheme, which enables the model to achieve high communication accuracy in both encrypted and unencrypted modes. Under the support of ESCS, at high SNRs, the BLEU score of Bob (receiver) approaches 1, while that of Eve (attacker) is less than 0.2. \cite{arxiv42} proposes a StyleGAN-based semantic communication framework, which utilizes a privacy filter and a KB to replace private information with natural features in the KB, ensuring communication security. Compared to the benchmark (a kind of DeepJSCC), the model has a smaller and more stable LPIPS value under different SNRs.
	
	Some new security-based metrics used in semantic communications are also proposed in the literature. In \cite{arxiv72}, the authors examine and contrast traditional security methods, such as physical layer security, covert communications, and encryption, in terms of semantic information security. They introduce two metrics, which are semantic secrecy outage probability (SOP) for physical layer security techniques, and detection failure probability (DFP) for covert communication techniques. With these metrics as the ultimate goal, the security-related design becomes clearer.
	\begin{remark}
		Finally, the following works, not classified into the above four categories, have also made significant contributions to semantic communications. Many of these have been used as benchmarks of the aforementioned references. We categorize them based on the source types: text \cite{DeepJSCC, SentenceSimilarity, arxiv43, arxiv59, arxiv60, arxiv93}, image/video \cite{dyimage1, dyimage2, dyimage5, IEEE3, arxiv21, arxiv27, arxiv28, arxiv41, arxiv57, arxiv58, arxiv81, arxiv107}, and speech \cite{dyspeech1, IEEE4, arxiv26, arxiv91, arxiv117}. Readers can delve into the details.
	\end{remark}
	
	\subsection{Lessons Learned from This Section}
	This section provides a detailed review of the meaning-oriented semantic communication system. The metrics reflecting the ``meaning similarity" between the original and reconstructed data serve as key guidelines for the system design, and it is worth noting that the accurate representation of the ``meaning similarity" for the utilized evaluation metrics is the basis to improve the system performance. As the fundamental method for system design, DNNs are considered as the absolute core of the meaning-oriented communication system. When designing the system, considering the characteristics of SAGSIN such as highly dynamic channels and limited device capabilities, incorporating meaning-enhancement techniques into DNNs is a powerful approach to further improve the system performance. As a summary, this section provides an idea for designing the meaning-oriented semantic communication system: starting from maximum meaning similarity metrics, establishing DNN-related models, and then proposing targeted meaning-enhancement techniques.
	
	It is worth noting that, the NNs introduced in the meaning-oriented semantic communications will not deteriorate the processing delay performance despite a high computing complexity. Since most of the meaning-oriented semantic communication systems are trained offline, i.e., pre-trained at edge/cloud servers before implementation, the high computation costs in the training phases can be neglected in the communication phases. Moreover, some existing works have quantitatively demonstrated that the processing delay of DL-based semantic communications is shorter than traditional communications. For instance, in \cite{dyimage1}, the authors do extensive experiments and simulation results show that the image processing delay under DeepJSCC method is only 18 ms in average, while this value under traditional joint photographic experts group (JPEG) source coding and LDPC channel coding method is larger than 30 ms. Therefore, DL-based meaning-oriented semantic communications will be a promising option in implementing task-critical applications requiring for lower delay.
	
	\section{Perception-Communication-Computing-Actuation-Integrated Semantic Communications in the SAGSIN: A Task Effectiveness Perspective}\label{sec5}
	In this section, we interpret the term ``semantics" as ``effectiveness-related information" and review a newly proposed approach of semantic communications called effectiveness-oriented semantic communications (also called task-oriented semantic communications), which focus on the effects yielded by the task actuation process by adopting a joint reconstructor \& decision-maker design.\footnote{In this section, the actuation process mentioned in the references is in fact only decision process mentioned in Section \ref{sec2}.}

	Before the review, we firstly elaborate on the reason for introduction of task-oriented communications. Traditionally, both  technical communications and  meaning-oriented semantic communications adopt separate reconstructor and decision-maker design, where the receiver are expected to firstly reconstruct the estimation of source data by inverse process of transmitter, and secondly generate task decision results based on the estimation of data. The reason why reconstruction process is imperative is that the receiver cannot distinguish which part of received data is task-oriented, and thus all the source data must be firstly reconstructed for further decision. In stark comparison, task-oriented communications jointly design the reconstructor and decision-maker, where an intelligent module at the receiver implicitly processes the received data and directly produces task decision results. 
	
	This joint design has a salient characteristic of non-symmetry, since no explicit reconstruction process  corresponding to pre-processing  (e.g., encoding or modulating) occurs at the receiver. Joint reconstructor \& decision-maker design is feasible for receiver side to generate moderate decision results, because the received data contain all the information that is needed for tasks, and thus reconstruction is unnecessary for decision process except when the task itself is to reconstruct source data.  Moreover, such non-symmetric design of joint reconstructor \& decision-maker module is proven to be  simpler in computation than separate design by state-of-the-art literature to be elaborated in Section \ref{sec5B}, which perfectly addresses the issue of limited computing resources in the SAGSIN.
	
	Moreover, since the goal of  task-oriented communications is just yielding perfect decision at receiver side for task actuation at the terminal, semantic extractor at the transmitter can be further refined to be also task-oriented. Specifically, the semantic extractor introduced in Section \ref{sec4} can extract all of the semantic information contained by source data. However, the extracted semantic information is not all useful for task decision. Therefore, a more intelligent task-oriented semantic extractor can be introduced, which extracts only effectiveness-related information (i.e., task-related information) of source data to generate effectiveness-related semantic representation for further transmission. Since the decision-maker at  the receiver needs only task-related information for decision, this effectiveness-related information extractor is feasible for task-oriented communications. Furthermore, the amount of data gets thoroughly declined as compared to extracting all the semantic information, which also tackles the challenge of limited communication capabilities in the SAGSIN. Similar to meaning-oriented semantic communication system introduced in Section \ref{sec4}, distributed learning methods are utilized for joint training and knowledge sharing. Through the design of the effectiveness-related information extractor and joint reconstructor \& decision-maker, the computing complexity gets significantly decreased compared to the meaning-oriented semantic communication system, resulting in a lightweight distributed learning-based effectiveness-oriented communication system.

\begin{figure*}[!t]
		\centering
		\includegraphics[width = 0.96\textwidth]{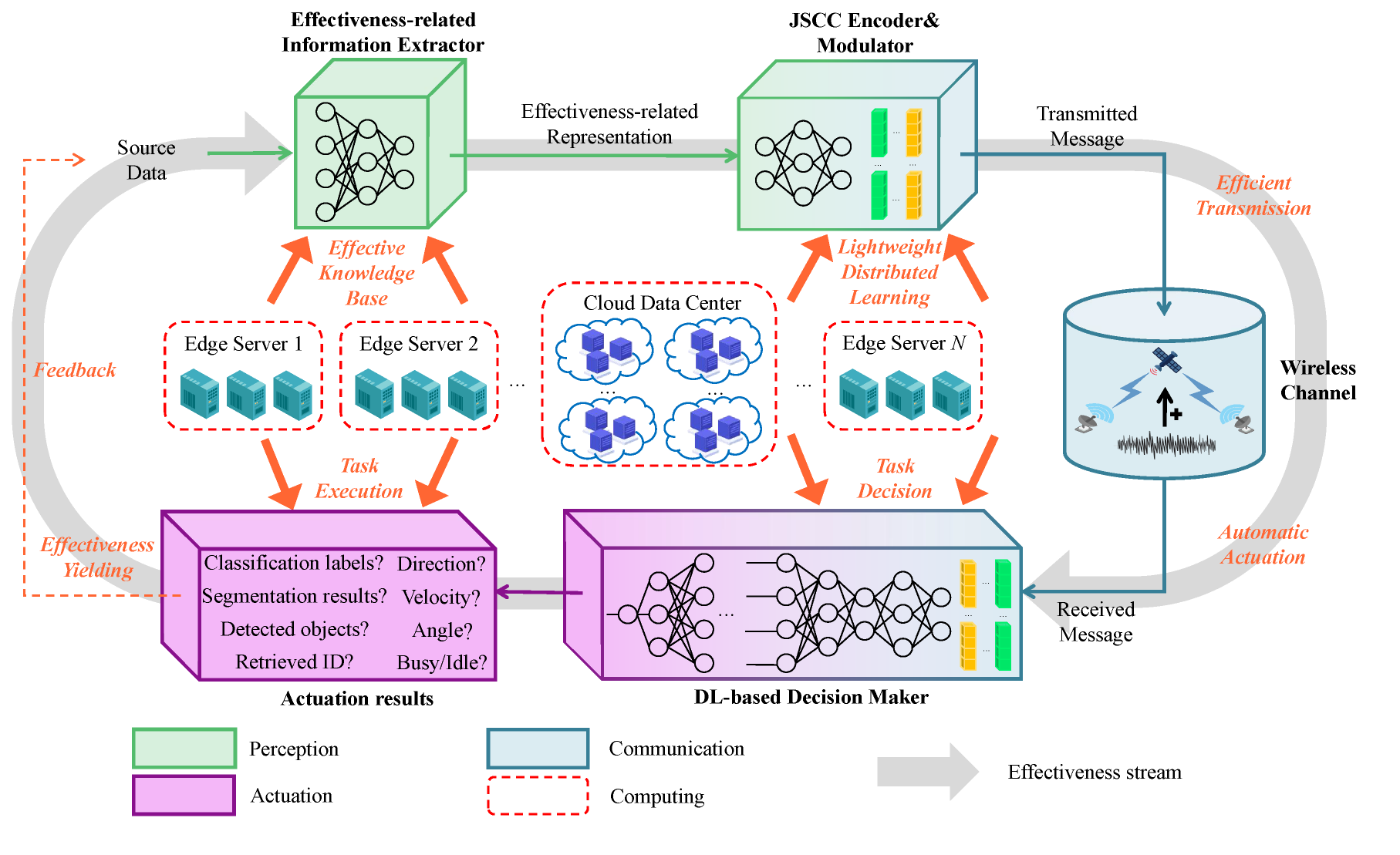}
		\caption{The framework of perception-communication-computing-actuation-integrated semantic communications in the SAGSIN.}
		\label{fig5}
	\end{figure*}

	In task-oriented communications, a radical departure of communication goal from reconstruction to enhancing ultimate effectiveness represents that the actuation process at the terminal along with abovementioned perception, communication, and computing processes is taken into account in network design, by which a new PCCAIP is developed as demonstrated in Fig. \ref{fig5}.  Specifically, in order to reduce the redundancy caused by effectiveness-unrelated information, we utilize an effectiveness-related information extractor as semantic extractor at the transmitter to remove unrelated information and generate effectiveness-related representation. The representation of effectiveness-related information will be further coded and modulated by JSCC encoder and modulator to generate transmitted messages. At the receiver, different from Section \ref{sec3} and Section \ref{sec4} where received messages are  reconstructed or recovered by demodulator, decoder,  and reconstructor, only one intelligent DL-based actuator receives messages as input and outputs directly the task decision results. Similarly, edge servers and cloud data centers provide knowledge bases and support the distributed learning process of the whole network. By intelligent actuator,  the actuation of tasks will affect the physical world (e.g., the variation of source data) at the transmitter via terminal execution and effectiveness is thus yielded. The task decision results will be also transmitted back through feedback channel, which will further affect the behavior of transmitter side. Therefore, the effectiveness-oriented semantic communication system is a closed-loop framework.
	
	In the following subsections, we firstly introduce  metrics to measure the ultimate effectiveness of task-oriented communications; then, we detail the EYTs for the task-oriented semantic communication system according to specific task categories, including image processing services, remote control services, and other computing-intensive services. 
	
	\subsection{Metrics: Towards Ultimate Effectiveness}\label{sec5A}
	Since reconstruction of symbols or semantics no longer exists in task-oriented communications, we cannot use end-to-end metrics, which are based on the distance between reconstructed messages/semantics and the original ones, to measure the performance. Instead, metrics that directly evaluate the ultimate performance of the tasks should be considered, such as the accuracy and  preciseness.
	
	\subsubsection{Accuracy of tasks}
	The tasks related to classification, such as image/video classification, image retrieval, and audio recognition, necessitate the correctness or accuracy of output results. For these classification tasks, the samples can be coarsely divided into four categories, which are true positive (TP), true negative (TN), false positive (FP), and false negative (FN). Denote the number of classified samples as $n$, and a specific category of samples as $n_\text{TP}$, $n_\text{TN}$, $n_\text{FP}$, or $n_\text{FN}$, respectively. Generally, the precision rate $\mathcal{R}_\text{precision}$ is defined as 
	\begin{equation}
		\mathcal{R}_\text{precision}=\frac{n_\text{TP}}{n_\text{TP}+n_\text{FP}},
	\end{equation}
	and the recall rate $\mathcal{R}_\text{precision}$ is defined as 
	\begin{equation}
		\mathcal{R}_\text{recall}=\frac{n_\text{TP}}{n_\text{TP}+n_\text{FN}}.
	\end{equation}
	
	The classification accuracy $\mathcal{R}_\text{acc}$ is defined as the ratio of correctly classified samples including TPs and FNs, that is
	\begin{equation}\label{acc}
		\mathcal{R}_\text{acc}=\frac{n_\text{TP}+n_\text{FN}}{n_\text{total}},
	\end{equation}
	where $n_\text{total}=n_\text{TP}+n_\text{TN}+n_\text{FP}+n_\text{FN}$ is the number of total samples. 
	
	The classification accuracy defined in equation (\ref{acc}) describes how many samples of one category is correctly classified, which is also called top-1 accuracy. As a general case of top-1 accuracy, top-$k$ accuracy is defined as the ratio of samples whose real category (also called ground truth) belong to the categories with highest $k$ degree of confidence.\footnote{The degree of confidence is usually the direct output of DL-based classifier, which represents the probabilities of all the categories. In the cases where the number of categories is small, top-1 accuracy is usually adopted. When the number of categories is large, in order to avoid feature overlapping among categories (a sample has similar degree of confidence on several categories), top-$k$ accuracy is more common.}
	
	In specific classification tasks where the negative impact of FPs is comparable to that of TNs,  $\mathcal{R}_\text{acc}$ is sufficient for serving as ultimate effectiveness metric. However,  in the SAGSIN scenario, the costs yielded by different classification results may be remarkably distinct, implying that the impact of FPs and TNs should be considered as different in more general situations. By adopting this idea,  the F-measure metric is proposed\cite{Fmeasure} to endow different weights for FPs and TNs by a hyper parameter $\beta$, defined as 
	\begin{equation}
		\mathcal{R}_\text{Fmeasure}=\frac{(1+\beta^2)\mathcal{R}_\text{precision}\mathcal{R}_\text{recall}}{\beta^2\mathcal{R}_\text{precision}+\mathcal{R}_\text{recall}}.
	\end{equation}

	F1-score is defined as a special F-measure when $\beta=1$, that is 
	\begin{equation}\label{f1}
		\mathcal{R}_\text{F1score}=\frac{2\mathcal{R}_\text{precision}\mathcal{R}_\text{recall}}{\mathcal{R}_\text{precision}+\mathcal{R}_\text{recall}}.
	\end{equation}
	
	As a more comprehensive accuracy metric, the weighted F1-score combines the accuracy of all categories of samples, which is expressed as
	\begin{equation}
		\mathcal{R}_\text{wF1}=\sum_{i=1}^{m} p_i\mathcal{R}_{\text{F1score},i},
	\end{equation}
	where $m$ is the number of categories, $p_i$ is the ratio of category $i$ among all samples, and $\mathcal{R}_{\text{F1score},i}$ is the F1-score of category $i$ calculated by (\ref{f1}). $\mathcal{R}_\text{wF1}$ can serve as an ultimate effectiveness metric for multi-classification tasks.
	
	For object detection and image segmentation tasks, we are interested in specific classified results in the pixel level. That is, the accuracy of these tasks is determined by the ratio of correctly classified pixels in each image, i.e., whether the detected area or segmented area overlaps the ground truth area. The intersection over union (IoU) metric is defined as the ratio of the intersected area by the union area between the task output results $n_\text{output}$ and the ground truth $n_\text{gt}$, that is
	\begin{equation}
		\mathcal{R}_\text{IoU}=\frac{S_{n_\text{output}\cap n_\text{gt}}}{S_{n_\text{output}\cup n_\text{gt}}},
	\end{equation}
	where $S$ represents the area (i.e., the number of pixels). 
	
	It is worth noting that the ground truth is usually labeled manually, which represents the perception of human on the task results. Because these detection and segmentation tasks necessitate evaluation from human during task decision, the IoU between classification and real  results (i.e., approximate perception of human) is competent for these tasks.
	
	\subsubsection{Preciseness of tasks} The tasks related to operation and control, such as automatic driving and UAV route tracking,  are subject to the preciseness of the commands generated by the decision-maker, for only the precise and exact commands based on the environment and current situation will yield positive effect on the tasks, while  inappropriate commands will hinder the effectiveness of tasks. For these operation tasks, the most determinant factor is system stability, which means that the generated commands must render the state of the system convergent. 
	\begin{figure}[!t]
		\centering
		\includegraphics[width = 0.48\textwidth]{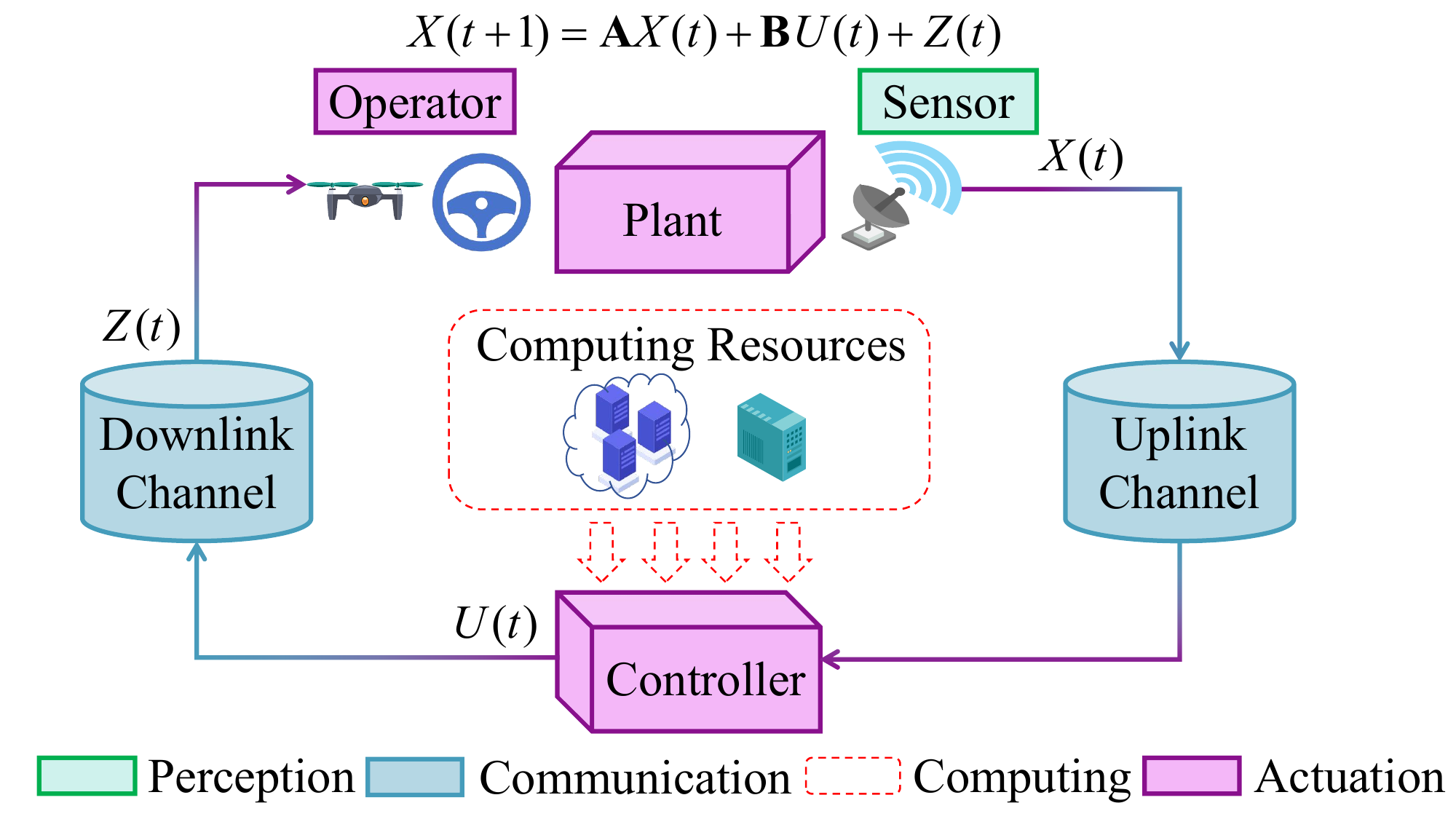}
		\caption{An illustration of WNCS supported by computing resources distributed in the SAGSIN. The WCNS is a typical closed-loop framework, where sensors, controllers, and operators cooperate to maintain the  stability of the processes at the plant. The sensor  captures the state of a process $X(t)$ and transmit it to remote controller through uplink channel. After that, the controller analyzes the state, generates a control command $U(t)$, and sends the command back to the operator through downlink channel. Finally, the operator executes the command interrupted by noises, and the state $X(t)$ will vary according to the execution processes, as shown in equation (\ref{wncs}).}
		\label{figcontrol}
	\end{figure}
	
	Here we take wireless network control system (WNCS), as illustrated in Fig. \ref{figcontrol}, as an example to introduce the preciseness metrics that measure the ultimate effectiveness. Consider a discrete-time process denoted by $X(t)\in \mathbb{R}^n$ which is controlled by command $U(t)\in \mathbb{R}^n$ and interrupted by a noise $Z(t)$ normally distributed with zero mean and covariance matrix $\mathbf{N}$. The process $X(t)$ is assumed as a linear time-invariant system, that is\cite{qiuyifei, liuwanchun}
	\begin{equation}\label{wncs}
		X(t+1)=\mathbf{A}X(t)+\mathbf{B}U(t)+Z(t),
	\end{equation}
	where $\mathbf{A},\mathbf{B}\in \mathbb{R}^{n\times n}$ are the state transition matrix and control coefficient matrix, respectively.
	
	One of the most common preciseness metric is  the MSE of process $X(t)$, with a tacit assumption that the goal of the control task is maintaining $X(t)$ as close as zero. Usually we utilize the long-term average MSE as preciseness metric of the WNCS, denoted by
	\begin{equation}
		\mathcal{R}_\text{MSE}=\frac{1}{T}\sum_{t=0}^{T}Q(t),
	\end{equation}
	where $Q(t)=E(X(t)X(t)^H)$ represents the MSE of each time slot, and $(\cdot)^H$ represents conjugate transpose operation.

	\subsection{Techniques: Towards Effectiveness Yielding}\label{sec5B}
	In the task-oriented semantic communication system, the extraction, dissemination, and ultimate actuation processes are remarkably distinct from the meaning-oriented semantic communication system discussed in Section \ref{sec4}. Task-oriented semantic communication systems are designed towards task actuation at the terminal, and thus the resource consumption of perception, communication, and computing gets reduced, which satisfies the resource-efficient demand in the SAGSIN. In this subsection, we review the EYTs applied in the task-oriented communications which facilitate typical services in the SAGSIN, such as image processing, remote control, and other computing-intensive services. A summary of these EYTs can be seen in Table \ref{V-B}.
	
	\begin{figure*}[!t]
		\centering
		\includegraphics[width = 0.96\textwidth]{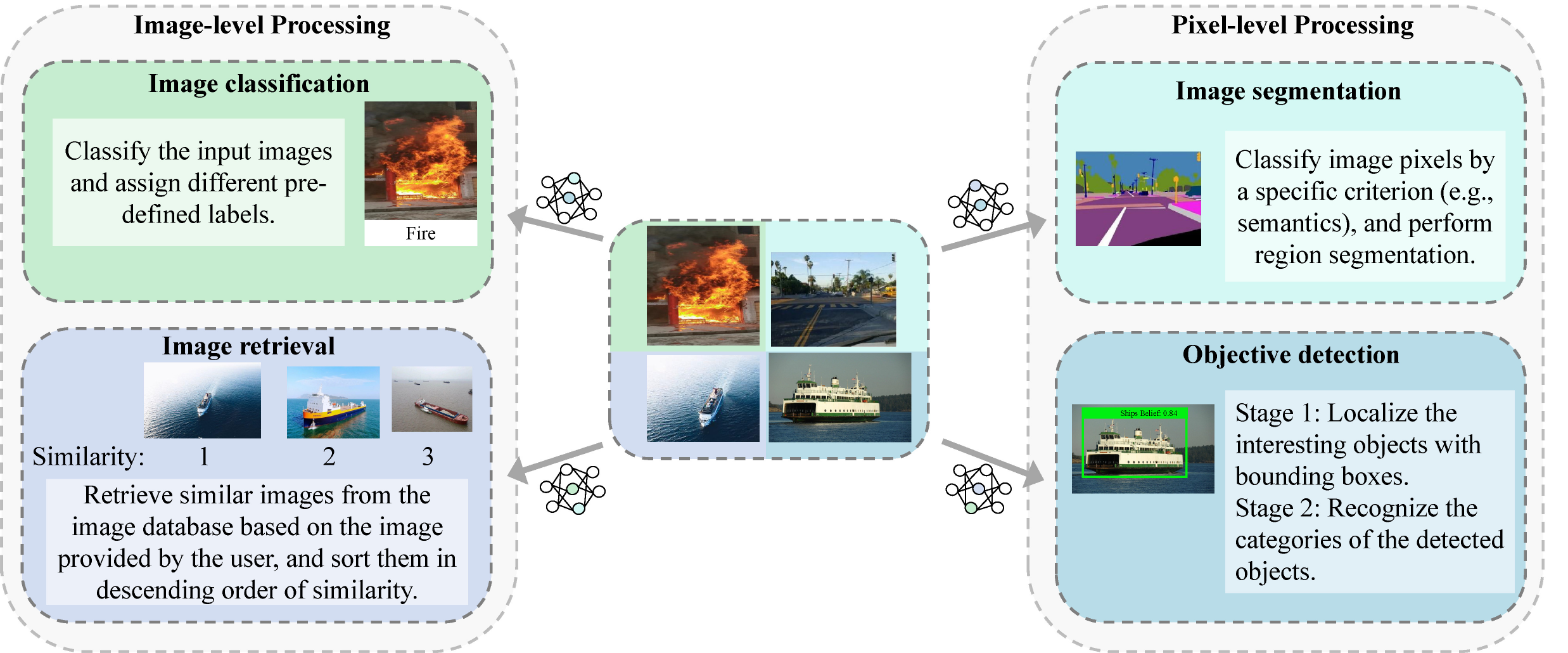}
		\caption{A comparison of typical image processing tasks that can be facilitated by SAGSIN.}
		\label{fig_image_processing1}
	\end{figure*}
	
	\begin{table*}[!t]\normalsize 
		\renewcommand{\arraystretch}{1.3} 
		\caption{SUMMARY OF EFFECTIVENESS YIELDING TECHNIQUES FOR DIFFERENT TASKS}\label{V-B}
		\begin{tabular}[b]{|c|c|c|c|c|c|} 
			\hline 
			$\textbf{Tasks}$ &\renewcommand{\arraystretch}{1.2} \begin{tabular}[c]{@{}c@{}}$\textbf{\normalsize SAGSIN}$\\ $\textbf{\normalsize Issue}$ \\ $\textbf{\normalsize Addressment}$ \end{tabular} & \renewcommand{\arraystretch}{1.2}  \begin{tabular}[c]{@{}c@{}}$\textbf{Related}$\\$\textbf{Metrics}$\end{tabular} & $\textbf{Specific Services}$ & $\textbf{Technical Details}$ & \renewcommand{\arraystretch}{1.2}  \begin{tabular}[c]{@{}c@{}}$\textbf{Related}$\\$\textbf{References}$\end{tabular} \\ 
			\hline
			
			\multirow{13.2}{*}{\begin{tabular}[c]{@{}c@{}}\normalsize $\textbf{Image}$\\ \normalsize $\textbf{processing}$\end{tabular}} & \multirow{13.2}{*}{\begin{tabular}[c]{@{}c@{}} Limited device\\ capabilities\end{tabular}} & \multirow{6.2}{*}{\begin{tabular}[c]{@{}c@{}}Accuracy \\ of tasks\end{tabular}} & \multirow{6.2}{*}{\begin{tabular}[c]{@{}c@{}}Image \\ classification\end{tabular}} & \begin{tabular}[c]{@{}c@{}}AoTI-based rate-accuracy joint \\ optimization\end{tabular} & \small \cite{task1} \\ \cline{5-6}
			& & & & Information bottleneck theory &\small \cite{task10} \\ \cline{5-6}
			& & & & Adaptive compression method   &\small \cite{arxiv10} \\ \cline{5-6}
			& & & & SNN-based digital modulation  &\small \cite{arxiv95} \\ \cline{5-6}
			& & & & \begin{tabular}[c]{@{}c@{}}GAN to resist semantic noise,  \\ semantic codebook method\end{tabular} &\small \cite{arxiv73,arxiv74} \\ \cline{3-6}
			& & \begin{tabular}[c]{@{}c@{}}Accuracy, \\ IoU \end{tabular} & Object detection & Temporal-spatial integrated model &\small \cite{task11} \\ \cline{3-6}
			& & IoU & \begin{tabular}[c]{@{}c@{}} Image (semantic) \\ segmentation \end{tabular}  & \begin{tabular}[c]{@{}c@{}} Collaborative task decision in  \\ internet of vehicles \end{tabular} & \cite{arxiv45} \\ \cline{3-6}
			& & \multirow{3}{*}{\begin{tabular}[c]{@{}c@{}}Accuracy \\ of tasks\end{tabular}} & \multirow{3}{*}{Image retrieval} & JSCC encoder and FCN actuator &\small \cite{arxiv116} \\ \cline{5-6}
			& & & & \begin{tabular}[c]{@{}c@{}}Collaborative task decision for \\ multi-device scenario\end{tabular} &\small \cite{arxiv20} \\
			\hline
			
			\multirow{8.2}{*}{\begin{tabular}[c]{@{}c@{}}\normalsize $\textbf{Remote}$\\ \normalsize $\textbf{control}$\end{tabular}} & \multirow{8.2}{*}{\begin{tabular}[c]{@{}c@{}} Highly dynamic\\ channels\end{tabular}} & \begin{tabular}[c]{@{}c@{}}Accuracy \\ of tasks \end{tabular} & \begin{tabular}[c]{@{}c@{}} Transmission \\ control \end{tabular} & \begin{tabular}[c]{@{}c@{}}Multi-device joint communication- \\ control design \end{tabular} &\small \cite{task5} \\ \cline{3-6}
			& & \multirow{3.2}{*}{\begin{tabular}[c]{@{}c@{}}Accuracy,\\ throughput,\\ or MSE \end{tabular}} & \multirow{3.2}{*}{\begin{tabular}[c]{@{}c@{}}Multi-device\\ scheduling \end{tabular}} & \begin{tabular}[c]{@{}c@{}}Task-oriented data compression\\ based on Dec-POMDP \end{tabular} &\small \cite{task15} \\ \cline{5-6}
			& & & & \begin{tabular}[c]{@{}c@{}}Scheduling for multi-modal tasks\\  based on MDP \end{tabular} & \small \cite{task3} \\ \cline{3-6}
			& & MSE & \begin{tabular}[c]{@{}c@{}}Practical \\ implementation \\ of remote control \end{tabular} & CartPole problem solution & \small \cite{arxiv76} \\
			\hline
			
			\multirow{9.2}{*}{\begin{tabular}[c]{@{}c@{}}\normalsize $\textbf{Other}$\\ \normalsize $\textbf{computing-}$ \\ \normalsize $\textbf{intensive }$ \\ \normalsize $\textbf{tasks}$\end{tabular}} & \multirow{9.2}{*}{\begin{tabular}[c]{@{}c@{}} Limited device\\ capabilities\end{tabular}} & \multirow{9.2}{*}{\begin{tabular}[c]{@{}c@{}} Accuracy, \\ traditional \\ metrics (e.g., \\ rate, error \\ probability) \end{tabular}} & Mixed reality & \begin{tabular}[c]{@{}c@{}}Scene construction according to \\  specific angle of view \end{tabular} & \cite{arxiv15} \\ \cline{4-6}
			& & & \begin{tabular}[c]{@{}c@{}}UAV-based image\\  transmission \end{tabular} & \begin{tabular}[c]{@{}c@{}}Personal-saliency-related semantic\\triplets \end{tabular} & \small \cite{arxiv65} \\ \cline{4-6}
			& & & Identification & \begin{tabular}[c]{@{}c@{}}Centralized identification in \\ multi-device scenario \end{tabular} & \small \cite{arxiv79} \\ \cline{4-6}
			& & & \multirow{2}{*}{\begin{tabular}[c]{@{}c@{}} Question \\ answering \end{tabular}} & {\begin{tabular}[c]{@{}c@{}}Multi-user VQA \end{tabular}}  & \small \cite{VQA,VQA2}  \\
			\cline{5-6}& & & & Question answering with memory  &\small \cite{newjsac5} \\ 
			\cline{4-6}
			& & & \begin{tabular}[c]{@{}c@{}} Sentiment  \\ analysis \end{tabular}  & Semantic triplets  & \small \cite{arxiv97} \\ 
			\hline
		\end{tabular}\centering
	\end{table*}

	\subsubsection{EYTs for image processing services} Image processing tasks include image classification, objective detection, image segmentation (semantic segmentation included), and image retrieval, etc, and the main differences among these tasks are illustrated in Fig. \ref{fig_image_processing1}.
	These services require only for accurate results at the terminal  instead of exact restoration of image itself. Therefore, a joint transmitter-actuator training method can perfectly and efficiently conduct these image processing tasks in the SAGSIN with limited computing resources, benefiting from lite model size and high computing speed. 
	
	\textit{Image classification services.} In \cite{task1}, the authors study a task-oriented image classification system from a freshness perspective. In this system, the transmitter encodes the images  as analog code symbols (i.e., code symbols in a float type) for transmission, and the intelligent classifier deployed at the receiver directly outputs the category of a specific image based on received symbols without reconstruction of image. Taking the age of task information (AoTI) as objective function, the optimal arrival rate and analog code length are respectively solved to achieve highest classification accuracy. The authors in \cite{task10} theoretically analyze a task-oriented communication system for classification based on robust information bottleneck theory, by which both high relevance of information on tasks and low data distortion are guaranteed, achieving high inference accuracy on images. From the resource allocation perspective, an adaptive compression method of the task-oriented communication system for classification is proposed in \cite{arxiv10}. According to the relevance on tasks, the compression ratio and resource consumption are jointly optimized by a two-stage iteration algorithm. Simulation results demonstrate that compared to baseline schemes, the proposed adaptive compression scheme saves 80\% transmission data while maintaining the classification accuracy, implying that both high ultimate effectiveness and high efficiency is achieved.
	
	For the sake of realization of task-oriented communications in the SAGSIN scenarios based on digital modulation,  \cite{arxiv95} proposes a possible solution which utilizes a spiking-neural-network-based (SNN-based) semantic communication system for image classification tasks. Compared with other NNs such as CNN, RNN which have been introduced in Section \ref{sec42}, a remarkable feature of SNN \cite{SNN} is that the output of the last layer is only a vector $\mathbf{z}$ consisting of binary elements, i.e., $\mathbf{z}\in \{0,1\}^n$, where $n$ is the number of neurons at the last layer of SNN. This remarkable feature implies that SNN directly outputs digital symbols that can be modulated by constellations such as BPSK without quantization procedure, reducing distortion of  effectiveness-related information caused by extra quantization. \cite{arxiv95} considers a task-oriented semantic codec equipped with SNN over the BSC or BEC, and provides simulation results on the classification accuracy with regards to transition/erasure probability.
	
	To enhance the robustness of task-oriented communication system against adversarial attacks, the authors in  \cite{arxiv73,arxiv74} focus on anti-semantic-noise semantic communications for robust image classification. The GAN is utilized for generating image data set polluted by semantic noise which cannot be distinguished by human but may mislead the intelligent classifier to output wrong category results. The authors introduce a mask auto-encoder to encounter the negative effect of semantic noise, and propose a semantic codebook method similar to \cite{arxiv8,arxiv9} in order to extract only the most significant features (i.e., the most task-relevant semantics) for transmission. By these means the classification accuracy does not decline because of semantic noise compared with noise-free scenarios.

	\textit{Objective detection services.} \cite{task11} considers task-oriented edge video transmission and detection. Specifically, a temporal entropy model and a temporal-spatial integrated model are respectively proposed to extract effectiveness-related information from both current frame and previous frame. The reason why the proposed method is feasible is that video sources are temporally continuous and thus the information involved by the previous frame can be neglected for feature extraction in the current frame. The ultimate effectiveness in specific tasks, e.g., precision (based on IoU thresholds) of pedestrian distribution detection and multi-device joint objective detection, gets enhanced by task-oriented model design while achieving better rate-effectiveness trade-off. 
	
	\textit{Image segmentation services.}  \cite{arxiv45} studies Internet-of-vehicles-based (IoV-based) task-oriented communication networks for real-time image segmentation. The cars in the front row capture, process, and transmit environment images in real time by swin Transformer encoder, and the cars in the back row conduct image segmentation tasks for received images. The intelligent task-oriented transmitter-actuator pair achieves higher mean IoU performance than traditional data-oriented codec pair, showing the superiority of ultimate effectiveness of task-oriented design. 
	
	\textit{Image retrieval services.} The authors in \cite{arxiv116} consider image retrieval issue at the edge, where one edge device captures and transmits images to the edge server, and the server conducts retrieval task to predict the identification of the images. Specifically, two task-oriented schemes namely SSCC and JSCC schemes are utilized for effectiveness-oriented  feature extraction,  coding, and modulation. Moreover, an fully-connected-network-based classifier at the receiver directly conducts retrieval tasks without image reconstruction. Numerical results show that JSCC scheme achieves better task accuracy in a higher actuation speed. Further considering the multi-device scenario at the edge,  \cite{arxiv20} studies collaborative image retrieval problem based on edge computing. Taking two-device case as an example, edge devices capture images and conduct JSCC-based effectiveness-oriented feature extraction independently, and the JSCC outputs (i.e., effectiveness-related information representation) are transmitted over a shared multiple access channel using NOMA or OMA modulation. The task actuator at the edge server conducts joint image retrieval task by simultaneously processing information from two edge devices. Such collaborative task actuation achieves performance gain in total retrieval accuracy as compared to separate task actuation design.
	
	\subsubsection{EYTs for remote control services} In remote control applications under the support of WNCS shown in Fig. \ref{figcontrol},  the sensor should transmit the real-time status messages through uplink channel to  the controller for command generation, and the commands will be transmitted back to the operator through downlink channel. The controlling process may be corrupted by noise and interference of highly dynamic channels in the SAGSIN. In order to resist the effect of dynamic erroneous channels and yield high ultimate effectiveness, the commands generated based on the received messages are expected to be precise at each time slot, i.e., commands should maintain the stability and convergence of the controlled process.

	From a transmission control perspective,  \cite{task5} constructs a task-oriented multi-device communication system with multi-modal sources and transmission control. Specifically, in the considered human activity recognition task, sources  including room environment and human acceleration information are firstly captured by several end devices and then transmitted over wireless channel. Then, a DL-based processor  generates  commands of whether to empower the monitoring cameras for video transmission based on the captured sources. By controlling the dissemination process of video sources, the task-oriented system achieves both high total transmission rate and moderate human activity recognition accuracy, which implies a positive ultimate effectiveness.
	
	For multi-device control services, the authors in  \cite{task15} aim at task-oriented data compression in a multi-agent scenario, conducting a joint communication and control optimization. The optimization problem is modeled as a decentralized partially observable MDP (Dec-POMDP) to minimize the distortion caused by compression (a mean absolute error (MAE) metric which is similar to MSE) under constraint of transmission rate.  A state-aggregation for information compression algorithm  is proposed as near-optimal solution for the optimization problem, yielding low MAE and thus high ultimate effectiveness of the task-oriented communications. Moreover, \cite{task3} considers scheduling problem among multi-modal users over time-varying channel, where some of the users conduct task-oriented status updating while others conduct traditional data transmission. In order to jointly optimize the ultimate effectiveness of such multi-user system with multi-modal tasks (i.e., both data reconstruction and status updating tasks), both the significance of information (i.e., AoII) for task-oriented user data and the throughput of data-oriented user data should be considered. The joint AoII-throughput optimization problem is formulated as an MDP, and solution verifies a better trade-off between the status prediction accuracy of status updating tasks and the throughput of data recovery tasks.
	
	As a practical example of remote control services which can be deployed in the SAGSIN,  \cite{arxiv76} studies both semantic-level (meaning-level) and effectiveness-level communications in remote control over unreliable channel based on dynamic feature compression. The authors consider a classic CartPole control task actuated by two agents, where one agent serves as observer which captures and encodes the CartPole status image into symbols implicitly representing semantic features, and the other agent serves as actuator which generates a binary command of ``left" and ``right" based on the received semantic symbols. The problem can be modeled as a POMDP, and different levels of communications can be achieved using different reward functions. The authors point out that by setting the reward function as minus MSE, the communication system is semantics-oriented; by setting the reward function as the Q-function learned by the actuator, the communication system is  effectiveness-oriented, since the Q-function comprehensively reflects the cost of each decision.
	
	\subsubsection{EYTs for other computing-intensive services} In this part, we review some recent work on task-oriented communication systems designed for other  tasks that also consume considerably large amount of computing resources. By a task-oriented design, the resource consumption from communication, computing, and task actuation can get further reduced while ensuring effectiveness of tasks, satisfying the energy-efficiency need  for the SAGSIN with limited computing resources.
	
	For instance,  \cite{arxiv15} studies information sharing between mixed reality (MR) devices (e.g., wearable devices) supported by task-oriented communications. The transmitter user captures and transmits the scene information from his angle of view via semantic extractor, and the receiver user receives the information and utilizes a generative network to reproduce the scene from the angle of view of receiver user. Simulation results show that the transmission rate and successful transmission probability at the receiver side get enhanced by task-oriented designs. Another example is  \cite{arxiv65}, in which the personal saliency of users in task-oriented communication networks is  considered. UAVs conduct semantic extraction for captured images to generate a semantic triplet, and personalized users  decide whether to download the whole image based on the preference score of the received semantic triplet. By formulating and solving multi-user resource allocation problem, high comprehensive utility values (i.e., the match scores between personal query and received triplets) of all the users can be achieved, hence yielding high ultimate effectiveness. Moreover, an identification-task-oriented multi-user communication system is studied in \cite{arxiv79}. Cars in the IoV extract the environment semantics and transmit to the central server, and the server conducts identification task for each car based on the integrated semantics from all the cars to ensure high identification accuracy. There are also some efforts on task-oriented communications for question asking and answering tasks.  \cite{VQA,VQA2} propose  multi-user DL-enabled semantic communication systems for visual question answering (VQA) tasks, both of which increase correct answering ratio of the tasks. \cite{newjsac5} also proposes a context-based semantic communication system with memory to enhance the correction ratio of question answering tasks. Besides, \cite{arxiv97} proposes a semantic-triplet-based semantic communication system to complete  sentiment analysis and question answering tasks, and the proposed system achieves at least 43.5\% and 52\% accuracy gains respectively on the two tasks.
	
	\subsection{Lessons Learned from This Section}
	This section elaborates on how a task-oriented communication system is designed by non-symmetric transceivers. Specifically, the core of such non-symmetric transceiver design is to reduce the resource consumption of source processing, data transmission and task decision. An effectiveness-related information extractor is adopted at transmitter side to release the burden of data amount, while an intelligent actuator is utilized to substitute traditional separate reconstructor-actuator design and thus reduce computing resource consumption as compared to reconstruction-based methods. We believe that the task-oriented communication is a promising  idea for semantic communications in the SAGSIN for its energy efficiency and brief design. 
	
	\section{Future Challenges on Semantic Communications in the SAGSIN}\label{sec6}
	In this section, we point out some major challenges and future directions on implementation and enhancement for semantic communications in the SAGSIN scenarios.
	
	\subsection{Precise and Unified Metrics Describing Semantics}
	
	Currently, there still lack precise and unified semantic metrics that can accurately and comprehensively measure multi-dimensional significance/meaning/effectiveness of the source data. On the one hand, the current semantic metrics, especially meaning similarity metrics, can only describe the structural or learned implicit similarity between original and received messages, which is rather inaccurate in ``semantic domain" description. One of the major reasons is that semantic information theory is still in its infant age. Although there have been some state-of-the-art works\cite{2011} to refine classical semantic information theory\cite{1953Semantic}, semantic communication theory is naturally abstract and vague in semantic representation, since it is based on the logical probability which cannot be explicitly calculated. Therefore, scholars are forced to seek alternative metrics that describe semantics from various perspectives, which indirectly while partly define how much semantic information is received by the receiver.
	
	On the other hand, the existing semantic metrics are only suitable for a certain type/modal of data, which lack unity and generality. For instance, AoI and nonlinear AoI reflect only the freshness aspect of semantics, and meaning similarity metrics such as BERT and SSIM are customized for specific modal of sources. Also, it is an indisputable fact that massiveness and multi-modality are fundamental characteristics of data in SAGSIN scenario. The existing one-dimensional metrics are not sufficient for describing the significance/meaning/effectiveness of multi-modal and multi-dimensional source data. 
	Therefore, it is imperative to seek a class of precise and unified semantic metrics for semantic communication system design of multi-modal and multi-dimensional data. 
	Our previous works \cite{beyondAoI,GoTmagazine} initially make some efforts on unifying the significance metrics to propose the GoT metric as a comprehensive goal-oriented semantic metric, which describes the significance by involving the effect of all the natural cost, actuation cost, and the potential cost caused by actuation. In order to propose comprehensive meaning similarity or effectiveness metrics, some problems should be further solved. For instance, a distance-based metric in a certain domain can be considered to unify current meaning-related metrics that respectively reflect the similarity of image, text, and speech sources. Moreover, to measure the effectiveness yielded by source data, we may quantify the potential cost in the life cycle of a certain message, including cost of total transmission delay, cost of total energy consumption, and cost reduced by a precise actuation (or cost increased by a wrong actuation), and a weighted ``unified effectiveness metric" can be finely designed.
	
	\subsection{Semantic Extraction of Multi-modal Data}
	Semantic extraction is always the initial and essential step in DL-based semantic communications, since the performance of extractor directly affects the performance of semantic transmission and further source reconstruction. Nevertheless, current research focuses on semantic extraction for only one specific modal of source data, which cannot directly extract the multi-dimensional semantics from multi-modal data widely distributed in the SAGSIN. In fact, scholars have recently discussed the semantic extraction for multi-modal image-text-fused source data in VQA task\cite{VQA}, while they separately extract the image semantics and text semantics by ResNet and LSTM network, respectively. This separate extractor design for different modals of sources must adopt two or more exclusive NNs. Training such semantic extractor will consume considerably large computing resources, which may not be feasible for end devices and edge servers that have access to limited computing resources in the SAGSIN. Therefore, the issue of extracting semantics of massive multi-modal data with only one intelligent NN has to be addressed.
	
	For instance,  videos are a typical type of multi-modal data with both image and audio sources. Instead of separating audio information from image frames and extracting the semantics respectively, we may consider image frames and audio information at the same time as semantic-related. We can cut audio signal frame by frame, and utilize state-of-the-art DL techniques such as Transformer to simultaneously extract semantic information of both audio signal and image frame in the current time slot. Because of semantic-related audio and image sources, the parameters of NNs will get reduced as compared to separate design, for semantic representation includes the relation of image and audio of current frame, and thus representation length the for current frame is shorter. By integrated extractor design for multi-modal data, a lite semantic coder can be realized and may be directly trained and deployed on small IoT devices in the SAGSIN. This promising idea has been initially explored by \cite{adding3_5}, where a joint image-audio generative diffusion model is utilized for video reconstruction. In their work, the authors quantitatively draw the conclusion that the computational complexity of the proposed model gets significantly decreased compared to baseline models. Such GAN-based model can be potentially utilized for end-to-end semantic communication systems that have smaller parameter sizes.

	\subsection{Joint Decision-Execution Design for the Actuator}
	In the SAGSIN, task execution at the terminals will directly affect the environment statuses, which is the direct effectiveness yielded by the decision commands generated from actuator. However, the current execution modules (or techniques) are principally assumed as separated from the decision-maker design, which is still way from the envisioned PCCAIP in Fig. \ref{fig2} of Section \ref{sec2}. For instance, the current task-oriented image classification decision-maker only gives the category results on a certain image without considering how the classification results facilitate task execution to affect the external environment. The only exception is \cite{arxiv76}, where the intelligent actuator both gives decision results (i.e., left or right) and conducts task execution to complete CartPole game. The real tasks facilitated by the SAGSIN are far more complex and heterogeneous than simple CartPole control, which necessitates more intelligent and comprehensive techniques of joint decision-execution realization for the actuator in order to directly yield ultimate effectiveness.
	
	Here we study an abnormal status observation and rescue task to demonstrate the advantage of such joint decision-actuation design. The observers distributed in the network monitor the environment, and when detecting abnormal statuses,  the observer will transmit them using SPTs, METs, or EYTs. The terminals receive the statuses using a joint reconstruction \& task-decision \& task-execution technique to directly conduct rescue task, e.g., sending rescuers to dangerous areas for sinking ships. The state-of-the-art works reviewed in Section \ref{sec5} mainly focus on the (remote) monitoring steps of this task in order to enhance the accuracy of observation. However, in joint decision-execution design, we are not only interested in the accuracy of observation but also the effect of the decision (which is based on observation results) caused by task execution, e.g., recovery of abnormal statuses due to timely rescue. When several observed objects are detected as abnormal at the same time but the terminal can only support one of the object to be rescued, the actuator should be finely designed to find out the decision yielding the most ultimate effectiveness via considering the different levels of significance of all the objects. In such cases, we must resort to joint design instead of only enhancing the accuracy of observation.
	
	\subsection{Heterogeneous Task-oriented Communication Networks Based on Edge Intelligence}
	In SAGSIN, massive tasks are actuated by heterogeneous nodes in a distributed manner by edge devices. Given the diverse nature of tasks, the relations among nodes may have two principal manners. For nodes which aim at the same task, they should cooperate with each other for the identical task. Conversely, for nodes which target different tasks, they may compete with each other to preempt for limited resources. Therefore, in the task-oriented communication network, it is of vital importance to study intelligent multi-dimensional resource scheduling  policies  among heterogeneous nodes based on edge intelligence.
	In the current literature, task-oriented communication network design usually consider homogeneous task executed cooperatively by edge devices\cite{task17,task19}, which cannot be directly applied for heterogeneous tasks scenario, since the former design considers each device as achieving the same goal. The only exceptions are  \cite{arxiv43, newjsac2}, where both semantic communication users and technical communication users work to achieve heterogeneous goals (i.e., both semantic and bit-level reconstruction). However, in the practical applications, tasks are more complex and different users may have distinct perception qualities and computing resources, which calls for more intelligent and comprehensive design for each users based on edge intelligence.
	
	For nodes which cooperate with each other for the same task, we may consider a multi-UAV tracking scenario where several heterogeneous UAVs track one moving object. Because of limited device capabilities of each UAV, the computing resources and perceived information are insufficient for each UAV to independently maintain the tracking task. As a result, they should resort to remote edge/cloud servers that can aggregate the perceived information from all the nodes to generate global status information (e.g. the absolute position and speed of the object). Next, the remote servers can compute moderate actuation commands (e.g. the speed and direction) for each UAV. Also, each UAV can compute their own actuation command based on local perception of the object. In order to better track the object while avoiding collision between UAVs,  command merging algorithm can be utilized at each UAV to integrate the commands from both itself and the remote server. In this case, both centralized and decentralized policies are adopted, which rely highly on the proposed PCCAIP in this survey.
	
	However, for nodes which compete with each other due to  the pursuit of  distinct tasks, it is worth noting that the shared communication/actuation resources must be moderately allocated to ensure high effectiveness for each task. Since the nodes serve different tasks,  the feasibility of implementing centralized policies becomes physically constrained. In this case, a decentralized resource scheduling policy based on game theory can be adopted, such that the nodes can reach equilibrium in utilizing shared resources. We may consider NOMA-based multi-UAV tracking scenario where each UAV tracks its own object and communicate with a remote server using the same spectrum. UAVs must adjust their own transmitting power to mitigate interference among themselves, thereby facilitating the reliable transmission of status information to the remote server. This in turn enables the remote server to make  more effective task decision for each UAV. Here, decentralized gaming among UAVs also relies highly on the proposed PCCAIP in the manuscript because of limited device capabilities in SAGSIN.
	
	\section{Conclusion}\label{sec7}
	In this survey, we have comprehensively reviewed the implementation of three types of semantic communication systems in the SAGSIN according to the PCCAIP. Starting with significance-oriented semantic communication systems, we introduce metrics measuring the significance of source data and review SPTs including sampling, coding, and modulation policy designs aiming at enhancing the performance in capturing data significance. Then, we  review meaning-oriented semantic communication systems by elaborating on meaning similarity metrics, DNN-based methods, and METs focusing on semantic reconstruction performance improvement. Next, we discuss a newly emerging semantic communication approach called effectiveness-oriented semantic communication systems by introducing effectiveness metrics and studying EYTs intended to increase ultimate effectiveness of task actuation. Finally, we propose several research challenges on implementation of semantic communications in the SAGSIN which will draw further insight research interests. 
	
	\bibliographystyle{IEEEtran}
	\bibliography{reference}

\begin{thebibliography}{100}
\providecommand{\url}[1]{#1}
\csname url@samestyle\endcsname
\providecommand{\newblock}{\relax}
\providecommand{\bibinfo}[2]{#2}
\providecommand{\BIBentrySTDinterwordspacing}{\spaceskip=0pt\relax}
\providecommand{\BIBentryALTinterwordstretchfactor}{4}
\providecommand{\BIBentryALTinterwordspacing}{\spaceskip=\fontdimen2\font plus
\BIBentryALTinterwordstretchfactor\fontdimen3\font minus
  \fontdimen4\font\relax}
\providecommand{\BIBforeignlanguage}[2]{{%
\expandafter\ifx\csname l@#1\endcsname\relax
\typeout{** WARNING: IEEEtran.bst: No hyphenation pattern has been}%
\typeout{** loaded for the language `#1'. Using the pattern for}%
\typeout{** the default language instead.}%
\else
\language=\csname l@#1\endcsname
\fi
#2}}
\providecommand{\BIBdecl}{\relax}
\BIBdecl

\bibitem{6Gwhite}
B.~Aazhang, M.~Juntti, R.~Kantola, P.~Kyösti, S.~LaValle, C.~Lima,
  M.~Matinmikko-Blue, T.~Ojala, A.~Pouttu, A.~Pärssinen, S.~Yrjola,
  P.~Ahokangas, H.~Alves, M.-S. Alouini, J.~Beek, H.~Benn, M.~Bennis,
  J.~Belfiore, E.~Strinati, and E.~Peltonen, ``Key drivers and research
  challenges for {6G} ubiquitous wireless intelligence (white paper),''
  \emph{6G Flagship University of Oulu Finland}, Sep. 2019.

\bibitem{56Gsurvey}
X.~Cheng, Z.~Huang, and L.~Bai, ``Channel nonstationarity and consistency for
  beyond {5G and 6G}: A survey,'' \emph{IEEE Commun. Surveys Tuts.}, vol.~24,
  no.~3, pp. 1634--1669, 3rd Quart., 2022.

\bibitem{IoTsurvey}
F.~Guo, F.~R. Yu, H.~Zhang, X.~Li, H.~Ji, and V.~C.~M. Leung, ``Enabling
  massive {IoT} toward {6G}: A comprehensive survey,'' \emph{IEEE Internet
  Things J.}, vol.~8, no.~15, pp. 11\,891--11\,915, Aug. 2021.

\bibitem{SAGsurvey}
J.~Liu, Y.~Shi, Z.~M. Fadlullah, and N.~Kato, ``Space-air-ground integrated
  network: A survey,'' \emph{IEEE Commun. Surveys Tuts.}, vol.~20, no.~4, pp.
  2714--2741, 4th Quart., 2018.

\bibitem{sagsin_railway}
J.~Sheng, X.~Cai, Q.~Li, C.~Wu, B.~Ai, Y.~Wang, M.~Kadoch, and P.~Yu,
  ``Space-air-ground integrated network development and applications in
  high-speed railways: A survey,'' \emph{IEEE Trans. Intell. Transp. Syst.},
  vol.~23, no.~8, pp. 10\,066--10\,085, Aug. 2022.

\bibitem{blockchain}
Y.~Wang, Z.~Su, J.~Ni, N.~Zhang, and X.~Shen, ``Blockchain-empowered
  space-air-ground integrated networks: Opportunities, challenges, and
  solutions,'' \emph{IEEE Commun. Surveys Tuts.}, vol.~24, no.~1, pp. 160--209,
  1st Quart. 2022.

\bibitem{surveySAGSIN}
H.~Guo, J.~Li, J.~Liu, N.~Tian, and N.~Kato, ``A survey on space-air-ground-sea
  integrated network security in {6G},'' \emph{IEEE Commun. Surveys Tuts.},
  vol.~24, no.~1, pp. 53--87, 1st Quart., 2022.

\bibitem{NTN}
M.~M. Azari, S.~Solanki, S.~Chatzinotas, O.~Kodheli, H.~Sallouha, A.~Colpaert,
  J.~F. Mendoza~Montoya, S.~Pollin, A.~Haqiqatnejad, A.~Mostaani, E.~Lagunas,
  and B.~Ottersten, ``Evolution of non-terrestrial networks from {5G to 6G}: A
  survey,'' \emph{IEEE Commun. Surveys Tuts.}, vol.~24, no.~4, pp. 2633--2672,
  4th Quart., 2022.

\bibitem{Shannon}
C.~E. Shannon, ``A mathematical theory of communication,'' \emph{Bell Syst.
  Tech. J}, vol.~27, no.~3, pp. 379--423, Jul. 1948.

\bibitem{turbo}
C.~Berrou, A.~Glavieux, and P.~Thitimajshima, ``Near {Shannon} limit
  error-correcting coding and decoding: {Turbo-codes. 1},'' in \emph{Proc. IEEE
  Int. Conf. Commun. (ICC)}, vol.~2, May 1993, pp. 1064--1070.

\bibitem{LDPC}
R.~Gallager, ``Low-density parity-check codes,'' \emph{IRE Trans. Inf. Theory},
  vol.~8, no.~1, pp. 21--28, Jan. 1962.

\bibitem{polar}
E.~Arikan, ``Channel polarization: A method for constructing capacity-achieving
  codes,'' in \emph{Proc. IEEE Int. Symp. Inf. Theor. (ISIT)}, Jul. 2008, pp.
  1173--1177.

\bibitem{spinal}
J.~Perry, P.~A. Lannucci, K.~Fleming, H.~Balakrishnan, and D.~Shah, ``Spinal
  codes,'' in \emph{Proc. ACM SIGCOMM Comput. Commun. Rev.}, vol.~42, no.~4,
  Aug. 2012, pp. 49--60.

\bibitem{weaver}
W.~Weaver, ``Recent contributions to the mathematical theory of
  communication,'' \emph{ETC Rev. Gen. Semant.}, vol.~10, no.~4, pp. 261--281,
  1953.

\bibitem{sagsin_magazine_siot}
J.~Jiao, S.~Wu, R.~Lu, and Q.~Zhang, ``Massive access in space-based {Internet}
  of things: Challenges, opportunities, and future directions,'' \emph{IEEE
  Wireless Commun.}, vol.~28, no.~5, pp. 118--125, Oct. 2021.

\bibitem{RASAG}
L.~Bai, R.~Han, J.~Liu, J.~Choi, and W.~Zhang, ``Relay-aided random access in
  space-air-ground integrated networks,'' \emph{IEEE Wireless Commun.},
  vol.~27, no.~6, pp. 37--43, Dec. 2020.

\bibitem{sagsin_magazine_ntn}
M.~Giordani and M.~Zorzi, ``Non-terrestrial networks in the {6G} era:
  Challenges and opportunities,'' \emph{IEEE Netw.}, vol.~35, no.~2, pp.
  244--251, Mar./Apr. 2021.

\bibitem{sagsin_magazine_3D}
G.~Geraci, D.~López-Pérez, M.~Benzaghta, and S.~Chatzinotas, ``Integrating
  terrestrial and non-terrestrial networks: {3D} opportunities and
  challenges,'' \emph{IEEE Commun. Mag.}, vol.~61, no.~4, pp. 42--48, Apr.
  2023.

\bibitem{sagsin_magazine_software}
N.~Zhang, S.~Zhang, P.~Yang, O.~Alhussein, W.~Zhuang, and X.~S. Shen,
  ``Software defined space-air-ground integrated vehicular networks: Challenges
  and solutions,'' \emph{IEEE Commun. Mag.}, vol.~55, no.~7, pp. 101--109, Jul.
  2017.

\bibitem{sagsin_magazine_software2}
Z.~Zhou, J.~Feng, C.~Zhang, Z.~Chang, Y.~Zhang, and K.~M.~S. Huq, ``{SAGECELL}:
  Software-defined space-air-ground integrated moving cells,'' \emph{IEEE
  Commun. Mag.}, vol.~56, no.~8, pp. 92--99, Aug. 2018.

\bibitem{sagsin_magazine_fso}
R.~Samy, H.-C. Yang, T.~Rakia, and M.-S. Alouini, ``Space-air-ground {FSO}
  networks for high-throughput satellite communications,'' \emph{IEEE Commun.
  Mag.}, vol.~61, no.~3, pp. 82--87, Mar. 2023.

\bibitem{6GSAGS}
T.~Hong, M.~Lv, S.~Zheng, and H.~Hong, ``Key technologies in {6G SAGS IoT}:
  Shape-adaptive antenna and radar-communication integration,'' \emph{IEEE
  Netw.}, vol.~35, no.~5, pp. 150--157, Sep./Oct. 2021.

\bibitem{sagsin_magazine_ai}
N.~Kato, Z.~M. Fadlullah, F.~Tang, B.~Mao, S.~Tani, A.~Okamura, and J.~Liu,
  ``Optimizing space-air-ground integrated networks by artificial
  intelligence,'' \emph{IEEE Wireless Commun.}, vol.~26, no.~4, pp. 140--147,
  Aug. 2019.

\bibitem{sagsin_magazine_ai2}
H.~Li, K.~Ota, and M.~Dong, ``{AI} in {SAGIN}: Building deep learning
  service-oriented space-air-ground integrated networks,'' \emph{IEEE Netw.},
  vol.~37, no.~2, pp. 154--159, Mar./Apr. 2023.

\bibitem{sagsin_magazine_vehicle}
F.~Tang, C.~Wen, M.~Zhao, and N.~Kato, ``Machine learning for
  space–air–ground integrated network assisted vehicular network: A novel
  network architecture for vehicles,'' \emph{IEEE Veh. Technol. Mag.}, vol.~17,
  no.~3, pp. 34--44, Sep. 2022.

\bibitem{DLSAGSIN}
D.~Liu, J.~Zhang, J.~Cui, S.-X. Ng, R.~G. Maunder, and L.~Hanzo, ``Deep
  learning aided routing for space-air-ground integrated networks relying on
  real satellite, flight, and shipping data,'' \emph{IEEE Wireless Commun.},
  vol.~29, no.~2, pp. 177--184, Apr. 2022.

\bibitem{gushushi}
S.~Gu, Q.~Zhang, and W.~Xiang, ``Coded storage-and-computation: A new paradigm
  to enhancing intelligent services in space-air-ground integrated networks,''
  \emph{IEEE Wireless Commun.}, vol.~27, no.~6, pp. 44--51, Dec. 2020.

\bibitem{sagsin_magazine_simu}
N.~Cheng, W.~Quan, W.~Shi, H.~Wu, Q.~Ye, H.~Zhou, W.~Zhuang, X.~Shen, and
  B.~Bai, ``A comprehensive simulation platform for space-air-ground integrated
  network,'' \emph{IEEE Wireless Commun.}, vol.~27, no.~1, pp. 178--185, Feb.
  2020.

\bibitem{magazinesignificance}
E.~Uysal, O.~Kaya, A.~Ephremides, J.~Gross, M.~Codreanu, P.~Popovski,
  M.~Assaad, G.~Liva, A.~Munari, B.~Soret, T.~Soleymani, and K.~H. Johansson,
  ``Semantic communications in networked systems: A data significance
  perspective,'' \emph{IEEE Netw.}, vol.~36, no.~4, pp. 233--240, Jul./Aug.
  2022.

\bibitem{magazineniukai}
K.~Niu, J.~Dai, S.~Yao, S.~Wang, Z.~Si, X.~Qin, and P.~Zhang, ``A paradigm
  shift toward semantic communications,'' \emph{IEEE Commun. Mag.}, vol.~60,
  no.~11, pp. 113--119, Nov. 2022.

\bibitem{magazineshiguangming}
G.~Shi, Y.~Xiao, Y.~Li, and X.~Xie, ``From semantic communication to
  semantic-aware networking: Model, architecture, and open problems,''
  \emph{IEEE Commun. Mag.}, vol.~59, no.~8, pp. 44--50, Aug. 2021.

\bibitem{magazinesemantic}
X.~Luo, H.-H. Chen, and Q.~Guo, ``Semantic communications: Overview, open
  issues, and future research directions,'' \emph{IEEE Wireless Commun.},
  vol.~29, no.~1, pp. 210--219, Feb. 2022.

\bibitem{what2021qiao}
Q.~Lan, D.~Wen, Z.~Zhang, Q.~Zeng, X.~Chen, P.~Popovski, and K.~Huang, ``What
  is semantic communication? a view on conveying meaning in the era of machine
  intelligence,'' \emph{J. Commun. Inf. Netw.}, vol.~6, pp. 336--371, Dec.
  2021.

\bibitem{arxivprinciple}
\BIBentryALTinterwordspacing
Z.~Qin, X.~Tao, J.~Lu, and G.~Y. Li, ``Semantic communications: Principles and
  challenges,'' 2021. [Online]. Available:
  \url{http://arxiv.org/abs/2201.01389}
\BIBentrySTDinterwordspacing

\bibitem{arxivless}
\BIBentryALTinterwordspacing
C.~Chaccour, W.~Saad, M.~Debbah, Z.~Han, and H.~V. Poor, ``Less data, more
  knowledge: Building next generation semantic communication networks,'' 2022.
  [Online]. Available: \url{http://arxiv.org/abs/2211.14343}
\BIBentrySTDinterwordspacing

\bibitem{beyond}
D.~Gündüz, Z.~Qin, I.~E. Aguerri, H.~S. Dhillon, Z.~Yang, A.~Yener, K.~K.
  Wong, and C.-B. Chae, ``Beyond transmitting bits: Context, semantics, and
  task-oriented communications,'' \emph{IEEE J. Sel. Areas Commun.}, vol.~41,
  no.~1, pp. 5--41, Jan. 2023.

\bibitem{2023survey}
W.~Yang, H.~Du, Z.~Q. Liew, W.~Y.~B. Lim, Z.~Xiong, D.~Niyato, X.~Chi, X.~Shen,
  and C.~Miao, ``Semantic communications for future {Internet}: Fundamentals,
  applications, and challenges,'' \emph{IEEE Commun. Surveys Tuts.}, vol.~25,
  no.~1, pp. 213--250, 1st Quart., 2023.

\bibitem{edgesemantic}
W.~Xu, Z.~Yang, D.~W.~K. Ng, M.~Levorato, Y.~C. Eldar, and M.~Debbah, ``Edge
  learning for {B5G} networks with distributed signal processing: Semantic
  communication, edge computing, and wireless sensing,'' \emph{IEEE J. Sel.
  Topics Signal Process.}, vol.~17, no.~1, pp. 9--39, Jan. 2023.

\bibitem{huawei}
M.~Chafii, L.~Bariah, S.~Muhaidat, and M.~Debbah, ``Twelve scientific
  challenges for {6G}: Rethinking the foundations of communications theory,''
  \emph{IEEE Commun. Surveys Tuts.}, vol.~25, no.~2, pp. 868--904, 2nd Quart.,
  2023.

\bibitem{Proposal35}
Z.~Wang, Z.~Zhou, H.~Zhang, G.~Zhang, H.~Ding, and A.~Farouk, ``{AI}-based
  cloud-edge-device collaboration in {6G} space-air-ground integrated power
  {IoT},'' \emph{IEEE Wireless Commun.}, vol.~29, no.~1, pp. 16--23, Feb. 2022.

\bibitem{Proposal36}
S.~Yu, X.~Gong, Q.~Shi, X.~Wang, and X.~Chen, ``{EC-SAGINs}:
  Edge-computing-enhanced space-air-ground-integrated networks for {Internet}
  of vehicles,'' \emph{IEEE Internet Things J.}, vol.~9, no.~8, pp. 5742--5754,
  Apr. 2022.

\bibitem{Proposal38}
S.~Gu, Y.~Wang, N.~Wang, and W.~Wu, ``Intelligent optimization of availability
  and communication cost in satellite-{UAV} mobile edge caching system with
  fault-tolerant codes,'' \emph{IEEE Trans. Cogn. Commun. Netw.}, vol.~6,
  no.~4, pp. 1230--1241, Dec. 2020.

\bibitem{Proposal39}
Z.~Zhang, W.~Zhang, and F.-H. Tseng, ``Satellite mobile edge computing:
  Improving {QoS} of high-speed satellite-terrestrial networks using edge
  computing techniques,'' \emph{IEEE Netw.}, vol.~33, no.~1, pp. 70--76,
  Jan./Feb. 2019.

\bibitem{arxiv76}
P.~Talli, F.~Pase, F.~Chiariotti, A.~Zanella, and M.~Zorzi, ``Semantic and
  effective communication for remote control tasks with dynamic feature
  compression,'' in \emph{Proc. IEEE Conf. Comput. Commun. Workshops (INFOCOM
  WKSHPS)}, May 2023, pp. 1--6.

\bibitem{ITU-T}
{ITU-T}, ``Representative use cases and key network requirements for network
  2030,'' FG-NET2030-Sub-G1, Tech. Rep. FG-NET2030-Sub-G1, Jan. 2020.

\bibitem{isac}
A.~Liu, Z.~Huang, M.~Li, Y.~Wan, W.~Li, T.~X. Han, C.~Liu, R.~Du, D.~K.~P. Tan,
  J.~Lu, Y.~Shen, F.~Colone, and K.~Chetty, ``A survey on fundamental limits of
  integrated sensing and communication,'' \emph{IEEE Commun. Surveys Tuts.},
  vol.~24, no.~2, pp. 994--1034, 2nd Quart., 2022.

\bibitem{iscc}
H.~Xing, G.~Zhu, D.~Liu, H.~Wen, K.~Huang, and K.~Wu, ``Task-oriented
  integrated sensing, computation and communication for wireless edge {AI},''
  \emph{IEEE Netw.}, vol.~37, no.~4, pp. 135--144, Jul./Aug. 2023.

\bibitem{AoI1}
S.~Kaul, M.~Gruteser, V.~Rai, and J.~Kenney, ``Minimizing age of information in
  vehicular networks,'' in \emph{Proc. 8th Annu. IEEE Commun. Soc. Conf.
  Sensor, Mesh Ad-Hoc Commun. Netw. (SECOM)}, Jun. 2011, pp. 350--358.

\bibitem{AoI2}
S.~Kaul, R.~Yates, and M.~Gruteser, ``Real-time status: How often should one
  update?'' in \emph{Proc. IEEE Conf. Comput. Commun. (INFOCOM)}, Mar. 2012,
  pp. 2731--2735.

\bibitem{voi}
A.~Kosta, N.~Pappas, A.~Ephremides, and V.~Angelakis, ``Age and value of
  information: Non-linear age case,'' in \emph{Proc. IEEE Int. Symp. Inf.
  Theor. (ISIT)}, Jun. 2017, pp. 326--330.

\bibitem{AoI6}
Y.~Sun and B.~Cyr, ``Sampling for data freshness optimization: Non-linear age
  functions,'' \emph{J. Commun. Netw.}, vol.~21, no.~3, pp. 204--219, Jun.
  2019.

\bibitem{aos}
J.~Zhong, R.~D. Yates, and E.~Soljanin, ``Two freshness metrics for local cache
  refresh,'' in \emph{Proc. IEEE Int. Symp. Inf. Theor. (ISIT)}, Jun. 2018, pp.
  1924--1928.

\bibitem{AoII2}
A.~Maatouk, S.~Kriouile, M.~Assaad, and A.~Ephremides, ``The age of incorrect
  information: A new performance metric for status updates,'' \emph{IEEE/ACM
  Trans. Netw.}, vol.~28, no.~5, pp. 2215--2228, Oct. 2020.

\bibitem{UoI1}
X.~Zheng, S.~Zhou, and Z.~Niu, ``Urgency of information for context-aware
  timely status updates in remote control systems,'' \emph{IEEE Trans. Wireless
  Commun.}, vol.~19, no.~11, pp. 7237--7250, Nov. 2020.

\bibitem{beyondAoI}
\BIBentryALTinterwordspacing
A.~Li, S.~Wu, S.~Sun, and J.~Cao, ``Goal-oriented tensor: Beyond age of
  information towards semantics-empowered goal-oriented communications,'' 2023.
  [Online]. Available: \url{https://arxiv.org/abs/2307.00535}
\BIBentrySTDinterwordspacing

\bibitem{qaoi}
F.~Chiariotti, J.~Holm, A.~E. Kalør, B.~Soret, S.~K. Jensen, T.~B. Pedersen,
  and P.~Popovski, ``Query age of information: Freshness in pull-based
  communication,'' \emph{IEEE Trans. Commun.}, vol.~70, no.~3, pp. 1606--1622,
  Mar. 2022.

\bibitem{AoI5}
R.~D. Yates, Y.~Sun, D.~R. Brown, S.~K. Kaul, E.~Modiano, and S.~Ulukus, ``Age
  of information: An introduction and survey,'' \emph{IEEE J. Sel. Areas
  Commun.}, vol.~39, no.~5, pp. 1183--1210, May 2021.

\bibitem{updateorwait}
Y.~Sun, E.~Uysal-Biyikoglu, R.~D. Yates, C.~E. Koksal, and N.~B. Shroff,
  ``Update or wait: How to keep your data fresh,'' \emph{IEEE Trans. Inf.
  Theory}, vol.~63, no.~11, pp. 7492--7508, Nov. 2017.

\bibitem{sampleiot}
B.~Zhou and W.~Saad, ``Optimal sampling and updating for minimizing age of
  information in the {Internet} of things,'' in \emph{Proc. IEEE Glob. Commun.
  Conf. (GLOBECOM)}, Dec. 2018, pp. 1--6.

\bibitem{wiener}
Y.~Sun, Y.~Polyanskiy, and E.~Uysal, ``Sampling of the {Wiener} process for
  remote estimation over a channel with random delay,'' \emph{IEEE Trans. Inf.
  Theory}, vol.~66, no.~2, pp. 1118--1135, Feb. 2020.

\bibitem{OUprocess}
T.~Z. Ornee and Y.~Sun, ``Sampling and remote estimation for the
  {Ornstein-Uhlenbeck} process through queues: Age of information and beyond,''
  \emph{IEEE/ACM Trans. Netw.}, vol.~29, no.~5, pp. 1962--1975, Oct. 2021.

\bibitem{multisourcesample}
A.~M. Bedewy, Y.~Sun, S.~Kompella, and N.~B. Shroff, ``Age-optimal sampling and
  transmission scheduling in multi-source systems,'' in \emph{Proceedings of
  the 20th ACM International Symposium On Mobile Ad Hoc Networking and
  Computing (MOBIHOC `19)}, Jul. 2019, pp. 121--130.

\bibitem{aoiipower}
Y.~Chen and A.~Ephremides, ``Minimizing age of incorrect information for
  unreliable channel with power constraint,'' in \emph{Proc. IEEE Glob. Commun.
  Conf. (GLOBECOM)}, Dec. 2021, pp. 1--6.

\bibitem{semanticaoii}
A.~Maatouk, M.~Assaad, and A.~Ephremides, ``Semantics-empowered communications
  through the age of incorrect information,'' in \emph{Proc. IEEE Int. Conf.
  Commun. (ICC)}, May 2022, pp. 3995--4000.

\bibitem{AoIIdelay}
\BIBentryALTinterwordspacing
Y.~Chen and A.~Ephremides, ``Minimizing age of incorrect information in the
  presence of timeout,'' 2022. [Online]. Available:
  \url{http://arxiv.org/abs/2207.02926}
\BIBentrySTDinterwordspacing

\bibitem{AoIIHARQ}
\BIBentryALTinterwordspacing
K.~Bountrogiannis, A.~Ephremides, P.~Tsakalides, and G.~Tzagkarakis, ``Age of
  incorrect information with hybrid {ARQ} under a resource constraint for
  {N}-ary symmetric {Markov} sources,'' 2023. [Online]. Available:
  \url{http://arxiv.org/abs/2303.18128}
\BIBentrySTDinterwordspacing

\bibitem{RateAlloc8}
S.~C. Bobbili, P.~Parag, and J.-F. Chamberland, ``Real-time status updates with
  perfect feedback over erasure channels,'' \emph{IEEE Trans. Commun.},
  vol.~68, no.~9, pp. 5363--5374, Sep. 2020.

\bibitem{RateAlloc9}
A.~Arafa, K.~Banawan, K.~G. Seddik, and H.~V. Poor, ``On timely channel coding
  with hybrid {ARQ},'' in \emph{Proc. IEEE Glob. Commun. Conf. (GLOBECOM)},
  Dec. 2019, pp. 1--6.

\bibitem{RateAlloc7}
M.~Xie, Q.~Wang, J.~Gong, and X.~Ma, ``Age and energy analysis for {LDPC} coded
  status update with and without {ARQ},'' \emph{IEEE Internet Things J.},
  vol.~7, no.~10, pp. 10\,388--10\,400, Oct. 2020.

\bibitem{RateAlloc1}
J.~You, S.~Wu, Y.~Deng, J.~Jiao, and Q.~Zhang, ``An age optimized hybrid {ARQ}
  scheme for {Polar} codes via {Gaussian} approximation,'' \emph{IEEE Wireless
  Commun. Lett.}, vol.~10, no.~10, pp. 2235--2239, Oct. 2021.

\bibitem{RateAlloc3}
A.~Li, S.~Wu, J.~Jiao, N.~Zhang, and Q.~Zhang, ``Age of information with
  hybrid-{ARQ}: A unified explicit result,'' \emph{IEEE Trans. Commun.},
  vol.~70, no.~12, pp. 7899--7914, Dec. 2022.

\bibitem{RateAlloc4}
Y.~Wang, S.~Wu, D.~Li, J.~Jiao, and Q.~Zhang, ``Age-optimal {IR-HARQ} design in
  the presence of non-trivial propagation delay,'' in \emph{Proc. IEEE 11th
  Int. Conf. Wireless Commun. Signal Process. (WCSP)}, Oct. 2019, pp. 1--6.

\bibitem{RateAlloc5}
D.~Li, S.~Wu, Y.~Wang, J.~Jiao, and Q.~Zhang, ``Age-optimal {HARQ} design for
  freshness-critical satellite-{IoT} systems,'' \emph{IEEE Internet Things J.},
  vol.~7, no.~3, pp. 2066--2076, Mar. 2020.

\bibitem{RateAlloc6}
D.~Li, S.~Wu, J.~Jiao, N.~Zhang, and Q.~Zhang, ``Age-oriented transmission
  protocol design in space-air-ground integrated networks,'' \emph{IEEE Trans.
  Wireless Commun.}, vol.~21, no.~7, pp. 5573--5585, Jul. 2022.

\bibitem{RateAlloc2}
S.~Meng, S.~Wu, A.~Li, J.~Jiao, N.~Zhang, and Q.~Zhang, ``Analysis and
  optimization of the {HARQ}-based {Spinal} coded timely status update
  system,'' \emph{IEEE Trans. Commun.}, vol.~70, no.~10, pp. 6425--6440, Oct.
  2022.

\bibitem{RateAlloc10}
Y.~Deng, S.~Wu, J.~You, J.~Jiao, N.~Zhang, and Q.~Zhang, ``Optimizing age of
  information in polar-coded status update system,'' \emph{IEEE Internet Things
  J.}, vol.~11, no.~1, pp. 1285--1300, Jan. 2024.

\bibitem{PowerAlloc4}
B.~R. Sharan, S.~Deshmukh, S.~R.~B. Pillai, and B.~Beferull-Lozano, ``Energy
  efficient {AoI} minimization in opportunistic {NOMA/OMA} broadcast wireless
  networks,'' \emph{IEEE Trans. Green Commun. Netw.}, vol.~6, no.~2, pp.
  1009--1022, Jun. 2022.

\bibitem{PowerAlloc1}
S.~Wu, C.~Guo, Z.~Deng, J.~Jiao, N.~Zhang, and Q.~Zhang, ``Optimizing age of
  information in adaptive {NOMA/OMA/}cooperative-{SWIPT-NOMA} system,''
  \emph{IEEE Trans. Wireless Commun.}, vol.~21, no.~12, pp. 11\,125--11\,138,
  Dec. 2022.

\bibitem{PowerAlloc2}
S.~Wu, Z.~Deng, A.~Li, J.~Jiao, N.~Zhang, and Q.~Zhang, ``Minimizing
  age-of-information in {HARQ}-{CC} aided {NOMA} systems,'' \emph{IEEE Trans.
  Wireless Commun.}, vol.~22, no.~2, pp. 1072--1086, Feb. 2023.

\bibitem{PowerAlloc5}
S.~Liao, J.~Jiao, S.~Wu, R.~Lu, and Q.~Zhang, ``Age-optimal power allocation
  scheme for {NOMA}-based {S-IoT} downlink network,'' in \emph{Proc. IEEE Int.
  Conf. Commun. (ICC)}, 2021, pp. 1--6.

\bibitem{PowerAlloc6}
J.~Jiao, H.~Hong, Y.~Wang, S.~Wu, R.~Lu, and Q.~Zhang, ``Age-optimal downlink
  {NOMA} resource allocation for satellite-based {IoT} network,'' \emph{IEEE
  Trans. Veh. Technol.}, vol.~72, no.~9, pp. 11\,575--11\,589, Sep. 2023.

\bibitem{compareHARQ}
Z.~Shi, H.~Ding, S.~Ma, K.-W. Tam, and S.~Pan, ``Inverse moment matching based
  analysis of cooperative {HARQ-IR} over time-correlated {Nakagami} fading
  channels,'' \emph{IEEE Trans. Veh. Technol.}, vol.~66, no.~5, pp. 3812--3828,
  May 2017.

\bibitem{caaoi}
A.~A. Al-Habob, O.~A. Dobre, and H.~V. Poor, ``Age- and correlation-aware
  information gathering,'' \emph{IEEE Wireless Commun. Lett.}, vol.~11, no.~2,
  pp. 273--277, Feb. 2022.

\bibitem{aoiixr}
J.~Chen, J.~Wang, C.~Jiang, and J.~Wang, ``Age of incorrect information in
  semantic communications for {NOMA} aided {XR} applications,'' \emph{IEEE J.
  Sel. Topics Signal Process.}, vol.~17, no.~5, pp. 1093--1105, Sep. 2023.

\bibitem{BLEU}
K.~Papineni, S.~Roukos, T.~Ward, and W.-J. Zhu, ``{BLEU}: a method for
  automatic evaluation of machine translation,'' in \emph{Proc. Annu. Meet.
  Assoc. Comput. Linguist. (ACL)}, Jul. 2002, pp. 311--318.

\bibitem{CIDEr}
R.~Vedantam, C.~Lawrence~Zitnick, and D.~Parikh, ``{CIDEr}: Consensus-based
  image description evaluation,'' in \emph{Proc. IEEE Conf. Comput. Vis.
  Pattern Recognit. (CVPR)}, Jun. 2015, pp. 4566--4575.

\bibitem{arxiv64}
Y.~Wang, M.~Chen, T.~Luo, W.~Saad, D.~Niyato, H.~V. Poor, and S.~Cui,
  ``Performance optimization for semantic communications: An attention-based
  reinforcement learning approach,'' \emph{IEEE J. Sel. Areas Commun.},
  vol.~40, no.~9, pp. 2598--2613, Sep. 2022.

\bibitem{BERT_score}
\BIBentryALTinterwordspacing
T.~Zhang, V.~Kishore, F.~Wu, K.~Q. Weinberger, and Y.~Artzi, ``{BERTSCORE}:
  Evaluating text generation with {BERT},'' 2019. [Online]. Available:
  \url{https://arxiv.org/abs/1904.09675}
\BIBentrySTDinterwordspacing

\bibitem{SentenceSimilarity}
H.~Xie, Z.~Qin, G.~Y. Li, and B.-H. Juang, ``Deep learning enabled semantic
  communication systems,'' \emph{IEEE Trans. Signal Process.}, vol.~69, pp.
  2663--2675, Apr. 2021.

\bibitem{SSIMandMSSSIM}
Z.~Wang, E.~P. Simoncelli, and A.~C. Bovik, ``Multiscale structural similarity
  for image quality assessment,'' in \emph{The Thrity-Seventh Asilomar
  Conference on Signals, Systems \& Computers, 2003}, vol.~2, Nov. 2003, pp.
  1398--1402.

\bibitem{LPIPS}
R.~Zhang, P.~Isola, A.~A. Efros, E.~Shechtman, and O.~Wang, ``The unreasonable
  effectiveness of deep features as a perceptual metric,'' in \emph{Proc. IEEE
  Conf. Comput. Vis. Pattern Recognit. (CVPR)}, Jun. 2018, pp. 586--595.

\bibitem{FID}
M.~Heusel, H.~Ramsauer, T.~Unterthiner, B.~Nessler, and S.~Hochreiter, ``{GANs}
  trained by a two time-scale update rule converge to a local {Nash}
  equilibrium,'' in \emph{Proc. Adv. Neural Inf. Process. Syst. (NIPS)}, Dec.
  2017, pp. 6627--6638.

\bibitem{KID}
\BIBentryALTinterwordspacing
M.~Bi{\'n}kowski, D.~J. Sutherland, M.~Arbel, and A.~Gretton, ``Demystifying
  {MMD GANs},'' 2018. [Online]. Available:
  \url{https://arxiv.org/abs/1801.01401}
\BIBentrySTDinterwordspacing

\bibitem{arxiv111}
P.~Jiang, C.-K. Wen, S.~Jin, and G.~Y. Li, ``Wireless semantic communications
  for video conferencing,'' \emph{IEEE J. Sel. Areas Commun.}, vol.~41, no.~1,
  pp. 230--244, Jan. 2023.

\bibitem{SDR}
E.~Vincent, R.~Gribonval, and C.~F{\'e}votte, ``Performance measurement in
  blind audio source separation,'' \emph{IEEE Trans. Audio, Speech, Language
  Process.}, vol.~14, no.~4, pp. 1462--1469, Jul. 2006.

\bibitem{PESQ}
``Perceptual evaluation of speech quality ({PESQ}): An objective method for
  end-to-end speech quality assessment of narrow-band telephone networks and
  speech codecs,'' \emph{ITU-T recommendation P.862}, Feb. 2001.

\bibitem{MCD}
R.~Kubichek, ``Mel-cepstral distance measure for objective speech quality
  assessment,'' in \emph{Proc. IEEE Pacific Rim Conf. Commun. Comput. Signal
  Process.}, vol.~1, May 1993, pp. 125--128.

\bibitem{FDSDKDSD}
\BIBentryALTinterwordspacing
M.~Bi{\'n}kowski, J.~Donahue, S.~Dieleman, A.~Clark, E.~Elsen, N.~Casagrande,
  L.~C. Cobo, and K.~Simonyan, ``High fidelity speech synthesis with
  adversarial networks,'' 2019. [Online]. Available:
  \url{https://arxiv.org/abs/1909.11646}
\BIBentrySTDinterwordspacing

\bibitem{BERT_model}
J.~Devlin, M.-W. Chang, K.~Lee, and K.~Toutanova, ``{BERT}: Pre-training of
  deep bidirectional {Transformers} for language understanding,'' in
  \emph{Proc. Conf. North Amer. Chapt. Assoc. Comput. Linguist. Human Lang.
  Technol. (NAACL HLT)}, Jun. 2019, pp. 4171--4186.

\bibitem{arxiv117}
Z.~Weng and Z.~Qin, ``Semantic communication systems for speech transmission,''
  \emph{IEEE J. Sel. Areas Commun.}, vol.~39, no.~8, pp. 2434--2444, Aug. 2021.

\bibitem{arxiv69}
\BIBentryALTinterwordspacing
K.~Lu, R.~Li, X.~Chen, Z.~Zhao, and H.~Zhang, ``Reinforcement learning-powered
  semantic communication via semantic similarity,'' 2021. [Online]. Available:
  \url{https://arxiv.org/abs/2108.12121}
\BIBentrySTDinterwordspacing

\bibitem{arxiv71}
K.~Lu, Q.~Zhou, R.~Li, Z.~Zhao, X.~Chen, J.~Wu, and H.~Zhang, ``Rethinking
  modern communication from semantic coding to semantic communication,''
  \emph{IEEE Wireless Commun.}, vol.~30, no.~1, pp. 158--164, Feb. 2023.

\bibitem{DeepJSCC}
N.~Farsad, M.~Rao, and A.~Goldsmith, ``Deep learning for joint source-channel
  coding of text,'' in \emph{Proc. IEEE Int. Conf. Acoust., Speech Signal
  Process. (ICASSP)}, Apr. 2018, pp. 2326--2330.

\bibitem{LeNet5}
Y.~LeCun, L.~Bottou, Y.~Bengio, and P.~Haffner, ``Gradient-based learning
  applied to document recognition,'' \emph{Proc. IEEE}, vol.~86, no.~11, pp.
  2278--2324, Nov. 1998.

\bibitem{AlexNet}
A.~Krizhevsky, I.~Sutskever, and G.~E. Hinton, ``Imagenet classification with
  deep convolutional neural networks,'' in \emph{Proc. Adv. Neural Inf.
  Process. Syst.}, 2012, pp. 1097--1105.

\bibitem{FCN}
J.~Long, E.~Shelhamer, and T.~Darrell, ``Fully convolutional networks for
  semantic segmentation,'' in \emph{Proc. IEEE Conf. Comput. Vis. Pattern
  Recognit. (CVPR)}, Jun. 2015, pp. 3431--3440.

\bibitem{ResNet}
K.~He, X.~Zhang, S.~Ren, and J.~Sun, ``Deep residual learning for image
  recognition,'' in \emph{Proc. IEEE Conf. Comput. Vis. Pattern Recognit.
  (CVPR)}, Jun. 2016, pp. 770--778.

\bibitem{ZFNet}
M.~D. Zeiler and R.~Fergus, ``Visualizing and understanding convolutional
  networks,'' in \emph{European Conference on Computer Vision (ECCV)}, Sep.
  2014, pp. 818--833.

\bibitem{VGGNet}
\BIBentryALTinterwordspacing
K.~Simonyan and A.~Zisserman, ``Very deep convolutional networks for
  large-scale image recognition,'' 2014. [Online]. Available:
  \url{https://arxiv.org/abs/1409.1556}
\BIBentrySTDinterwordspacing

\bibitem{GoogLeNet}
C.~Szegedy, W.~Liu, Y.~Jia, P.~Sermanet, S.~Reed, D.~Anguelov, D.~Erhan,
  V.~Vanhoucke, and A.~Rabinovich, ``Going deeper with convolutions,'' in
  \emph{Proc. IEEE Conf. Comput. Vis. Pattern Recognit. (CVPR)}, Jun. 2015, pp.
  1--9.

\bibitem{DenseNet}
G.~Huang, Z.~Liu, L.~Van Der~Maaten, and K.~Q. Weinberger, ``Densely connected
  convolutional networks,'' in \emph{Proc. IEEE Conf. Comput. Vis. Pattern
  Recognit. (CVPR)}, Jul. 2017, pp. 2261--2269.

\bibitem{PSPNet}
H.~Zhao, J.~Shi, X.~Qi, X.~Wang, and J.~Jia, ``Pyramid scene parsing network,''
  in \emph{Proc. IEEE Conf. Comput. Vis. Pattern Recognit. (CVPR)}, Jul. 2017,
  pp. 6230--6239.

\bibitem{PatchGAN}
P.~Isola, J.-Y. Zhu, T.~Zhou, and A.~A. Efros, ``Image-to-image translation
  with conditional adversarial networks,'' in \emph{Proc. IEEE Conf. Comput.
  Vis. Pattern Recognit. (CVPR)}, Jul. 2017, pp. 5967--5976.

\bibitem{Hopfield}
J.~J. Hopfield, ``Neural networks and physical systems with emergent collective
  computational abilities,'' \emph{Proceedings of the National Academy of
  Sciences of the United States of America}, vol.~79, no.~8, pp. 2554--2558,
  Apr. 1982.

\bibitem{Jordan}
M.~Jordan, ``Serial order: a parallel distributed processing approach.{
  Technical report, June 1985-March 1986},'' California Univ., San Diego, La
  Jolla (USA). Inst. for Cognitive Science, Tech. Rep., May 1986.

\bibitem{Elman}
J.~L. Elman, ``Finding structure in time,'' \emph{Cognitive Science}, vol.~14,
  no.~2, pp. 179--211, Mar. 1990.

\bibitem{LSTM}
S.~Hochreiter and J.~Schmidhuber, ``Long short-term memory,'' \emph{Neural
  Comput.}, vol.~9, no.~8, pp. 1735--1780, Nov. 1997.

\bibitem{BiRNN}
M.~Schuster and K.~K. Paliwal, ``Bidirectional recurrent neural networks,''
  \emph{IEEE Trans. Signal Process.}, vol.~45, no.~11, pp. 2673--2681, Nov.
  1997.

\bibitem{BiLSTM}
A.~Graves and J.~Schmidhuber, ``Framewise phoneme classification with
  bidirectional {LSTM} networks,'' in \emph{Proc. IEEE Int. Joint Conf. Neural
  Netw.}, vol.~4, Jul. 2005, pp. 2047--2052.

\bibitem{DeepBiLSTM}
A.~Graves, A.-r. Mohamed, and G.~Hinton, ``Speech recognition with deep
  recurrent neural networks,'' in \emph{Proc. IEEE Int. Conf. Acoust., Speech
  Signal Process. (ICASSP)}, May 2013, pp. 6645--6649.

\bibitem{GRU}
\BIBentryALTinterwordspacing
K.~Cho, B.~Van~Merri{\"e}nboer, C.~Gulcehre, D.~Bahdanau, F.~Bougares,
  H.~Schwenk, and Y.~Bengio, ``Learning phrase representations using {RNN}
  encoder-decoder for statistical machine translation,'' 2014. [Online].
  Available: \url{https://arxiv.org/abs/1406.1078}
\BIBentrySTDinterwordspacing

\bibitem{Transformer}
A.~Vaswani, N.~Shazeer, N.~Parmar, J.~Uszkoreit, L.~Jones, A.~N. Gomez,
  {\L}.~Kaiser, and I.~Polosukhin, ``Attention is all you need,'' in
  \emph{Proc. Adv. Neural Inf. Process. Syst.}, pp. 6000--6010.

\bibitem{ViT}
\BIBentryALTinterwordspacing
A.~Dosovitskiy, L.~Beyer, A.~Kolesnikov, D.~Weissenborn, X.~Zhai,
  T.~Unterthiner, M.~Dehghani, M.~Minderer, G.~Heigold, S.~Gelly \emph{et~al.},
  ``An image is worth 16x16 words: {Transformers} for image recognition at
  scale,'' 2020. [Online]. Available: \url{https://arxiv.org/abs/2010.11929}
\BIBentrySTDinterwordspacing

\bibitem{DETR}
N.~Carion, F.~Massa, G.~Synnaeve, N.~Usunier, A.~Kirillov, and S.~Zagoruyko,
  ``End-to-end object detection with transformers,'' in \emph{European
  Conference on Computer Vision (ECCV)}, Aug. 2020, pp. 213--229.

\bibitem{swin}
Z.~Liu, Y.~Lin, Y.~Cao, H.~Hu, Y.~Wei, Z.~Zhang, S.~Lin, and B.~Guo, ``Swin
  {Transformer}: Hierarchical vision transformer using shifted windows,'' in
  \emph{Proc. IEEE/CVF Int. Conf. Comput. Vis. (ICCV)}, Oct. 2021, pp.
  9992--10\,002.

\bibitem{GAN}
I.~J. Goodfellow, J.~Pouget-Abadie, M.~Mirza, B.~Xu, D.~Warde-Farley, S.~Ozair,
  A.~Courville, and Y.~Bengio, ``Generative adversarial nets,'' in \emph{Proc.
  Advances Neural Inf. Process. Syst.}, Dec. 2014, pp. 2672--2680.

\bibitem{CGAN}
\BIBentryALTinterwordspacing
M.~Mirza and S.~Osindero, ``Conditional generative adversarial nets,'' 2014.
  [Online]. Available: \url{https://arxiv.org/abs/1411.1784}
\BIBentrySTDinterwordspacing

\bibitem{DCGAN}
\BIBentryALTinterwordspacing
A.~Radford, L.~Metz, and S.~Chintala, ``Unsupervised representation learning
  with deep convolutional generative adversarial networks,'' 2015. [Online].
  Available: \url{https://arxiv.org/abs/1511.06434}
\BIBentrySTDinterwordspacing

\bibitem{WGAN}
M.~Arjovsky, S.~Chintala, and L.~Bottou, ``Wasserstein generative adversarial
  networks,'' in \emph{Proc. Int. Conf. Mach. Learn.}, Aug. 2017, pp. 214--223.

\bibitem{SAGAN}
H.~Zhang, I.~Goodfellow, D.~Metaxas, and A.~Odena, ``Self-attention generative
  adversarial networks,'' in \emph{International Conference on Machine Learning
  (ICML)}, Jun. 2019, pp. 7354--7363.

\bibitem{BigGAN}
\BIBentryALTinterwordspacing
A.~Brock, J.~Donahue, and K.~Simonyan, ``Large scale {GAN} training for high
  fidelity natural image synthesis,'' 2018. [Online]. Available:
  \url{https://arxiv.org/abs/1809.11096}
\BIBentrySTDinterwordspacing

\bibitem{StyleGAN}
T.~Karras, S.~Laine, and T.~Aila, ``A style-based generator architecture for
  generative adversarial networks,'' in \emph{Proc. IEEE Conf. Comput. Vis.
  Pattern Recognit. (CVPR)}, Jun. 2019, pp. 4396--4405.

\bibitem{HiFiGAN}
J.~Kong, J.~Kim, and J.~Bae, ``{Hifi-GAN}: Generative adversarial networks for
  efficient and high fidelity speech synthesis,'' in \emph{Proc. Adv. Neural
  Inf. Process. Syst.}, Dec. 2020, pp. 17\,022--17\,033.

\bibitem{1962}
D.~H. Hubel and T.~N. Wiesel, ``Receptive fields, binocular interaction and
  functional architecture in the cat's visual cortex,'' \emph{The Journal of
  Physiology}, vol. 160, no.~1, p. 106, Jan. 1962.

\bibitem{1980}
K.~Fukushima, ``Neocognitron: A self-organizing neural network model for a
  mechanism of pattern recognition unaffected by shift in position,''
  \emph{Biological Cybernetics}, vol.~36, no.~4, pp. 193--202, Apr. 1980.

\bibitem{GCN}
\BIBentryALTinterwordspacing
T.~N. Kipf and M.~Welling, ``Semi-supervised classification with graph
  convolutional networks,'' 2016. [Online]. Available:
  \url{https://arxiv.org/abs/1609.02907}
\BIBentrySTDinterwordspacing

\bibitem{arxiv13}
Z.~Lu, Y.~Xiao, Z.~Sun, Y.~Li, G.~Shi, X.~Chen, M.~Bennis, and H.~V. Poor,
  ``Adversarial learning for implicit semantic-aware communications,'' in
  \emph{Proc. IEEE Int. Conf. Commun. (ICC)}, May 2023, pp. 4063--4069.

\bibitem{arxiv57}
C.~K. Thomas and W.~Saad, ``Neuro-symbolic artificial intelligence ({AI}) for
  intent based semantic communication,'' in \emph{Proc. IEEE Glob. Commun.
  Conf. (GLOBECOM)}, Dec. 2022, pp. 2698--2703.

\bibitem{arxiv58}
------, ``Neuro-symbolic causal reasoning meets signaling game for emergent
  semantic communications,'' \emph{IEEE Trans. Wireless Commun.}, early access,
  Oct. 4, 2023, doi: {\color{blue} \href
  {https://ieeexplore.ieee.org/document/10272264}{10.1109/TWC.2023.3319981}}.

\bibitem{arxiv102}
D.~Huang, F.~Gao, X.~Tao, Q.~Du, and J.~Lu, ``Toward semantic communications:
  Deep learning-based image semantic coding,'' \emph{IEEE J. Sel. Areas
  Commun.}, vol.~41, no.~1, pp. 55--71, Jan. 2023.

\bibitem{arxiv50}
J.~Huang, D.~Li, C.~Huang, X.~Qin, and W.~Zhang, ``Joint task and data-oriented
  semantic communications: A deep separate source-channel coding scheme,''
  \emph{IEEE Internet Things J.}, vol.~11, no.~2, pp. 2255--2272, Jan. 2024.

\bibitem{arxiv93}
J.-H. Lee, D.-H. Lee, E.~Sheen, T.~Choi, and J.~Pujara, ``{SEQ2SEQ-SC}:
  End-to-end semantic communication systems with pre-trained language model,''
  in \emph{2023 57th Asilomar Conference on Signals, Systems, and Computers},
  Oct. 2023, pp. 260--264.

\bibitem{arxiv99}
\BIBentryALTinterwordspacing
F.~Liu, W.~Tong, Y.~Yang, Z.~Sun, and C.~Guo, ``Task-oriented image semantic
  communication based on rate-distortion theory,'' 2022. [Online]. Available:
  \url{http://arxiv.org/abs/2201.10929}
\BIBentrySTDinterwordspacing

\bibitem{dytext4}
P.~Jiang, C.-K. Wen, S.~Jin, and G.~Y. Li, ``Deep source-channel coding for
  sentence semantic transmission with {HARQ},'' \emph{IEEE Trans. Commun.},
  vol.~70, no.~8, pp. 5225--5240, Aug. 2022.

\bibitem{arxiv34}
X.~Luo, Z.~Chen, M.~Tao, and F.~Yang, ``Encrypted semantic communication using
  adversarial training for privacy preserving,'' \emph{IEEE Commun. Lett.},
  vol.~27, no.~6, pp. 1486--1490, Jun. 2023.

\bibitem{arxiv91}
T.~Han, Q.~Yang, Z.~Shi, S.~He, and Z.~Zhang, ``Semantic-preserved
  communication system for highly efficient speech transmission,'' \emph{IEEE
  J. Sel. Areas Commun.}, vol.~41, no.~1, pp. 245--259, Jan. 2023.

\bibitem{arxiv36}
S.~Imran, G.~Charan, and A.~Alkhateeb, ``Environment semantic aided
  communication: A real world demonstration for beam prediction,'' in
  \emph{Proc. IEEE Int. Conf. Commun. Workshops (ICC Workshops)}, May 2023, pp.
  48--53.

\bibitem{arxiv37}
Y.~Yang, F.~Gao, X.~Tao, G.~Liu, and C.~Pan, ``Environment semantics aided
  wireless communications: A case study of {mmWave} beam prediction and
  blockage prediction,'' \emph{IEEE J. Sel. Areas Commun.}, vol.~41, no.~7, pp.
  2025--2040, Jul. 2023.

\bibitem{arxiv1}
Z.~Qin, F.~Gao, B.~Lin, X.~Tao, G.~Liu, and C.~Pan, ``A generalized semantic
  communication system: From sources to channels,'' \emph{IEEE Wireless
  Commun.}, vol.~30, no.~3, pp. 18--26, Jun. 2023.

\bibitem{arxiv2}
X.~Peng, Z.~Qin, D.~Huang, X.~Tao, J.~Lu, G.~Liu, and C.~Pan, ``A robust deep
  learning enabled semantic communication system for text,'' in \emph{Proc.
  IEEE Glob. Commun. Conf. (GLOBECOM)}, Dec. 2022, pp. 2704--2709.

\bibitem{arxiv12}
\BIBentryALTinterwordspacing
J.~Dai, S.~Wang, K.~Yang, K.~Tan, X.~Qin, Z.~Si, K.~Niu, and P.~Zhang,
  ``Adaptive semantic communications: Overfitting the source and channel for
  profit,'' 2022. [Online]. Available: \url{https://arxiv.org/abs/2211.04339}
\BIBentrySTDinterwordspacing

\bibitem{arxiv114}
K.~Yang, S.~Wang, J.~Dai, K.~Tan, K.~Niu, and P.~Zhang, ``{WITT}: A wireless
  image transmission transformer for semantic communications,'' in \emph{Proc.
  IEEE Int. Conf. Acoust., Speech Signal Process. (ICASSP)}, Jun. 2023, pp.
  1--5.

\bibitem{arxiv47}
S.~Wang, J.~Dai, X.~Qin, Z.~Si, K.~Niu, and P.~Zhang, ``Improved nonlinear
  transform source-channel coding to catalyze semantic communications,''
  \emph{IEEE J. Sel. Topics Signal Process.}, vol.~17, no.~5, pp. 1022--1037,
  Sep. 2023.

\bibitem{arxiv25}
J.~Xu, T.-Y. Tung, B.~Ai, W.~Chen, Y.~Sun, and D.~Gündüz, ``Deep joint
  source-channel coding for semantic communications,'' \emph{IEEE Commun.
  Mag.}, vol.~61, no.~11, pp. 42--48, Nov. 2023.

\bibitem{arxiv54}
Y.~Bo, Y.~Duan, S.~Shao, and M.~Tao, ``Learning based joint coding-modulation
  for digital semantic communication systems,'' in \emph{Proc. IEEE 14th Int.
  Conf. Wireless Commun. Signal Process. (WCSP)}, Nov. 2022, pp. 1--6.

\bibitem{dytext3}
H.~Xie and Z.~Qin, ``A lite distributed semantic communication system for
  {Internet} of things,'' \emph{IEEE J. Sel. Areas Commun.}, vol.~39, no.~1,
  pp. 142--153, Jan. 2021.

\bibitem{arxiv11}
Q.~Zhou, R.~Li, Z.~Zhao, Y.~Xiao, and H.~Zhang, ``Adaptive bit rate control in
  semantic communication with incremental knowledge-based {HARQ},'' \emph{IEEE
  Open J. Commun. Soc}, vol.~3, pp. 1076--1089, Jul. 2022.

\bibitem{dyimage4}
D.~B. Kurka and D.~G{\"u}nd{\"u}z, ``{DeepJSCC}-f: Deep joint source-channel
  coding of images with feedback,'' \emph{IEEE J. Sel. Areas Inf. Theory},
  vol.~1, no.~1, pp. 178--193, May 2020.

\bibitem{dyimage6}
M.~Yang and H.-S. Kim, ``Deep joint source-channel coding for wireless image
  transmission with adaptive rate control,'' in \emph{Proc. IEEE Int. Conf.
  Acoust., Speech Signal Process. (ICASSP)}, May 2022, pp. 5193--5197.

\bibitem{dyimage7}
J.~Dai, S.~Wang, K.~Tan, Z.~Si, X.~Qin, K.~Niu, and P.~Zhang, ``Nonlinear
  transform source-channel coding for semantic communications,'' \emph{IEEE J.
  Sel. Areas Commun.}, vol.~40, no.~8, pp. 2300--2316, Aug. 2022.

\bibitem{IEEE1}
S.~Wang, J.~Dai, Z.~Liang, K.~Niu, Z.~Si, C.~Dong, X.~Qin, and P.~Zhang,
  ``Wireless deep video semantic transmission,'' \emph{IEEE J. Sel. Areas
  Commun.}, vol.~41, no.~1, pp. 214--229, Jan. 2023.

\bibitem{arxiv86}
B.~Zhang, Z.~Qin, and G.~Y. Li, ``Semantic communications with variable-length
  coding for extended reality,'' \emph{IEEE J. Sel. Topics Signal Process.},
  vol.~17, no.~5, pp. 1038--1051, Sep. 2023.

\bibitem{arxiv19}
F.~Zhou, Y.~Li, X.~Zhang, Q.~Wu, X.~Lei, and R.~Q. Hu, ``Cognitive semantic
  communication systems driven by knowledge graph,'' in \emph{Proc. IEEE Int.
  Conf. Commun. (ICC)}, May 2022, pp. 4860--4865.

\bibitem{arxiv51}
B.~Wang, R.~Li, J.~Zhu, Z.~Zhao, and H.~Zhang, ``Knowledge enhanced semantic
  communication receiver,'' \emph{IEEE Commun. Lett.}, vol.~27, no.~7, pp.
  1794--1798, Jul. 2023.

\bibitem{arxiv101}
S.~Seo, J.~Park, S.-W. Ko, J.~Choi, M.~Bennis, and S.-L. Kim, ``Toward semantic
  communication protocols: A probabilistic logic perspective,'' \emph{IEEE J.
  Sel. Areas Commun.}, vol.~41, no.~8, pp. 2670--2686, Aug. 2023.

\bibitem{arxiv7}
J.~Choi, S.~W. Loke, and J.~Park, ``A unified approach to semantic information
  and communication based on probabilistic logic,'' \emph{IEEE Access},
  vol.~10, pp. 129\,806--129\,822, Dec. 2022.

\bibitem{arxiv80}
D.~Wheeler, E.~E. Tripp, and B.~Natarajan, ``Semantic communication with
  conceptual spaces,'' \emph{IEEE Commun. Lett.}, vol.~27, no.~2, pp. 532--535,
  Feb. 2023.

\bibitem{arxiv17}
Z.~Li, X.~Liu, G.~Nan, J.~Zhou, X.~Lyu, Q.~Cui, and X.~Tao, ``Boosting physical
  layer black-box attacks with semantic adversaries in semantic
  communications,'' in \emph{Proc. IEEE Int. Conf. Commun. (ICC)}, May 2023,
  pp. 5614--5619.

\bibitem{newjsac4}
G.~Nan, Z.~Li, J.~Zhai, Q.~Cui, G.~Chen, X.~Du, X.~Zhang, X.~Tao, Z.~Han, and
  T.~Q.~S. Quek, ``Physical-layer adversarial robustness for deep
  learning-based semantic communications,'' \emph{IEEE J. Sel. Areas Commun.},
  vol.~41, no.~8, pp. 2592--2608, Aug. 2023.

\bibitem{arxiv108}
Y.~E. Sagduyu, T.~Erpek, S.~Ulukus, and A.~Yener, ``Vulnerabilities of deep
  learning-driven semantic communications to backdoor ({Trojan}) attacks,'' in
  \emph{Proc. 57th Annu. Conf. Inf. Sci. Syst. (CISS)}, Mar. 2023, pp. 1--6.

\bibitem{arxiv24}
T.-Y. Tung and D.~Gündüz, ``Deep joint source-channel and encryption coding:
  Secure semantic communications,'' in \emph{Proc. IEEE Int. Conf. Commun.
  (ICC)}, May 2023, pp. 5620--5625.

\bibitem{arxiv42}
T.~Han, J.~Tang, Q.~Yang, Y.~Duan, Z.~Zhang, and Z.~Shi, ``Generative model
  based highly efficient semantic communication approach for image
  transmission,'' in \emph{Proc. IEEE Int. Conf. Acoust., Speech Signal
  Process. (ICASSP)}, Jun. 2023, pp. 1--5.

\bibitem{arxiv72}
H.~Du, J.~Wang, D.~Niyato, J.~Kang, Z.~Xiong, M.~Guizani, and D.~I. Kim,
  ``Rethinking wireless communication security in semantic internet of
  things,'' \emph{IEEE Wireless Commun.}, vol.~30, no.~3, pp. 36--43, Jun.
  2023.

\bibitem{arxiv75}
\BIBentryALTinterwordspacing
Z.~Yang, M.~Chen, G.~Li, Y.~Yang, and Z.~Zhang, ``Secure semantic
  communications: Fundamentals and challenges,'' 2023. [Online]. Available:
  \url{https://arxiv.org/abs/2301.01421}
\BIBentrySTDinterwordspacing

\bibitem{arxiv43}
X.~Mu, Y.~Liu, L.~Guo, and N.~Al-Dhahir, ``Heterogeneous semantic and bit
  communications: A semi-{NOMA} scheme,'' \emph{IEEE J. Sel. Areas Commun.},
  vol.~41, no.~1, pp. 155--169, Jan. 2023.

\bibitem{arxiv59}
W.~Li, H.~Liang, C.~Dong, X.~Xu, P.~Zhang, and K.~Liu, ``Non-orthogonal
  multiple access enhanced multi-user semantic communication,'' \emph{IEEE
  Trans. Cogn. Commun. Netw.}, vol.~9, no.~6, pp. 1438--1453, Dec. 2023.

\bibitem{arxiv60}
H.~Hu, X.~Zhu, F.~Zhou, W.~Wu, R.~Q. Hu, and H.~Zhu, ``One-to-many semantic
  communication systems: Design, implementation, performance evaluation,''
  \emph{IEEE Commun. Lett.}, vol.~26, no.~12, pp. 2959--2963, Dec. 2022.

\bibitem{dyimage1}
E.~Bourtsoulatze, D.~B. Kurka, and D.~G{\"u}nd{\"u}z, ``Deep joint
  source-channel coding for wireless image transmission,'' \emph{IEEE Trans.
  Cogn. Commun. Netw.}, vol.~5, no.~3, pp. 567--579, Sep. 2019.

\bibitem{dyimage2}
D.~B. Kurka and D.~G{\"u}nd{\"u}z, ``Successive refinement of images with deep
  joint source-channel coding,'' in \emph{Proc. IEEE 20th Int. Workshop Signal
  Process. Adv. Wireless Commun. (SPAWC)}, Jul. 2019, pp. 1--5.

\bibitem{dyimage5}
M.~Ding, J.~Li, M.~Ma, and X.~Fan, ``{SNR}-adaptive deep joint source-channel
  coding for wireless image transmission,'' in \emph{Proc. IEEE Int. Conf.
  Acoust., Speech Signal Process. (ICASSP)}, Jun. 2021, pp. 1555--1559.

\bibitem{IEEE3}
C.~Dong, H.~Liang, X.~Xu, S.~Han, B.~Wang, and P.~Zhang, ``Semantic
  communication system based on semantic slice models propagation,'' \emph{IEEE
  J. Sel. Areas Commun.}, vol.~41, no.~1, pp. 202--213, Jan. 2023.

\bibitem{arxiv21}
J.~Dai, P.~Zhang, K.~Niu, S.~Wang, Z.~Si, and X.~Qin, ``Communication beyond
  transmitting bits: Semantics-guided source and channel coding,'' \emph{IEEE
  Wireless Commun.}, vol.~30, no.~4, pp. 170--177, Aug. 2023.

\bibitem{arxiv27}
H.~Zhang, S.~Shao, M.~Tao, X.~Bi, and K.~B. Letaief, ``Deep learning-enabled
  semantic communication systems with task-unaware transmitter and dynamic
  data,'' \emph{IEEE J. Sel. Areas Commun.}, vol.~41, no.~1, pp. 170--185, Jan.
  2023.

\bibitem{arxiv28}
W.~Zhang, K.~Bai, S.~Zeadally, H.~Zhang, H.~Shao, H.~Ma, and V.~C.~M. Leung,
  ``{DeepMA}: End-to-end deep multiple access for wireless image transmission
  in semantic communication,'' \emph{IEEE Trans. Cogn. Commun. Netw.}, vol.~10,
  no.~2, pp. 387--402, Apr. 2024.

\bibitem{arxiv41}
S.~Ma, Z.~Zhang, Y.~Wu, H.~Li, G.~Shi, D.~Gao, Y.~Shi, S.~Li, and N.~Al-Dhahir,
  ``Features disentangled semantic broadcast communication networks,''
  \emph{IEEE Trans. Wireless Commun.}, early access, Nov. 29, 2023, doi:
  {\color{blue} \href
  {https://ieeexplore.ieee.org/document/10335608}{10.1109/TWC.2023.3334225}}.

\bibitem{arxiv81}
S.~Yang, H.~Pan, T.-T. Chan, and Z.~Wang, ``Semantic communication-empowered
  physical-layer network coding,'' in \emph{Proc. IEEE Wireless Commun. Netw.
  Conf. (WCNC)}, Mar. 2023, pp. 1--6.

\bibitem{arxiv107}
Q.~Fu, H.~Xie, Z.~Qin, G.~Slabaugh, and X.~Tao, ``Vector quantized semantic
  communication system,'' \emph{IEEE Wireless Commun. Lett.}, vol.~12, no.~6,
  pp. 982--986, Jun. 2023.

\bibitem{dyspeech1}
Z.~Weng, Z.~Qin, and G.~Y. Li, ``Semantic communications for speech signals,''
  in \emph{Proc. IEEE Int. Conf. Commun. (ICC)}, Jun. 2021, pp. 1--6.

\bibitem{IEEE4}
Y.~Tang, N.~Zhou, Q.~Yu, D.~Wu, C.~Hou, G.~Tao, and M.~Chen, ``Intelligent
  fabric enabled {6G} semantic communication system for in-cabin scenarios,''
  \emph{IEEE Trans. Intell. Transp. Syst.}, vol.~24, no.~1, pp. 1153--1162,
  Jan. 2023.

\bibitem{arxiv26}
Z.~Weng, Z.~Qin, X.~Tao, C.~Pan, G.~Liu, and G.~Y. Li, ``Deep learning enabled
  semantic communications with speech recognition and synthesis,'' \emph{IEEE
  Trans. Wireless Commun.}, vol.~22, no.~9, pp. 6227--6240, Sep. 2023.

\bibitem{Fmeasure}
N.~Chinchor, ``{MUC-4} evaluation metrics,'' in \emph{Proceedings of the 4th
  Conference on Message Understanding}, Jun. 1992, pp. 22--29.

\bibitem{qiuyifei}
Y.~Qiu, S.~Wu, Y.~Wang, J.~Jiao, N.~Zhang, and Q.~Zhang, ``On scheduling policy
  for multiprocess cyber–physical system with edge computing,'' \emph{IEEE
  Internet Things J.}, vol.~9, no.~19, pp. 18\,559--18\,572, Oct. 2022.

\bibitem{liuwanchun}
K.~Huang, W.~Liu, Y.~Li, B.~Vucetic, and A.~Savkin, ``Optimal downlink–uplink
  scheduling of wireless networked control for industrial {IoT},'' \emph{IEEE
  Internet Things J.}, vol.~7, no.~3, pp. 1756--1772, Mar. 2020.

\bibitem{task1}
Y.~E. Sagduyu, S.~Ulukus, and A.~Yener, ``Age of information in deep
  learning-driven task-oriented communications,'' in \emph{Proc. IEEE Conf.
  Comput. Commun. Workshops (INFOCOM WKSHPS)}, May 2023, pp. 1--6.

\bibitem{task10}
S.~Xie, S.~Ma, M.~Ding, Y.~Shi, M.~Tang, and Y.~Wu, ``Robust information
  bottleneck for task-oriented communication with digital modulation,''
  \emph{IEEE J. Sel. Areas Commun.}, vol.~41, no.~8, pp. 2577--2591, Aug. 2023.

\bibitem{arxiv10}
C.~Liu, C.~Guo, Y.~Yang, and N.~Jiang, ``Adaptable semantic compression and
  resource allocation for task-oriented communications,'' \emph{IEEE Trans.
  Cogn. Commun. Netw.}, early access, Dec. 28, 2023, doi: {\color{blue} \href
  {https://ieeexplore.ieee.org/document/10375768}{10.1109/TCCN.2023.3346481}}.

\bibitem{arxiv95}
\BIBentryALTinterwordspacing
M.~Wang, J.~Li, M.~Ma, and X.~Fan, ``{SNN-SC}: A spiking semantic communication
  framework for classification,'' 2023. [Online]. Available:
  \url{http://arxiv.org/abs/2210.06836}
\BIBentrySTDinterwordspacing

\bibitem{arxiv73}
Q.~Hu, G.~Zhang, Z.~Qin, Y.~Cai, G.~Yu, and G.~Y. Li, ``Robust semantic
  communications against semantic noise,'' in \emph{Proc. IEEE 96th Veh.
  Technol. Conf. (VTC-fall)}, Sep. 2022, pp. 1--6.

\bibitem{arxiv74}
------, ``Robust semantic communications with masked {VQ-VAE} enabled
  codebook,'' \emph{IEEE Trans. Wireless Commun.}, vol.~22, no.~12, pp.
  8707--8722, Dec. 2023.

\bibitem{task11}
J.~Shao, X.~Zhang, and J.~Zhang, ``Task-oriented communication for edge video
  analytics,'' \emph{IEEE Trans. Wireless Commun.}, early access, Sep. 20,
  2023, doi: {\color{blue} \href
  {https://ieeexplore.ieee.org/document/10258036}{10.1109/TWC.2023.3314888}}.

\bibitem{arxiv45}
Q.~Pan, H.~Tong, J.~Lv, T.~Luo, Z.~Zhang, C.~Yin, and J.~Li, ``Image
  segmentation semantic communication over {Internet} of vehicles,'' in
  \emph{Proc. IEEE Wireless Commun. Netw. Conf. (WCNC)}, Mar. 2023, pp. 1--6.

\bibitem{arxiv116}
M.~Jankowski, D.~Gündüz, and K.~Mikolajczyk, ``Wireless image retrieval at
  the edge,'' \emph{IEEE J. Sel. Areas Commun.}, vol.~39, no.~1, pp. 89--100,
  Jan. 2021.

\bibitem{arxiv20}
W.~F. Lo, N.~Mital, H.~Wu, and D.~Gündüz, ``Collaborative semantic
  communication for edge inference,'' \emph{IEEE Wireless Commun. Lett.},
  vol.~12, no.~7, pp. 1125--1129, Jul. 2023.

\bibitem{task5}
S.~Wan, Q.~Yang, Z.~Shi, Z.~Yang, and Z.~Zhang, ``Cooperative task-oriented
  communication for multi-modal data with transmission control,'' in
  \emph{Proc. IEEE Int. Conf. Commun. Workshops (ICC Workshops)}, May 2023, pp.
  1635--1640.

\bibitem{task15}
A.~Mostaani, T.~X. Vu, S.~Chatzinotas, and B.~Ottersten, ``Task-oriented data
  compression for multi-agent communications over bit-budgeted channels,''
  \emph{IEEE Open J. Commun. Soc}, vol.~3, pp. 1867--1886, Oct. 2022.

\bibitem{task3}
R.~Li, C.~Huang, X.~Qin, S.~Jiang, N.~Ma, and S.~Cui, ``Coexistence between
  task- and data-oriented communications: A {Whittle}’s index guided
  multiagent reinforcement learning approach,'' \emph{IEEE Internet Things J.},
  vol.~11, no.~2, pp. 2630--2647, Jan. 2024.

\bibitem{arxiv15}
H.~Du, J.~Wang, D.~Niyato, J.~Kang, Z.~Xiong, and D.~I. Kim, ``{AI}-generated
  incentive mechanism and full-duplex semantic communications for information
  sharing,'' \emph{IEEE J. Sel. Areas Commun.}, vol.~41, no.~9, pp. 2981--2997,
  Sep. 2023.

\bibitem{arxiv65}
J.~Kang, H.~Du, Z.~Li, Z.~Xiong, S.~Ma, D.~Niyato, and Y.~Li, ``Personalized
  saliency in task-oriented semantic communications: Image transmission and
  performance analysis,'' \emph{IEEE J. Sel. Areas Commun.}, vol.~41, no.~1,
  pp. 186--201, Jan. 2023.

\bibitem{arxiv79}
W.~Xu, Y.~Zhang, F.~Wang, Z.~Qin, C.~Liu, and P.~Zhang, ``Semantic
  communication for the {Internet} of vehicles: A multiuser cooperative
  approach,'' \emph{IEEE Veh. Technol. Mag.}, vol.~18, no.~1, pp. 100--109,
  Mar. 2023.

\bibitem{VQA}
H.~Xie, Z.~Qin, and G.~Y. Li, ``Task-oriented multi-user semantic
  communications for {VQA},'' \emph{IEEE Wireless Commun. Lett.}, vol.~11,
  no.~3, pp. 553--557, Mar. 2022.

\bibitem{VQA2}
H.~Xie, Z.~Qin, X.~Tao, and K.~B. Letaief, ``Task-oriented multi-user semantic
  communications,'' \emph{IEEE J. Sel. Areas Commun.}, vol.~40, no.~9, pp.
  2584--2597, Sep. 2022.

\bibitem{newjsac5}
H.~Xie, Z.~Qin, and G.~Y. Li, ``Semantic communication with memory,''
  \emph{IEEE J. Sel. Areas Commun.}, vol.~41, no.~8, pp. 2658--2669, Aug. 2023.

\bibitem{arxiv97}
C.~Liu, C.~Guo, S.~Wang, Y.~Li, and D.~Hu, ``Task-oriented semantic
  communication based on semantic triplets,'' in \emph{Proc. IEEE Wireless
  Commun. Netw. Conf. (WCNC)}, Mar. 2023, pp. 1--6.

\bibitem{SNN}
W.~Maass, ``Network of spiking neurons: the third generation of neural network
  models,'' \emph{Neural Netw.}, vol.~10, no.~9, pp. 1659--1671, Dec. 1997.

\bibitem{arxiv8}
G.~Zhang, Q.~Hu, Z.~Qin, Y.~Cai, G.~Yu, and X.~Tao, ``A unified multi-task
  semantic communication system for multimodal data,'' \emph{IEEE Trans.
  Commun.}, early access, Feb. 9, 2024, doi: {\color{blue} \href
  {https://ieeexplore.ieee.org/document/10431795}{10.1109/TCOMM.2024.3364990}}.

\bibitem{arxiv9}
G.~Zhang, Q.~Hu, Z.~Qin, Y.~Cai, and G.~Yu, ``A unified multi-task semantic
  communication system with domain adaptation,'' in \emph{Proc. IEEE Glob.
  Commun. Conf. (GLOBECOM)}, Dec. 2022, pp. 3971--3976.

\bibitem{2011}
J.~Bao, P.~Basu, M.~Dean, C.~Partridge, A.~Swami, W.~Leland, and J.~A. Hendler,
  ``Towards a theory of semantic communication,'' in \emph{Proc. IEEE Netw.
  Sci. Workshop}, Jun. 2011, pp. 110--117.

\bibitem{1953Semantic}
Y.~Bar-Hillel and R.~Carnap, ``Semantic information,'' \emph{The British
  Journal for the Philosophy of Science}, vol.~4, no.~14, pp. 147--157, Aug.
  1953.

\bibitem{GoTmagazine}
\BIBentryALTinterwordspacing
A.~Li, S.~Wu, S.~Meng, and Q.~Zhang, ``Towards goal-oriented semantic
  communications: New metrics, open challenges, and future research
  directions,'' 2023. [Online]. Available:
  \url{http://arxiv.org/abs/2304.00848}
\BIBentrySTDinterwordspacing

\bibitem{adding3_5}
L.~Ruan, Y.~Ma, H.~Yang, H.~He, B.~Liu, J.~Fu, N.~J. Yuan, Q.~Jin, and B.~Guo,
  ``{MM}-diffusion: Learning multi-modal diffusion models for joint audio and
  video generation,'' in \emph{Proc. IEEE Conf. Comput. Vis. Pattern Recognit.
  (CVPR)}, Jun. 2023, pp. 10\,219--10\,228.

\bibitem{task17}
D.~Wen, P.~Liu, G.~Zhu, Y.~Shi, J.~Xu, Y.~C. Eldar, and S.~Cui, ``Task-oriented
  sensing, computation, and communication integration for multi-device edge
  {AI},'' \emph{IEEE Trans. Wireless Commun.}, vol.~23, no.~3, pp. 2486--2502,
  Mar. 2024.

\bibitem{task19}
P.~Liu, G.~Zhu, S.~Wang, W.~Jiang, W.~Luo, H.~V. Poor, and S.~Cui, ``Toward
  ambient intelligence: Federated edge learning with task-oriented sensing,
  computation, and communication integration,'' \emph{IEEE J. Sel. Topics
  Signal Process.}, vol.~17, no.~1, pp. 158--172, Jan. 2023.

\bibitem{newjsac2}
X.~Mu and Y.~Liu, ``Exploiting semantic communication for non-orthogonal
  multiple access,'' \emph{IEEE J. Sel. Areas Commun.}, vol.~41, no.~8, pp.
  2563--2576, Aug. 2023.

\end{thebibliography}

	\ifCLASSOPTIONcaptionsoff
	\newpage
	\fi
\end{document}